%% file: main.tex
\documentclass[pdflatex,sn-mathphys-ay]{sn-jnl}

\usepackage{graphicx}%
\usepackage{multirow}%
\usepackage{amsmath,amssymb,amsfonts}%
\usepackage{amsthm}%
\usepackage{mathrsfs}%
\usepackage[title]{appendix}%
\usepackage{xcolor}%
\usepackage{textcomp}%
\usepackage{manyfoot}%
\usepackage{booktabs}%
\usepackage{algorithm}%
\usepackage{algorithmicx}%
\usepackage{algpseudocode}%
\usepackage{listings}%
\usepackage{placeins}%
\usepackage{hyperref}%

\algrenewcommand\algorithmicrequire{\textbf{Input:}}
\algrenewcommand\algorithmicensure{\textbf{Output:}}
\DeclareMathAlphabet{\mathbbold}{U}{bbold}{m}{n}
\graphicspath{{tmpplt/}}
\DeclareUnicodeCharacter{03B3}{\ensuremath{\gamma}}

\begin{document}
\title[Article Title]{Flexible Transformations for Bayesian Score Calibration}


\author*[1, 2]{\fnm{Adam} \sur{Bretherton}}\email{adam.bretherton@hdr.qut.edu.au}

\author[3]{\fnm{Joshua} \sur{J. Bon}}\email{joshua.bon@adelaide.edu.au}

\author[1, 2, 4]{\fnm{David} \sur{J. Warne}}\email{david.warne@qut.edu.au}

\author[5]{\fnm{David} \sur{J. Nott}}\email{standj@nus.edu.sg}

\author[1, 2, 4]{\fnm{Christopher} \sur{Drovandi}}\email{c.drovandi@qut.edu.au}

\affil*[1]{\orgdiv{School of Mathematical Sciences, Faculty of Science}, \orgname{Queensland University of Technology}, \orgaddress{\city{Brisbane}, \country{Australia}}}

\affil*[2]{\orgdiv{Centre for Data Science}, \orgname{Queensland University of Technology}, \orgaddress{\city{Brisbane}, \country{Australia}}}

\affil[3]{\orgdiv{School of Mathematical Sciences}, \orgname{Adelaide University}, \orgaddress{\city{Adelaide}, \country{Australia}}}

\affil[4]{\orgdiv{Centre of Excellence for the Mathematical Analysis of Cellular Systems}, \orgname{Queensland University of Technology}, \orgaddress{\city{Brisbane}, \country{Australia}}}

\affil[5]{\orgdiv{Department of Statistics and Data Science}, \orgname{National University of Singapore}, \country{Singapore}}

\abstract{Modern statistical models are growing increasingly complex in an effort to realistically capture system dynamics.  Using standard simulation-based inference, these models may be computationally prohibitive, necessitating the use of model calibration methods.  Bayesian score calibration is a computationally efficient framework for model calibration with strong theoretical guarantees.  This framework learns an appropriate correction for an approximate model using a small number of simulations from the data-generating process.  Currently, only a location-scale transformation has been explored, which may lack the flexibility to correct the complex error introduced by some approximate models.  In this paper, we develop two flexible transformations for use in the Bayesian score calibration framework.  The first is a polynomial extension, which can appropriately adjust approximate models with location-varying error.  The second is a sequential application of Bayesian score calibration, which can accommodate approximate models with posteriors that have low support for the true parameter values.  We also discuss an additional diagnostic for use with this framework.  We demonstrate the increased flexibility these two approaches provide over Bayesian score calibration in two illustrative simulation studies.}
\keywords{likelihood-free inference, simulation-based inference, scoring rules, posterior
correction, surrogate model, model calibration}

\maketitle

\section{Introduction}\label{sec::Intro}
    Modelling the complex dynamics of real systems is a central challenge across many scientific fields, for example, fluid dynamics \citep{Vinuesa2022}, cosmology \citep{MishraSharma2022}, and biology \citep{Barbers2024, Dingeldein2025}.  For sufficiently complex models, the likelihood function may be computationally prohibitive to evaluate or impossible to express analytically.  A common approach to enable Bayesian inference under such settings is to use \textit{simulation-based inference} (SBI).  SBI methods avoid explicitly evaluating the likelihood by instead simulating from the \textit{data-generating process} (DGP) and making comparisons with the observed data.  Popular SBI approaches include approximate Bayesian computation \citep[e.g.,][]{Sisson2018} and neural SBI \citep[e.g.,][]{Cranmer2020}.  Despite their flexibility, SBI methods typically require a significant number of model simulations \citep{Lueckmann2021}, and often only provide an approximation to the target posterior \citep{Hermans2022}.  Additionally, for complex systems, even a single simulation from the DGP can require significant computational resources \citep{Zenke2014, Hoppe2021}.

    An alternative approach to standard SBI techniques is to consider model calibration with the \textit{Bayesian score calibration} framework \citep[BSC,][]{Bon2025}.  BSC indirectly samples the target distribution by transforming samples from a computationally efficient approximate posterior distribution.  There are several convenient choices for forming an approximate posterior distribution.  One option is to approximate the complex model with a tractable surrogate model \citep[e.g.,][]{Rynn2019, Dyer2023}.  For a certain class of models, we can replace the intractable likelihood function with a tractable pseudo-likelihood or composite likelihood \citep{Varin2011, Gruner2023}.  Alternatively, we might consider using a short run of a standard SBI technique.  The chosen approximate posterior distribution is likely to exhibit bias or poor uncertainty quantification with respect to the target distribution \citep{Warne2021}, which the transformation within BSC aims to overcome.


    To learn a transformation that appropriately corrects the approximate posterior distribution in a computationally efficient manner, BSC uses a small number of expensive model simulations to generate training data.  Specifically, these training data consist of pairs of a `true' sample and samples from an approximate posterior distribution.  Each approximate posterior is conditioned on the calibration data set that is simulated using the DGP and its associated `true' sample.  BSC learns a transformation by optimising a strictly proper scoring rule over the training data, with the transformation applied to the approximate posterior samples.  Despite the general theory supporting the BSC framework, \citet{Bon2025} only explore a location-scale transformation.  This transformation provides a single correction across the parameter space and may not be sufficiently flexible to capture the complex differences between the target and approximate posterior distributions.
    
    In this work, we develop a richer class of transformations for use in the BSC framework, allowing BSC to produce more accurate posterior approximations on a wider range of problems.  In particular, we consider two flexible transformations; a polynomial extension of the location-scale transformation in \citet{Bon2025} and a sequential application of BSC.  The polynomial transformation can calibrate approximate models where the behaviour of the error varies across the parameter space by including polynomial terms that depend on the location of the sample being transformed.  Alternatively, the sequential application of BSC chains together multiple applications of the location-scale transformation, which we find can lead to more accurate posterior approximations, particularly when the initial posterior approximation is poor and does not cover the true parameter value.  The extra flexibility these two transformations provide permits calibration samples to be drawn from less informative distributions, for example, the prior distribution for amortised inference, or from approximate posteriors that are biased and overconfident.  We demonstrate the performance of both proposed transformations with two simulated examples.

    The remainder of this paper proceeds as follows.  In Section~\ref{sec:BG}, we provide background, including an overview of related calibration methods.  We introduce our two flexible extensions and discuss calibration diagnostics in Section~\ref{sec::NM}.  Section~\ref{sec:SS} contains the simulation studies and their results.  We conclude with a discussion on the strengths and limitations of the two proposed transformations, and potential directions for future work.

\section{Background}\label{sec:BG}
    We are interested in sampling from the target posterior distribution~$\Pi(\,\cdot\, | y)$ with corresponding probability density
    \begin{equation*}\label{eq:bayes}
        \pi(\theta |y) \propto p(y|\theta) \pi(\theta),   
    \end{equation*}
    for $\theta \in \Theta$, where $p(y|\theta)$ is the likelihood function for the observed data $y$ and $\pi$ is the prior density.  Under this setting, $\theta$ are the unknown parameters of interest, where $\Theta \subseteq \mathbb{R}^d$ is the parameter space and $d$ is the number of parameters.  For many complex statistical models, evaluating the likelihood function is not feasible, making standard Bayesian inference approaches, such as Markov chain Monte Carlo \citep[MCMC,][]{Brooks2011}, inaccessible.  Further, SBI methods can be computationally prohibitive when simulating from the complex model of interest is computationally costly \citep{Zenke2014, Hoppe2021}.  We therefore turn to model calibration methods that attempt to combine an approximate posterior distribution $\tilde{\Pi}(\,\cdot\,|y)$, which is a computationally efficient approximation of $\Pi(\,\cdot\,|y)$ over the same parameter space $\Theta$, with a small number of DGP simulations in an attempt to achieve fast and accurate inference for complex statistical models.

    Model calibration methods can help mitigate the error induced by approximating the inference technique or model.  Such calibration methods attempt to reduce this approximation error by adjusting the posterior distribution so that it is calibrated.  Several notions of calibration exist in the statistical literature; we focus on calibration as defined by \citet{Cockayne2022}.  That is, the posterior distribution $\Pi(\,\cdot\,|y)$ is calibrated to the DGP $P(\,\cdot\,|\theta)$ and prior distribution $\Pi$ when the data-generating parameters $\theta'$ are equivalent to samples from $\Pi(\,\cdot\,|y)$ where  
    \begin{equation*}
        \begin{split}
            \theta' &\sim \Pi, \\
            \ y &\sim P(\,\cdot\,|\theta'), \\
            \theta &\sim \Pi(\,\cdot\,|y).
        \end{split}
    \end{equation*}
    To test if a posterior is calibrated, we can consider the probability integral transform \citep{Dawid1984, Diebold1998}, with diagnostics such as simulation-based calibration checking \citep[SBCC,][]{Cook2006, Talts2018, Modrak2025}, or confirm that credible sets from the posterior distribution achieve nominal coverage, with diagnostics such as those introduced in \citet{Prangle2014}.
    
    The BSC framework \citep{Bon2025} adjusts samples from the approximate posterior distribution $\tilde{\theta} \sim \tilde{\Pi}(\,\cdot\,|y)$ such that they are calibrated using a deterministic transformation \mbox{$f:\Theta\to\Theta$}.  To learn~$f$, BSC uses $P(\,\cdot\,|\theta)$, $\tilde{\Pi}(\,\cdot\,|y)$, and a calibration distribution $\overline{\Pi}$, which focuses calibration on suitable regions of the parameter space, for example~$\tilde{\Pi}(\,\cdot\,|y)$ or the prior.  We use analogous notation for the observed data $y$, approximate data $\tilde{y} \sim P(\,\cdot\,| \tilde{\theta})$ and calibration data $\bar{y} \sim P(\,\cdot\, | \bar{\theta})$, where $\tilde{\theta}$ and $\bar{\theta}$ are samples from the approximate and calibration distributions, respectively.  BSC utilises both importance sampling \citep{Kahn1951, Kloek1978} and scoring rules \citep{Gneiting2007}; therefore, we provide a brief overview of both. 

    \subsection{Importance Sampling}
        Importance sampling is a Monte Carlo technique that can be used to approximate a target distribution by instead sampling from an importance distribution.  In the case of posterior sampling, we choose some $\overline\Pi$ instead of $\Pi(\,\cdot\,|y)$, which is easier to sample from.  Each sample $\theta \sim \overline\Pi$ can then be weighted with a correction~$r(\theta)$ so that the samples reflect $\Pi(\,\cdot\,|y)$ \citep{Rubinstein2016}.  When $\Pi(\,\cdot\,|y)$ and $\overline\Pi$ admit compatible densities $\pi(\,\cdot\,|y)$ and $\bar\pi$, the appropriate importance correction~$r(\theta)~=~\frac{\pi(\theta|y)}{\overline{\pi}(\theta)}$ is based on the identity
        \begin{equation} \label{eq:isid}
            \int_\Theta h(\theta)\pi(\theta|y) \mathrm{d}\theta = \int_\Theta h(\theta)\frac{\pi(\theta|y)}{\overline\pi(\theta)}\overline\pi(\theta) \mathrm{d}\theta = \int_\Theta h(\theta)r(\theta)\overline\pi(\theta) \mathrm{d}\theta,
        \end{equation}
        where $h(\theta)$ is some quantity of interest.  It follows from Eq. \eqref{eq:isid} that a necessary condition for importance sampling is that the support of $\overline\Pi$ contains the support of $\Pi(\,\cdot\,|y)$.  
        
        Using the importance correction, we can weight a set of $N$ samples from the importance distribution $\{({r(\theta^{(n)})}, \theta^{(n)})\}_{n=1}^N,$ with $\theta^{(n)} \sim \overline\Pi$, which can be used to estimate the density $\pi(\,\cdot\,|y)$ or to evaluate the Monte Carlo estimate of Eq. \eqref{eq:isid} as 
        \begin{equation}\label{eq:ismc}
            \mathbb{E}_{\theta \sim\Pi(\,\cdot\,|y)}\left[h(\theta)\right] = \mathbb{E}_{\theta \sim\overline\Pi}\left[h(\theta)r(\theta)\right] \approx \frac{1}{N}\sum_{n=1}^N h(\theta^{(n)})r(\theta^{(n)}).
        \end{equation}
        When $\overline\Pi$ and $\Pi(\,\cdot\,|y)$ are known only up to a multiplicative constant, the importance correction can be self-normalised by noting $\mathbb{E}_{\theta\sim\overline\Pi}[r(\theta)] = 1$.  Although Eq. \eqref{eq:ismc} is asymptotically unbiased when~$r(\theta)$ is self-normalised \citep{Chopin2020}, how closely the weighted samples reflect~$\Pi(\,\cdot\,|y)$ still depends on the variance of $r(\theta)$.  Unfortunately, the variance of $r(\theta)$ might be large or infinite when the probability density of $\theta$ under the importance distribution is much less than its probability density under the target distribution for any $\theta \in \Theta$.  
        
        
        


    \subsection{Scoring Rules}
        Scoring rules $S(U, \theta)$ are useful for evaluating the quality of a single sample $\theta$ against a predictive distribution $U$.  When we wish to evaluate the quality of a predictive distribution $V$, which is in the same class of probability distributions $\mathcal{P}$ as $U$, we can take the expectation of $S(U, \theta)$ over $V$ with
        \begin{equation*}
            S(U, V) := \mathbb{E}_{\theta\sim V}[S(U, \theta)]. 
        \end{equation*}
        A scoring rule is said to be \textit{proper} if $S(V, V) \geq S(U, V)$ for all $U, V \in \mathcal{P}$, and \textit{strictly proper} when equality holds if and only if $V = U$ \citep{Gneiting2007}.  The BSC framework can accommodate any strictly proper scoring rule, for example \citet{Bon2025} consider the energy score \citep{Gneiting2007} that is given by
        \begin{equation}\label{eq:ES}
            S(U, \theta) = \frac{1}{2}\mathbb{E}_{u, u' \sim U}||u-u'||_2^\beta - \mathbb{E}_{u\sim U}||u - \theta||_2^\beta,
        \end{equation}
        where $u$ and $u'$ are independent realisations of $U$, $\beta \in (0, 2)$, and $||\cdot ||_2$ is the L2 norm.  The energy score has several properties that make it attractive for use in the BSC framework.  First, the energy score is a strictly proper scoring rule when $\mathbb{E}_{u\sim U}||u||_2^\beta$ is finite.  Second, the energy score naturally facilitates Monte Carlo approximation.  Consider an empirical probability distribution~$\hat{U}~=~\frac{1}{N}\sum_{i=1}^N\delta_{u^{(i)}}$, where $\delta_{u^{(i)}}$ is the Dirac delta function at $u^{(i)}$, then the energy score is 
        \begin{equation}\label{eq:appES}
            S(\hat{U}, \theta) = \frac{1}{2}\frac{1}{N^2}\sum_{i=1}^N\sum_{j=1}^N||u^{(i)} - u^{(j)}||^\beta_2 - \frac{1}{N}\sum_{i=1}^N||u^{(i)} - \theta||^\beta_2.
        \end{equation}
        Finally, the energy score is a multivariate generalisation of the \textit{continuous ranked probability score} (CRPS) \citep{Matheson1976}.  The CRPS is a widely used univariate scoring rule that is a special case of Eq. \eqref{eq:ES} when~$\beta~=~1$ and $U$ has dimension $1$ \citep{Baringhaus2004, Szekely2005}.  Of particular interest to calibration, the CRPS can be expressed as the integral over all $\alpha \in (0, 1)$ level quantiles for the quantile score \citep{Koenker1999}
        \begin{equation*}\label{eq:scoreCov}
            \text{CRPS}(U, \theta) = 2\int_0^1\text{QS}_\alpha(U, \theta)d\alpha,
        \end{equation*}
        where QS$_\alpha(U, \theta)$ is the $\alpha$ level quantile score \citep{Gneiting2011, Ehm2016, BenBouallegue2018}. 
    
    \subsection{Bayesian Score Calibration}\label{subsec:BSC} 
        BSC uses a deterministic transform $f$, from some family of transformations $\mathcal{F}$, to calibrate samples~\mbox{$\tilde{\theta} \sim \tilde{\Pi}(\,\cdot\,|y)$} such that they are approximately distributed according to $\Pi(\,\cdot\,|y)$.  We represent the distribution of a transformed random variable as $f_\sharp\tilde{\Pi}(\,\cdot\,|y)$ to be the law of the transformed random variable $f(\tilde\theta)$ for $\tilde\theta \sim \tilde\Pi(\,\cdot\,|y)$.  In practice, a sufficiently expressive $\mathcal{F}$ that is computationally efficient to optimise over can be difficult to specify, necessitating the exploration of flexible parametric transformations.  For example, \citet{Bon2025} only explore location-scale transformation types.
        

        To learn an appropriate $f$ that reduces the error induced by $\tilde\Pi(\,\cdot\,|y)$, BSC first uses calibration samples $\bar\theta \sim \overline{\Pi}$ and the DGP to generate calibration data sets $\bar{y} \sim P(\,\cdot\,|\bar{\theta})$.  Next, for each~$\bar{y}$, an approximate posterior distribution $\tilde\Pi(\,\cdot\,|\bar{y})$ is estimated.  Finally, the transformed approximate posterior distribution $f_\sharp\tilde\Pi(\,\cdot\,|\bar{y})$ is compared with the calibration sample $\bar\theta$ that generated $\bar{y}$ using a strictly proper scoring rule $S(f_\sharp\tilde\Pi(\,\cdot\,|\bar{y}), \bar\theta)$.  The corresponding optimisation problem is given by
        \begin{equation}\label{eq:BSCOb}
            f^* = \underset{f \in \mathcal{F}}{\text{arg max}} \ \mathbb{E}_{\bar\theta \sim \overline{\Pi}}\mathbb{E}_{\bar{y}\sim P(\,\cdot\,|\bar\theta)}\left[w\left(\bar\theta,\bar{y}\right)S\left(f_\sharp\tilde\Pi(\,\cdot\,|\bar{y}), \bar\theta\right)\right],
        \end{equation}
        where $w(\bar\theta, \bar{y})$ is a weighting function. 
        
        The weighting function $w(\bar\theta,\bar{y}) = r(\bar\theta)v(\bar{y})$ consists of an importance correction $r(\bar\theta)$ and a stabilising function~$v(\bar{y})$.  The BSC framework concentrates calibration around suitable regions of the parameter space using $\overline\Pi$, which can be conditioned on the observed data $y$.  The calibration distribution is an importance distribution that requires an importance correction $r(\bar\theta)=\frac{\pi(\bar\theta)}{\overline{\pi}(\bar\theta)}$, where~$\pi$ is the prior density function.  \citet{Bon2025} provide theoretical justification for approximating~$w(\bar\theta,\bar{y})$ with unit weights~$\hat{w} = 1$.  In particular, they show for the correct choice of~$v(\bar{y})$ that~$\hat{w} = 1$ is an asymptotically consistent estimator of $w(\bar\theta,\bar{y})$.  Therefore,~$w(\bar\theta,\bar{y})~=~\hat{w}~=~1$ can be chosen without directly specifying $v(\bar{y})$.

        In practice, BSC learns a transformation by first drawing~$M$ calibration samples from the calibration distribution $\{\bar{\theta}^{(m)}\}_{m=1}^M \sim \overline{\Pi}$, typically $\overline\Pi$ is chosen to be $\tilde\Pi(\,\cdot\,|y)$ with inflated variance.  For each calibration sample $\bar{\theta}^{(m)}$, a calibration data set~$\bar{y}^{(m)}\sim P(\,\cdot\, |\bar{\theta}^{(m)})$ is simulated using the DGP.  Then, an associated set of $N$ approximate samples is drawn from the approximate posterior distribution~$\{\tilde{\theta}^{(n,m)}\}_{n=1}^N \sim \tilde\Pi(\,\cdot\,|\bar{y}^{(m)})$.  Finally, using unit weights and the energy score with $\beta=1$, Eq. \eqref{eq:BSCOb} can be optimised over the set of transformed approximate samples and each associated calibration sample $\{(\{f(\tilde{\theta}^{(n, m)})\}_{n=1}^N, \bar{\theta}^{(m)})\}_{m=1}^M$, with the updated optimisation problem given by 
        \begin{equation}
            \label{eq:BSCemob}
            f^* = \underset{f \in \mathcal{F}}{\text{arg max}} \ \frac{1}{M}\sum_{m=1}^M\left(\frac{1}{2}\frac{1}{N^2}\sum_{i=1}^N\sum_{j=1}^N||f(\tilde\theta^{(i, m)}) - f(\tilde\theta^{(j,m)})||_2 - \frac{1}{N}\sum_{i=1}^N||f(\tilde\theta^{(i,m)}) - \bar{\theta}^{(m)}||_2\right).
        \end{equation}
        After computing $f^*(\,\cdot\,)$ from Eq. \eqref{eq:BSCemob}, $f^*(\,\cdot\,)$ can be applied to samples from the approximate posterior distribution~$\theta~=~f^*(\tilde\theta)~\sim~f^*_\sharp\tilde\Pi(\,\cdot\,|y)$ so that the samples $\theta$ approximately reflect $\Pi(\,\cdot\,|y)$.

        The family of transformation functions explored by \citet{Bon2025} is parameterised by a location adjustment $b \in \mathbb{R}^d$ and a scale adjustment $A \in \mathbb{R}^{d\times d}$.  Here $A=VD^{1/2}$, which follows from the eigenvalue decomposition of the positive definite matrix $C=VDV^T=AA^T$ for a strictly positive diagonal matrix $D$.  The resulting transformation function is given by
        \begin{equation}\label{eq:famxform}
            f(\theta, \hat{\mu}(y); A, b) = A(\theta - \hat{\mu}(y)) + \hat{\mu}(y) + b, 
        \end{equation}
        where $\hat{\mu}(y)$ is the sample mean computed from the set of $N$ samples being transformed~\mbox{$\{\theta_i\}_{i=1}^N\sim\tilde\Pi(\,\cdot\,|\bar{y})$}.  To ensure~$C$ remains positive definite during optimisation, the property~$VV^T=I$ can be enforced, with $I$ being the identity matrix.  We provide an overview of learning the location-scale transformation in Eq. \eqref{eq:famxform} under the BSC framework, using unit weights and the energy score with $\beta = 1$, in Algorithm \ref{alg:BSC}.  Unfortunately, the location-scale transformation shown in Eq. \eqref{eq:famxform} only adjusts each of the approximate samples by a relative amount and may not be flexible enough to appropriately calibrate samples from some approximate posterior distributions.  We address this limitation in Section \ref{sec::NM}, where we propose two flexible transformations for use in the BSC framework.

        \begin{algorithm}
            \caption{Learning the location-scale transformation under the Bayesian score calibration framework of \citet{Bon2025} using unit weights and the energy score with $\beta = 1$.}
            \label{alg:BSC}
            \begin{algorithmic}
                \Require{Number of calibration data sets $M$, number of approximate posterior samples $N$, calibration distribution $\overline\Pi$, approximate distribution $\tilde\Pi(\,\cdot\,|y)$.}
                \Ensure{Parameters $A$ and $b$ for the estimated optimal transformation function $f^*(\,\cdot\,)$.}
                \For{$m \in \{1, \ldots, M\}$}
                    \State Draw calibration sample, $\bar\theta^{(m)} \sim \overline\Pi$.
                    \State Simulate calibration data set, $\bar{y}^{(m)} \sim P(\,\cdot\,|\bar\theta^{(m)})$.
                    \State Sample from the approximate posterior distribution $\tilde\theta^{(n, m)} \sim \tilde{\Pi}(\,\cdot\,|\bar{y}^{(m)})$ for $n \in \{1, \ldots, N\}$.
                \EndFor
                \State Find $A$ and $b$ by computing Eq. \eqref{eq:BSCemob} with $\{(\{\tilde{\theta}^{(n, m)}\}_{n=1}^N, \bar{\theta}^{(m)})\}_{m=1}^M$. \\
                \Return $A$ and $b$ for the estimated optimal transformation $f^*(\,\cdot\,)$.
            \end{algorithmic}
        \end{algorithm}
        \FloatBarrier
        
    \subsection{Related Calibration Methods}
        Similar to BSC, \citet{Cai2026} directly calibrate samples from the approximate model in order to achieve a better approximation of the true posterior.  They propose a method that calibrates the approximate posterior distribution using a fixed-width scaling adjustment.  To learn the adjustment, this method takes advantage of the observation that the prior and the data-averaged posterior are self-consistent under SBCC \citep{Cook2006}.  Specifically, they expand the uniformity check in SBCC to confirm instead that the transformed samples have appropriate coverage or that the first two moments of their $z$-scores are 0 and~1, respectively.  This method requires a similar number of simulations from the DGP as BSC. However, the proposed fixed-width scaling does not account for a location adjustment, which can lead to adjusted posteriors that overinflate their variance.  Additionally, the calibration process they propose only calibrates each marginal parameter independently and therefore does not account for any posterior correlation structure between parameters.

        An alternative calibration approach is to learn a post-process adjustment to improve the uncertainty quantification of the approximate posterior.  There are several examples of this post-process adjustment. For example, \citet{Yu2021} learn a location-scale transformation using the \textit{tower property of conditional expectation} (TPCE) and the \textit{law of total variance} (LTV) \citep{Blitzstein2019}.  Specifically, they relate the mean and covariance of the prior and approximate posterior using the TPCE and LTV, respectively.  They propose a location-scale transformation in which the location adjustment is found using the left-hand and right-hand sides of the TPCE.  Similarly, the scale adjustment uses a Cholesky decomposition of the left-hand and right-hand sides of the LTV.  Such a transformation does not have the same calibration guarantees as BSC and cannot correct for complex differences between the true and approximate posteriors. 

        \citet{Rodrigues2018} adjust each marginal \textit{cumulative distribution function} (CDF) based on the coverage property diagnostics introduced in \citet{Prangle2014}.  They only apply their methods when approximate Bayesian computation is used to approximate the posterior.  However, their approach can calibrate samples from other approximate posteriors.  Relatedly, \citet{Xing2020} learn an approximate distortion map to transform each marginal CDF of the approximate posterior.  Both of these methods require a significant number of simulations from the expensive DGP (typically more than 10000) \citep{Lueckmann2021}.  Additionally, these methods apply the post-process calibration to the marginal CDFs, which may be numerically unstable when the true data-generating parameter value has low posterior support under the approximate posterior.  The adjusted approximate sample will be infinite when the marginal CDF is numerically either 0 or 1.
        
        \citet{Vandeskog2024} propose a post-process correction when the approximate posterior is formed.  To correct this misspecification, they apply a scale transformation to approximate posterior samples centred around a suitable point estimate, for example, the mode of the approximate posterior.  They choose this transformation such that the transformed approximate samples are asymptotically Gaussian with covariance given by the inverse Godambe sandwich information matrix \citep{Godambe1960, Godambe2010}.  Despite the improved coverage of their transformed approximate samples, their method requires an analytical form for the computationally intractable likelihood of the true model, which is generally not achievable for the complex models of interest.  A similar kind of adjustment is applied in the context of \textit{Bayesian synthetic likelihood} (BSL), which uses a normal approximation of the model summary statistic likelihood in SBI.  Deliberately misspecifying the covariance matrix of the normal approximation can improve the computational efficiency of BSL, but the resulting credible intervals will not have the correct level of frequentist coverage.  \citet{Frazier2024} develop an adjustment to the misspecified BSL posterior to achieve valid uncertainty quantification.

        Rather than adjusting the samples directly, calibration methods can be used to adjust credible regions to improve uncertainty quantification.  \citet{Lee2019} and \citet{Xing2019} develop a computational framework for estimating the true coverage of approximate credible sets.  To estimate the true coverage, they generate training data using the DGP and an approximate model.  However, their framework does not directly adjust the approximate posterior samples and only corrects the approximate posterior coverage, which is used as a diagnostic tool.
        
        Conformal prediction \citep{Angelopoulos2023} can also be used to calibrate the credible regions of the approximate posterior.  \citet{Cabezas2025} propose two calibration approaches to learn a mapping between conformal scores and the cutoff values for $1-\alpha$ credible regions.  The first partitions the data using a regression tree \citep{Cabezas2024}, providing finite-sample local coverage guarantees.  The second uses a CDF-based conformity score \citep{Dheur2025} to achieve asymptotic conditional coverage.  \citet{Cabezas2025} show that both calibration approaches improve the uncertainty quantification of posteriors found using SBI methods.  Neither approach adjusts the posterior samples and only produces calibrated credible regions.
    
\section{More Flexible Transformations for Bayesian Score Calibration}\label{sec::NM}
    The quality of posterior correction found by BSC depends on the richness of the family of transformation functions $\mathcal{F}$.  The choice of $\mathcal{F}$ must balance the expressiveness of the family with the computational cost of searching the space for an optimal solution.  For a sufficiently flexible~$\mathcal{F}$, BSC can draw calibration samples directly from the prior distribution, enabling amortised inference \citep{Li2026}.  As a result, calibration on new data sets can be performed rapidly without re-optimising the transformation.  Additionally, no importance sampling approximation is required when drawing calibration samples from the prior distribution.  Despite the appeal of amortised inference, substantially more simulations may be required to train a flexible transformation, leading to high computational costs.  We seek to increase the flexibility of the transformations used in \citet{Bon2025}, without requiring a significant increase in the number of model simulations.

    To balance these competing demands, we consider two flexible transformations for use in BSC.  First, we detail a \textit{polynomial transformation} (PBSC) in Section~\ref{subsec:poly}.  This transformation can calibrate approximate models where the behaviour of the approximation error depends on the parameter values.  In Section~\ref{subsec:sbsc}, we explore a \textit{sequential application of BSC} (SBSC).  SBSC can calibrate approximate models even when the true parameter value is not in the high-probability region of the initial approximation (calibration distribution).  To better evaluate the quality of adjusted posteriors produced by these transformations in practice, we also discuss two diagnostics in Section~\ref{subsec:diag}.  The first diagnostic is the miscoverage plot explored in \citet{Bon2025}.  The second diagnostic we propose visualises the change in credible intervals for the approximate and adjusted posteriors, across the space of $\overline\Pi$.
    
    \subsection{Polynomial Bayesian Score Calibration}\label{subsec:poly}
        The first transformation we propose extends Eq. \eqref{eq:famxform} by updating the bias correction to depend on the location of the sample that is being transformed.  We propose expanding the bias correction with a polynomial of order $I \in \mathbb{N}$.  Incorporating this polynomial extension into Eq.~\eqref{eq:famxform} yields the following transformation
        \begin{equation}\label{eq:polytf}
            f(\theta, \hat{\mu}(y); A, \{B_i\}_{i=0}^I) = A(\theta - \hat{\mu}(y)) + \hat{\mu}(y) + \sum_{i=0}^I  B_i\theta^i,
        \end{equation}
        where $B_i$ is a diagonal matrix with the elements of each bias adjustment along its main diagonal.  Notably, when $I=0$, we recover the original location-scale transformation from Eq. \eqref{eq:famxform}.  To learn the parameters of Eq. \eqref{eq:polytf}, we solve the optimisation problem defined in Eq. \eqref{eq:BSCemob}.  The resulting transformation is then applied to the approximate posterior distribution~$\theta~=~f^*(\tilde\theta)~\sim~f^*_\sharp\tilde\Pi(\,\cdot\,|y)$.

\begin{figure}[h]
            \centering{
            \resizebox{0.95\textwidth}{!}{ 
                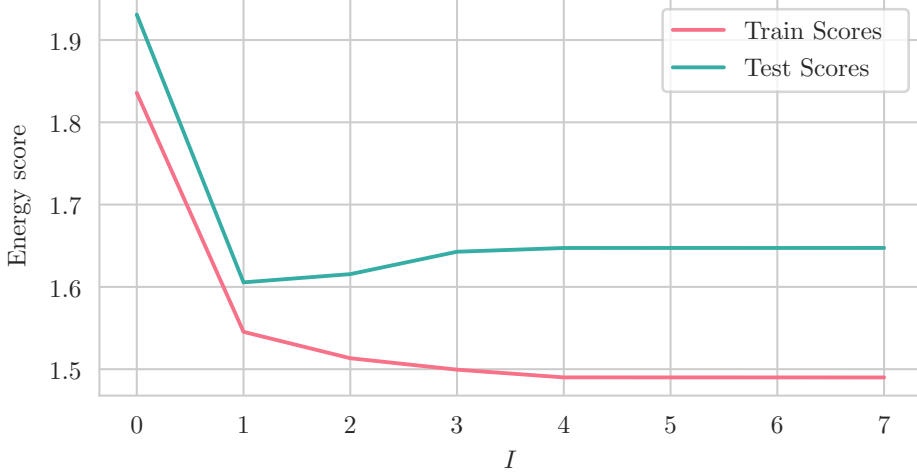
                }
            }
            \caption{Comparison of the energy scores for the training data set (Train Scores) and the test data set (Test Scores) as the order of the polynomial transformation increases.  We consider the polynomial transformation up to $I=7$ (PBSC $0$--$7$), where PBSC $0$ corresponds to standard BSC.  The energy scores are evaluated on the Ornstein-Uhlenbeck example we explore in Section~\ref{subsec:OU}.}
            \label{fig:espoly}
        \end{figure}

        The choice of $I$ is problem-dependent; thus, we suggest an adaptive approach.  This approach initialises at $I=0$ and iteratively increases $I$, while the energy score between the adjusted posteriors and their associated calibration sample continues to improve.  To enhance computational efficiency, we warm-start the optimisation at each iteration by initialising with the learned parameters from the previous iteration.  We suggest a stopping criterion based on the \textit{energy score for a test data set} (Test Scores).  This test data set is generated independently from the calibration distribution and is given by $\{(\bar\theta^{(m)}_\text{test}, \bar y^{(m)}_\text{test}, \{\tilde\theta^{(i, m)}_\text{test}\}_{i=1}^N)\}_{m=1}^M$ with $\bar\theta^{(m)}_\text{test} \sim \overline\Pi, \bar y^{(m)}_\text{test} \sim P(\,\cdot\,|\bar\theta^{(m)}_\text{test})$ and~$\tilde\theta^{(i, m)}_\text{test} \sim \tilde\Pi(\,\cdot\,|\bar y^{(m)}_\text{test})$.  We recommend stopping the incrementation of $I$ once the Test Scores between iterations start to increase.  However, the computational cost of generating a separate test data set may be prohibitive.  In this case, we instead recommend stopping once the \textit{energy score for the training data set} (Train Scores) stops substantially improving.  Figure~\ref{fig:espoly} displays the Train Scores and the Test Scores as we increase the polynomial order $I \in \{0, 1, \ldots, 7\}$ for the example we explore in Section~\ref{subsec:OU}.  This figure shows that after an initial sharp decrease, the change in energy score is minimal.  After order~$1$ of our polynomial transformation (PBSC 1), the Train Scores continue to decrease while the Test Scores start increasing.  This divergence indicates that higher orders are overfit for this example.  Algorithm~\ref{alg:BSCpoly} provides an overview of our adaptive approach to learning the polynomial transformation in Eq.~\eqref{eq:polytf} under the BSC framework.  It uses a test data set, unit weights, and the energy score with $\beta = 1$.

        \begin{algorithm}
            \caption{Learning the polynomial transformation under the Bayesian score calibration framework of \citet{Bon2025} using a test data set, unit weights, and the energy score with $\beta = 1$.}
            \label{alg:BSCpoly}
            \begin{algorithmic}
                \Require{Number of calibration data sets $M$, number of approximate posterior samples $N$, calibration distribution $\overline\Pi$, approximate distribution $\tilde\Pi(\,\cdot\,|y)$.}
                \Ensure{Parameters $A_I$ and $\{B_i\}_{i=0}^I$ for the estimated optimal transformation function $f^*(\,\cdot\,)$.}
                \For{$m \in \{1, \ldots, M\}$}
                    \State Draw calibration samples, $\bar\theta^{(m)}_{\text{train}}, \bar\theta^{(m)}_{\text{test}} \sim \overline\Pi$.
                    \State Simulate calibration data sets, $\bar{y}^{(m)}_{\text{train}} \sim P(\,\cdot\,|\bar\theta^{(m)}_{\text{train}}), \bar{y}^{(m)}_{\text{test}} \sim P(\,\cdot\,|\bar\theta^{(m)}_{\text{test}})$.
                    \State Sample from the approximate posterior distributions $\tilde\theta^{(n, m)}_{\text{train}} \sim \tilde{\Pi}(\,\cdot\,|\bar{y}^{(m)}_{\text{train}})$, $\tilde\theta^{(n, m)}_{\text{test}} \sim \tilde{\Pi}(\,\cdot\,|\bar{y}^{(m)}_{\text{test}})$
                    \State for $n \in \{1, \ldots, N\}$.  
                \EndFor
                \State Set $I=1$ and ES$_{\text{curr}} = \frac{1}{M}\sum_{m=1}^MS(\tilde\Pi(\,\cdot\,|\bar{y}^{(m)}_{\text{test}}), \bar\theta^{(m)}_{\text{test}})$.
                \Repeat{\ (Incrementing $I$)}
                    \State Find $A_I$ and $\{B_i\}_{i=0}^I$ by computing Eq. \eqref{eq:BSCemob} with $\{(\{\tilde{\theta}^{(n, m)}_{\text{train}}\}_{n=1}^N, \bar{\theta}^{(m)}_{\text{train}})\}_{m=1}^M$.
                    \State Update ES$_\text{prev}=$ ES$_\text{curr}$ and ES$_{\text{curr}} = \frac{1}{M}\sum_{m=1}^MS(f_\sharp^*\tilde\Pi(\,\cdot\,|\bar{y}^{(m)}_{\text{test}}), \bar\theta^{(m)}_{\text{test}})$.
                \Until{ES$_\text{prev} <$ ES$_\text{curr}$.} \\
                \Return $A_{I}$ and $\{B_i\}_{i=0}^{I}$ for the estimated optimal transformation $f^*(\,\cdot\,)$.
            \end{algorithmic}
        \end{algorithm}
        \FloatBarrier

    \subsection{Sequential Bayesian Score Calibration}\label{subsec:sbsc} 
        BSC uses $\overline\Pi$ to focus calibration on suitable regions of the parameter space.  When we are interested in calibrating an approximate model given a specific data set $y$, a natural choice for $\overline\Pi$ is the approximate posterior $\tilde\Pi(\,\cdot\,|y)$.  However, if $\tilde\Pi(\,\cdot\,|y)$ has low support for the true data-generating parameters $\theta^*$, then using it directly results in a misspecified calibration distribution.  To correct this misspecification, we first apply BSC to $\tilde\Pi(\,\cdot\,|y)$ using the initial calibration distribution.  We then use $f_\sharp^*\tilde\Pi(\,\cdot\,|y)$ as the updated calibration distribution in the subsequent round.  Applying the transformation amounts to extrapolation when the high-probability regions of the calibration and true distributions are disjoint.  In such cases, several rounds of BSC may be required until the calibration distribution is no longer misspecified.  To calibrate from the original approximate model, we chain the learned transformations into a composite function.  By iteratively updating the calibration distribution until it is well-specified, the composite transformation captures changes in error across the parameter space, while avoiding extrapolation.
        
        The sequential application of BSC that we consider updates the calibration distribution over $R$ rounds.  In each round $r \in \{0, 1, \ldots, R\}$, we apply the location-scale transformation from Eq. \eqref{eq:famxform}.  Therefore, when $R=0$, we recover the standard application of BSC with the transformation given by
        \begin{equation*}\label{eq:round0xform}
            f(\theta, \hat{\mu}(y); A_0, b_0) = A_0(\theta - \hat{\mu}(y)) + \hat{\mu}(y) + b_0.
        \end{equation*}
        For each successive round $r > 0$, the calibration distribution is updated by applying the composition of all transformations learned in earlier rounds.  We then use this updated calibration distribution to learn the parameters of the location-scale transformation in the next round.  Formally, the transformation for round $r > 0$ is defined recursively as the composition of the previous transformation with a new location-scale transformation:
        \begin{equation}\label{eq:roundrxform}
            f(\theta, \hat{\mu}(y); \{A_r\}_{r=0}^R, \{b_r\}_{r=0}^R) = A_R(f(\theta, \hat{\mu}(y); \{A_{r}\}_{r=0}^{R-1}, \{b_{r}\}_{r=0}^{R-1}) - \hat{\mu}(y)) + \hat{\mu}(y) + b_R. 
        \end{equation}  
        Although we consider only the location-scale transformation, SBSC can incorporate PBSC at each round.  The location-scale transformation alone is sufficient to achieve calibration for the example presented in Section~\ref{subsec:SIR}.  However, when the error introduced by $\tilde\Pi(\,\cdot\,|y)$ changes significantly across the parameter space, incorporating PBSC and choosing $\Pi$ for the initial calibration distribution might be appropriate.  For each round $r\in\{0,1,\ldots,R\}$, we learn~$A_r, b_r$ by solving the optimisation problem defined in Eq.~\eqref{eq:BSCemob}.  To ensure the current calibration samples reflect the updated calibration distribution, we draw a new set of samples in each round.  The resulting optimal transformation $f^{*}$ can be applied to samples from the approximate posterior distribution~$\theta~=~f^{*}(\tilde\theta)\sim f^{*}_\sharp\tilde\Pi(\,\cdot\,|y)$. 

\begin{figure}[h]
            \centering{
            \resizebox{0.95\textwidth}{!}{ 
                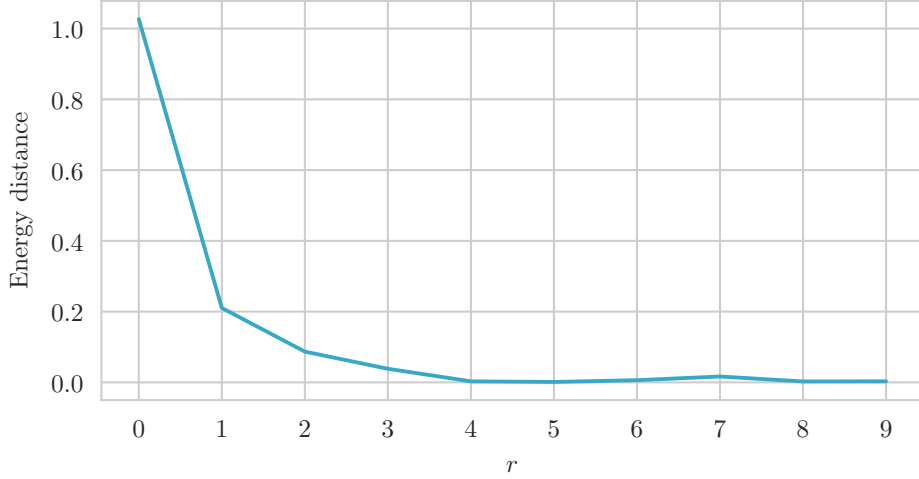
                }
            }
            \caption{Visualisation of the stopping criterion for SBSC as $r \in \{0, 1, \ldots, 9\}$ increases for the susceptible-infected-recovered example we explore in Section~\ref{subsec:SIR}.  This criterion is the energy distance between the current and previous adjusted approximate distribution (Energy Distance).}
            \label{fig:esseq}
        \end{figure}

        Because the required number of rounds depends on how the approximate model introduces error, we propose an adaptive approach to selecting the total number of rounds $R$.  Updating~$\overline\Pi$ in each round alters the probability distribution underlying the outer expectation of Eq. \eqref{eq:BSCOb}; hence, comparing the energy score between rounds of SBSC is not a valid stopping criterion.  In addition, an ideal criterion will detect when the adjusted approximate distribution ceases to change between rounds, thereby limiting the risk of extrapolation.  Accordingly, we suggest evaluating the distance between successive adjusted approximate distributions to detect convergence.  We propose the energy distance \citep{Rizzo2015} as the distance metric, which is rotation-invariant and sensitive to differences in both location and scale \citep{Fan2025}.  The energy distance between the empirical distributions~$\hat{U}~=~\frac{1}{N}\sum_{i=1}^N \delta_{u^{(i)}}$ and $\hat{V} = \frac{1}{N}\sum_{i=1}^N \delta_{v^{(i)}}$ takes the following form
        \begin{equation}\label{eq:edist}
            \text{ED}(\hat{U}, \hat{V}) = \frac{2}{N^2}\sum_{i=1}^N\sum_{j=1}^N||u^{(i)}-v^{(j)}||_2 - \frac{1}{N^2}\sum_{i=1}^N\sum_{j=1}^N||u^{(i)}-u^{(j)}||_2 - \frac{1}{N^2}\sum_{i=1}^N\sum_{j=1}^N||v^{(i)}-v^{(j)}||_2.
        \end{equation}
        Figure~\ref{fig:esseq} visualises the energy distance between successive rounds of SBSC as we increase $R$ for the example we explore in Section~\ref{subsec:SIR}.  Here, we consider the \textit{energy distance between the current and previous adjusted approximate distribution} (Energy Distance).  The Energy Distance appropriately identifies when the adjusted approximate distribution stops changing between rounds by decreasing monotonically until round~3, after which it plateaus near~$0$.  This trend indicates that round~3 is the appropriate stopping point for this example, matching the empirical results in Section~\ref{subsec:SIR}.  Algorithm~\ref{alg:SBSC} provides an overview of how SBSC adaptively learns the sequential transformation in Eq. \eqref{eq:roundrxform}.  It uses the Energy Distance stopping rule, unit weights, and the energy score with $\beta = 1$.

        \begin{algorithm} 
            \caption{Learning the sequential transformation under the Bayesian score calibration framework of \citet{Bon2025} using the Energy Distance stopping rule, unit weights, and the energy score with $\beta = 1$.}
            \label{alg:SBSC}
            \begin{algorithmic}
                \Require{Number of calibration data sets $M$, number of approximate posterior samples $N$, calibration distribution $\overline\Pi$, approximate distribution $\tilde\Pi(\,\cdot\,|y)$.}
                \Ensure{Parameters $\{A_r\}_0^{R-1}$ and $\{b_r\}_0^{R-1}$ for the estimated optimal transformation $f^*(\,\cdot\,)$.}
                \State Set $R=0$.
                \For{$m \in \{1, \ldots, M\}$}
                    \State Draw calibration sample, $\bar\theta^{(m, 0)} \sim \overline\Pi$.
                    \State Simulate calibration data set, $\bar{y}^{(m, 0)} \sim P(\,\cdot\,|\bar\theta^{(m, 0)})$.
                    \State Sample from the approximate posterior distribution $\tilde\theta^{(n, m, 0)} \sim \tilde{\Pi}(\,\cdot\,|\bar{y}^{(m, 0)})$ for $n \in \{1, \ldots, N\}$.
                \EndFor
                \State Find $A_0$ and $b_0$ by computing Eq. \eqref{eq:BSCemob} with $\{(\{\tilde{\theta}^{(n, m, 0)}\}_{n=1}^N, \bar{\theta}^{(m, 0)})\}_{m=1}^M$.
                \Repeat{\ (Incrementing $R$)}
                    \For{$m \in \{1, \ldots, M\}$}
                        \State Draw calibration sample, $\bar\theta^{(m,R)} \sim f^*_\sharp\overline\Pi$.
                        \State Simulate calibration data set, $\bar{y}^{(m,R)} \sim P(\,\cdot\,|\bar\theta^{(m,R)})$.
                        \State Sample from the adjusted approximate distribution $\tilde\theta^{(n, m,R)} \sim f^*_\sharp\tilde{\Pi}(\,\cdot\,|\bar{y}^{(m,R)})$
                        \State for $n \in \{1, \ldots, N\}$.
                    \EndFor
                    \State Find $A_R$ and $b_R$ by computing Eq. \eqref{eq:BSCemob} with $\{(\{\tilde{\theta}^{(n, m,R)}\}_{n=1}^N, \bar{\theta}^{(m,R)})\}_{m=1}^M$.
                \Until{ED$(f^{(R-1)}_\sharp\tilde\Pi(\,\cdot\,|y), f^{(R)}_\sharp\tilde\Pi(\,\cdot\,|y)) \approx 0$} \\
                \Return $\{A_r\}_0^{R-1}$ and $\{b_r\}_0^{R-1}$ for the estimated optimal transformation $f^*(\,\cdot\,)$.
            \end{algorithmic}
        \end{algorithm}
        \FloatBarrier

    \subsection{Diagnostics}\label{subsec:diag}
        Despite the increased flexibility of our two proposed transformations, the adjusted posteriors are not guaranteed to be calibrated in practice.  To determine whether the adjusted posterior is calibrated, we discuss an additional diagnostic tool that complements the diagnostic explored in \citet{Bon2025}.  These diagnostics can detect cases in which the credible regions of the adjusted posteriors fail to achieve nominal coverage \citep{Cockayne2022}.  These diagnostics also indicate whether the calibration process has adequately adjusted the approximate model. 
        
        Both diagnostic tools can be calculated using the training data~$\{(\bar\theta^{(m)}, \bar y^{(m)}, \{\tilde\theta^{(i, m)}\}_{i=1}^N)\}_{m=1}^M$ with $\bar y^{(m)} \sim P(\,\cdot\,|\bar\theta^{(m)})$ and $\tilde\theta^{(i, m)} \sim \tilde\Pi(\,\cdot\,|\bar y^{(m)})$.  By reusing the training data, we avoid the need for additional model simulations.  However, when there is a sufficient computational budget, we suggest using a test data set instead.  For simplicity and clarity, we recommend evaluating both diagnostics for each marginal parameter separately.  These diagnostics require two metrics: the~\mbox{\textit{$(100\cdot\alpha)\%$} frequentist coverage probability} (FCP$_{\alpha}$) and miscoverage.  For probability distribution $U$, the $(100\cdot\alpha)\%$ highest posterior credible region is given by $\text{CR}(U, \alpha)$, with the FCP$_{\alpha}$ given by 
        \begin{equation}\label{eq:fcp}
            \text{FCP}_{\alpha}\left(f_\sharp\tilde\Pi(\,\cdot\,|\bar y^{(m)}), \bar\theta^{(m)} \right) = \text{Pr}\left[\bar\theta^{(m)} \in \text{CR}(f_\sharp\tilde\Pi(\,\cdot\,|\bar y^{(m)}), \alpha)\right],
        \end{equation}
        where we estimate the probability empirically with the proportion of $\bar\theta^{(m)}$ that fall inside the~\mbox{$(100\cdot\alpha)\%$} credible region of their corresponding approximate posterior.  The miscoverage for a given $\alpha$ is
        \begin{equation}\label{eq:miscov}
            \text{MC}\left(f_\sharp\tilde\Pi(\,\cdot\,|\bar y^{(m)}), \bar\theta^{(m)}, \alpha \right) = \text{FCP}_{\alpha}\left(f_\sharp\tilde\Pi(\,\cdot\,|\bar y^{(m)}), \bar\theta^{(m)} \right) - \alpha,
        \end{equation}
        where positive values indicate overcoverage and negative values indicate undercoverage.

        The first diagnostic we suggest is the miscoverage diagnostic detailed in \citet{Bon2025}.  This diagnostic plots the average miscoverage for $\alpha \in (0, 1)$, providing a comparison between achieved and target coverage for all $\alpha$.  Examining this metric across all $\alpha$ can help identify calibration failures in the high-probability region of the calibration distribution.
        
        The second diagnostic compares the $(100\cdot\alpha)\%$ highest posterior credible region of the approximate and adjusted posteriors at a nominated $\alpha$ level.  This visualisation summarises how the approximation error varies across the parameter space; hence, the specific choice of $\alpha$ for this diagnostic is arbitrary, which we acknowledge by choosing $\alpha=0.89$.  Comparing the approximate and adjusted posteriors additionally indicates whether calibration has appropriately corrected this error.

\begin{figure}[h]
            \centering{
            \resizebox{0.95\textwidth}{!}{ 
                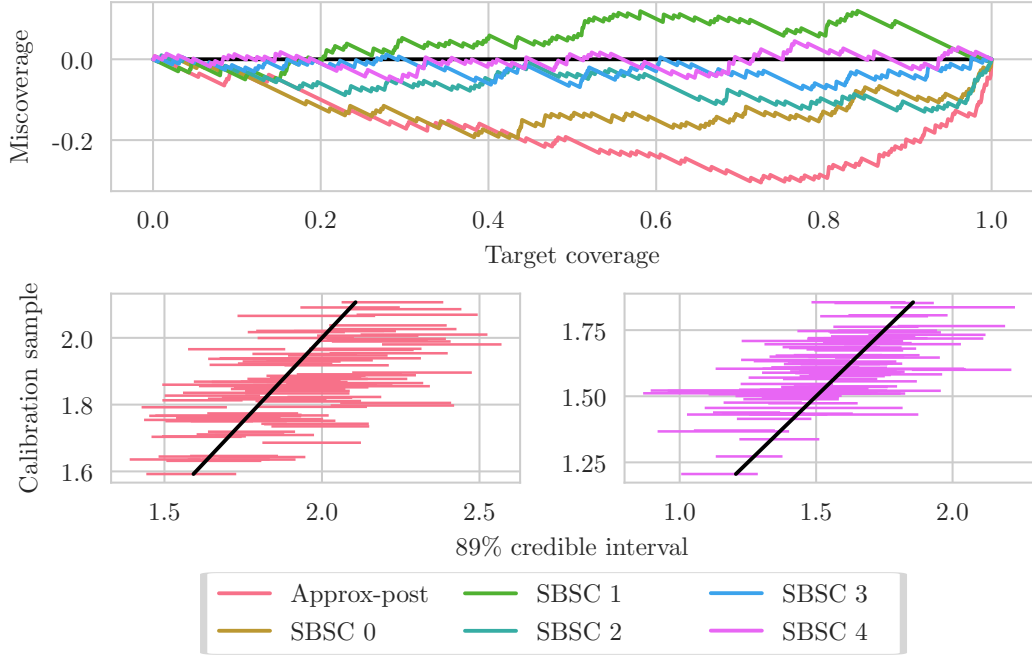
                } 
            }
            \caption{An example diagnostic plot combining both diagnostics for the example we explore in Section~\ref{subsec:SIR}, using test data.  We compare the original approximate model (Approx-post) with SBSC for $r \in \{0, 1, \ldots, 4\}$.  The top row shows the miscoverage diagnostic, and the bottoms row shows the credible region diagnostic with the 89\% credible regions for each posterior in the test data for Approx-post (left) and SBSC 4 (right).}
            \label{fig:diags}
        \end{figure}

        
        When diagnosing calibration for SBSC, we must account for the calibration distribution being updated at each round.  Furthermore, SBSC is intended for approximate models that assign low posterior probability to the true data-generating parameters.  As a result, an appropriate test data set cannot be drawn directly from the model, so diagnostics can only be applied to each round separately.  Fortunately, the adjusted posterior for each round becomes the calibration distribution for the next.  Examining the diagnostics at each round can thus help detect when SBSC does not appropriately correct the approximate model.  Additionally, the adjusted approximate distribution found using SBSC may be an extrapolation.  For such distributions, the adjusted posterior is sensitive to how the error introduced by the approximate model varies across the parameter space.  To detect when this error is affecting calibration, we suggest posterior predictive checks be carried out for each round.  Figure~\ref{fig:diags} displays an example of these diagnostics.

        \FloatBarrier

\section{Simulation Studies}\label{sec:SS}
    We evaluate the flexible transformations we introduced in Section \ref{sec::NM} using two illustrative examples.  The first is an Ornstein-Uhlenbeck (OU) process \citep{Uhlenbeck1930}, and the second is a susceptible-infected-recovered (SIR) model.  These examples are selected to highlight the flexibility of the polynomial transformation shown in Eq. \eqref{eq:polytf} and SBSC using the sequential transformation shown in Eq. \eqref{eq:roundrxform}, respectively.  Similarly to \citet{Bon2025}, we use the energy score with~$\beta=1$.  We choose the number of calibration data sets to be $M = 100$, use unit weighting, and learn a joint transformation on all unknown parameters.  As BSC and PBSC each optimise a single transformation, both operate under an identical simulation budget of $M=100$. SBSC instead learns a transformation per round, and so requires a further $M=100$ simulations at each round, giving a modestly larger cumulative budget for $R>0$.  In these simulation studies, we evaluate our new transformations against the location-scale transformation of \citet{Bon2025} (BSC), the approximate posterior (Approx-post), and the true posterior (True-post).  The example in Section~\ref{subsec:OU} has a tractable likelihood function that we use to estimate the true posterior via MCMC.  We estimate the true posterior for the example in Section~\ref{subsec:SIR} using a computationally expensive particle Markov chain Monte Carlo \citep[pMCMC,][]{Andrieu2010} sampler designed for state-space models.  We note that for many complex statistical models, sampling from the true posterior is typically computationally intractable.  We provide a Python repository that implements BSC, our flexible transformations, and reproduces our results; see Appendix~\ref{app:code} for details.
    
    To assess the performance of the competing methods, we report on the average of four comparison metrics over $1000$ independently generated data sets (test data), for each parameter component.  We compare these metrics against the parameters that generated the test data, with $\theta^*$ denoting the specific marginal component.  These four metrics are: posterior bias, posterior mean squared error (MSE), posterior standard deviation (SD), and the achieved FCP$_{0.89}$.  For a set of marginal posterior samples $\{\theta^{(n)}\}_{n=1}^N$, we approximate the bias with 
    \begin{equation*}\label{eq:bias}
        \text{Bias}\left(\{\theta^{(n)}\}_{n=1}^N, \theta^*\right) = \left|\theta^* - \frac{1}{N}\sum_{n=1}^N\theta^{(n)} \right|,
    \end{equation*}
    and the MSE with
    \begin{equation*}\label{eq:mse}
        \text{MSE}\left(\{\theta^{(n)}\}_{n=1}^N, \theta^*\right) = \frac{\sum_{n=1}^N\left(\theta^{(n)} - \theta^*\right)^2}{N}.
    \end{equation*}
    We also provide a plot of the average miscoverage for~$\alpha \in (0, 1)$.  By considering miscoverage for all~$\alpha \in (0, 1)$, the specific choice of $\alpha$ for FCP$_{\alpha}$ is arbitrary, as discussed in Section~\ref{subsec:diag}.


    \subsection{Ornstein-Uhlenbeck Process} \label{subsec:OU}
    We illustrate the flexibility of the polynomial transformation on the OU process example from \citet{Bon2025} and \citet{Warne2021}.  The OU process is a mean-reverting continuous-time stochastic process.  The stochastic differential equation that governs the mean-reverting behaviour is given by
    \begin{equation}\label{eq:OU}
        dX_t = \gamma(\mu-X_t)dt + \sigma dW_t,
    \end{equation}
    where $\{X_t\}_{t>0}$ for $X_t \in \mathbb{R}$ represents the OU process for time $t$, $\gamma$ is the rate of mean reversion, $\mu$ is the long-term mean, $\sigma$ is the volatility, and $W_t$ is a Wiener process.  For an initial condition~$X_0~=~x_0$ at $t=0$, the distribution of the state at a future time $T$ is tractable and is given by the solution to the forward Kolmogorov equation for Eq. \eqref{eq:OU},
    \begin{equation}\label{eq:OUtrue}
        X_T \sim \mathcal{N}\left(\mu + (x_0 - \mu)e^{-\gamma T}, \frac{D}{\gamma}\bigg(1 - e^{-2\gamma T}\bigg)\right), 
    \end{equation}
    where $D = \frac{\sigma^2}{2}$.
    
\begin{figure}[h]
        \centering{
        \resizebox{0.95\textwidth}{!}{ 
            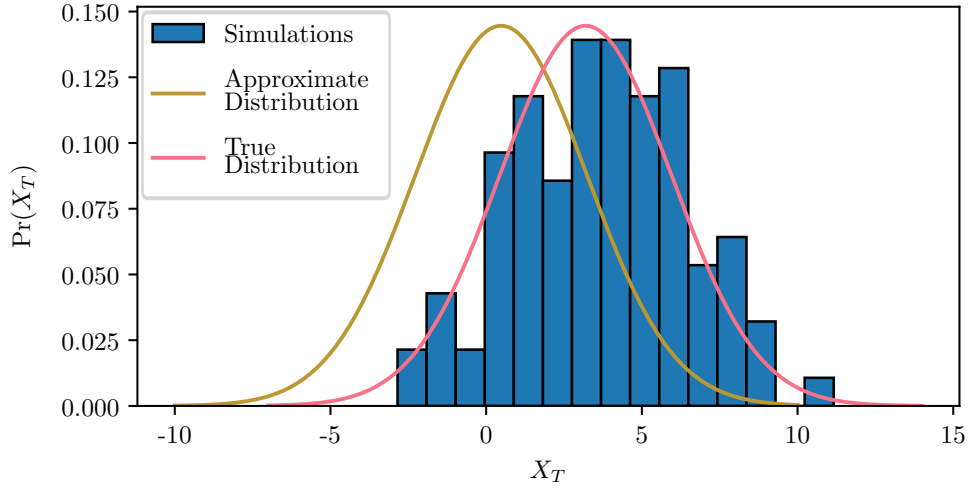
            }
        }
        \caption{Histogram for a simulated data set for $X_T$, with $T=1$, compared with the true and approximate distributions evaluated at the true parameter values.  The simulated data set contains $100$ realisations of the DGP and is generated with parameter values $\mu=1$, $\gamma=2$ and $D=10$.}
        \label{fig:obsou}
    \end{figure}

    For this example, we take Eq. \eqref{eq:OUtrue} to be the true DGP, drawing $100$ realisations for each data set.  The approximate model uses the likelihood associated with the limiting distribution as $T\to \infty$ given by 
    \begin{equation}\label{eq:OUapp}
        X_\infty \sim \mathcal{N}\left(\mu, \frac{D}{\gamma}\right).
    \end{equation}
    We assume $x_0 = 10$ and $T = 1$ are known, and we wish to infer the parameters $\mu, \gamma$ and~$D$.  In contrast, \citet{Warne2021} and \citet{Bon2025} simplify the inference task by treating $\gamma$ as fixed.  The mean and variance of the true model are directly affected by $\gamma$.  Allowing $\gamma$ to be random consequently increases the difficulty of calibration.  We use the same independent priors for~$\mu\sim \mathcal{N}(0, 10^2)$ and $D \sim \text{Exp}(\frac{1}{10})$ as \citet{Bon2025}, and choose $\gamma \sim \mathcal{N}(2, 0.4^2)$.  For the calibration distribution, we use the prior distribution to achieve amortised inference, and draw~$M=100$ independent calibration samples.  By targeting amortised inference, we further increase the calibration difficulty relative to the example in \citet{Bon2025}. To generate the test data, we simulate $1000$ independent data sets using the DGP with $\mu, \gamma$ and $D$ drawn from their respective priors.  Figure~\ref{fig:obsou} displays an example realisation from the DGP compared with the true and approximate distributions evaluated at the true parameter values.  To sample from the true and approximate posteriors, we use the No-U-Turn Hamiltonian Monte Carlo algorithm \citep{Hoffman2014}.

\begin{figure}[h]
        \centering{
        \resizebox{0.95\textwidth}{!}{
            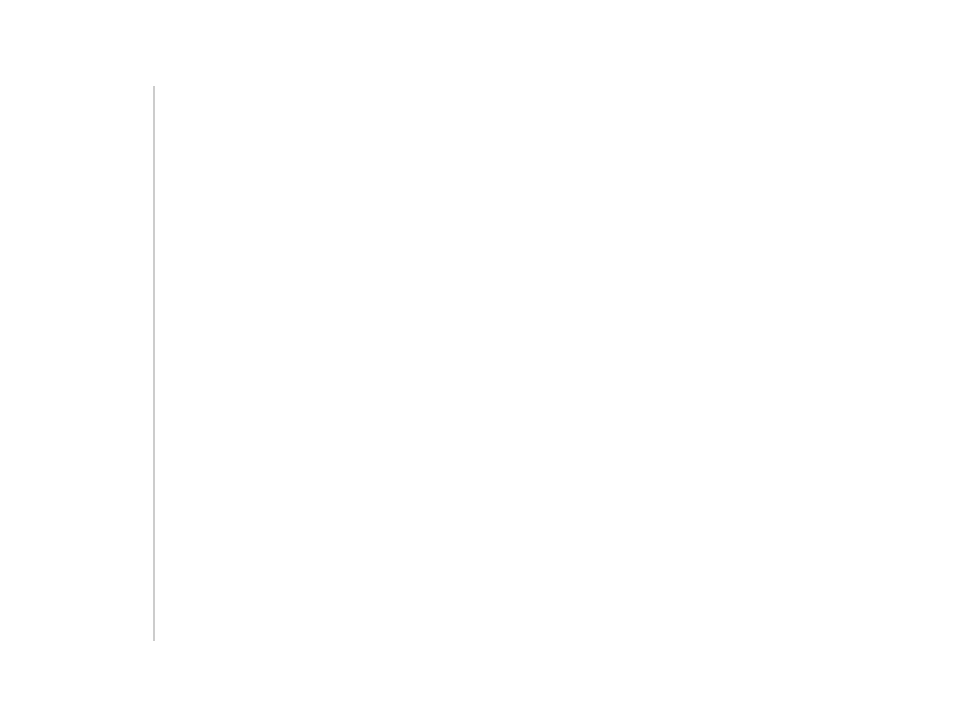
            }
        }
        \caption{Miscoverage for $\mu$ across coverage levels $\alpha \in (0, 1)$ over the 1000 independent test data sets for the OU example for the true posterior (True-post), the original approximate posterior (Approx-post), the adjusted posterior found using the location-scale transformation of \citet{Bon2025}, and the adjusted posteriors found using the polynomial transformation of order 1 (PBSC~1), order 2 (PBSC~2) and order 3 (PBSC~3).}
        \label{fig:oumuac}
    \end{figure}

\begin{figure}[h]
        \centering{
        \resizebox{\textwidth}{!}{
            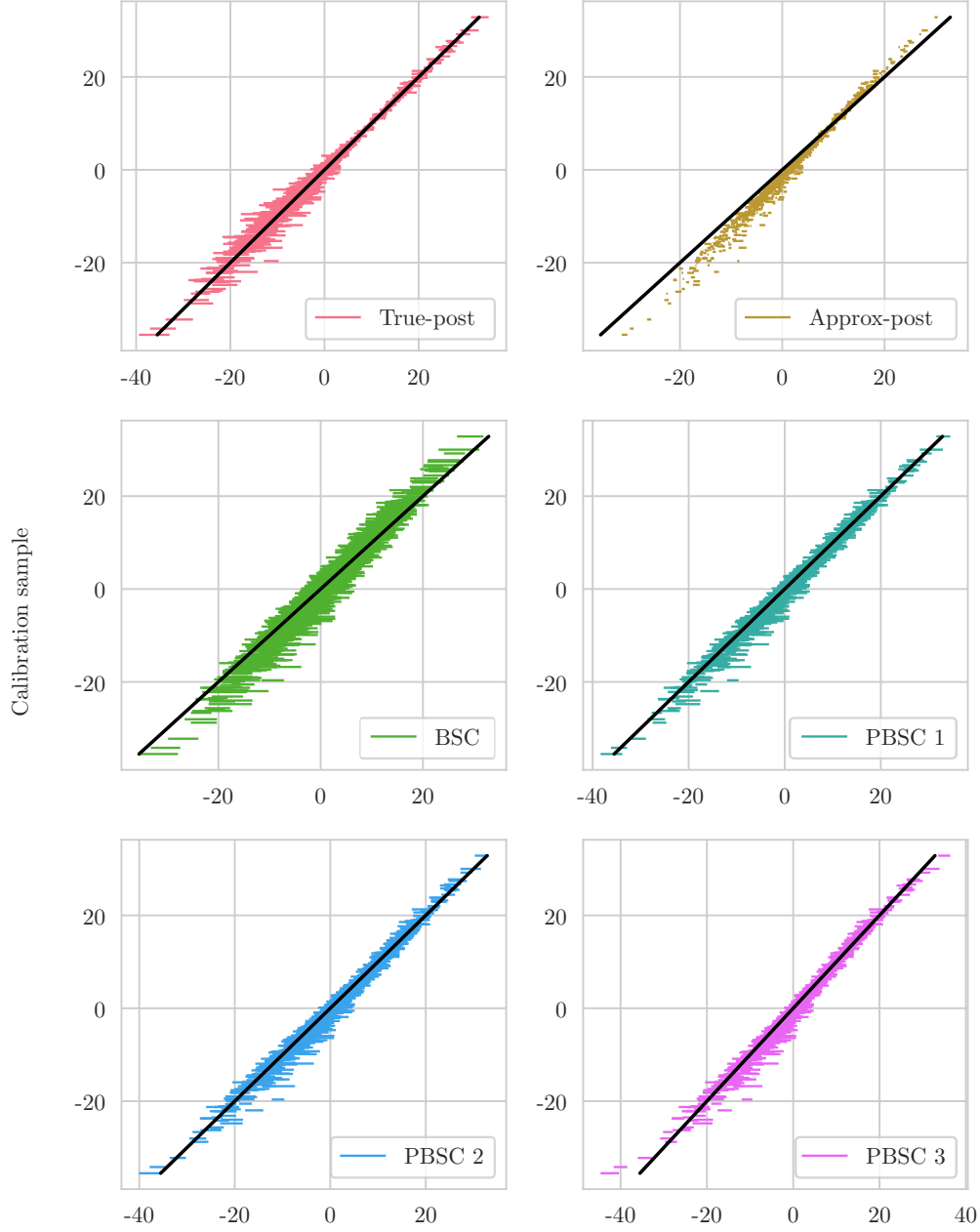
            }
        }
        \caption{89\% highest posterior credible intervals of each test data set for $\mu$ for the OU example.  The posteriors compared are the true posterior (True-post), the original approximate posterior (Approx-post), the adjusted posterior found using the location-scale transformation of \citet{Bon2025}, and the adjusted posteriors found using the polynomial transformation of order 1 (PBSC~1), order 2 (PBSC~2) and order 3 (PBSC~3).} 
        \label{fig:oumucr}
    \end{figure}

    For brevity, we report results only up to the third order of our polynomial transformation for~$\mu$ here; results for higher orders, and for $\gamma$ and $D$, appear in Appendix~\ref{app:resultsOU}.  Table \ref{tab:oumu} presents the results for the MSE, bias, SD, and FCP$_{0.89}$.  Comparing Approx-post and True-post reveals that \mbox{Approx-post} exhibits poor performance across all metrics.  The results for BSC highlight the limitation of applying a single correction across the parameter space.  To achieve good FCP$_{0.89}$ and low bias, in this example, BSC has to greatly increase the posterior SD, resulting in a larger MSE than Approx-post.  All three polynomial transformations outperform BSC, achieving significantly lower MSE, bias, and SD.  However, the polynomial of order~3 (PBSC~3) exhibits greater undercoverage for FCP$_{0.89}$.  We present the miscoverage for $\mu$ across coverage levels $\alpha \in (0, 1)$ for the test data in Figure~\ref{fig:oumuac}.  All four transformations have absolute miscoverage less than $0.1$ for all~$\alpha~\in~(0, 1)$, with BSC exhibiting the lowest absolute miscoverage.  Finally, we provide a visual comparison of the FCP$_{0.89}$ of each data set in the test data for $\mu$ in Figure~\ref{fig:oumucr}.  This figure displays how the error induced by Approx-post changes over the parameter space.  The figure also reveals how the additional flexibility of the polynomial transformations allows the adjusted posteriors to be calibrated without overinflating the posterior variance.
    The results for $\gamma$, and $D$ are similar to those of $\mu$, with the primary difference being that Approx-post has good coverage for $D$.  For both parameters, PBSC~1 and PBSC~2 outperform the adjusted posterior produced by BSC.  BSC, in turn, improves on Approx-post, and PBSC~3 has slightly lower coverage than BSC.
    
    \begin{table}[!hbtp]
        \normalsize
        \setlength{\tabcolsep}{12pt}
        \begin{tabular}{ lcccc }
            \hline \hline
            Method            & MSE           & Bias           & SD            & FCP$_{0.89}$ \\
            \hline
            True-post         & 1.880 (0.118) & -0.043 (0.032) & 0.802 (0.013) & 0.882\\
            \hline
            Approx-post       & 5.298 (0.300) & -1.478 (0.055) & 0.197 (0.003) & 0.097\\
            BSC               & 6.445 (0.247) & -0.694 (0.055) & 1.640 (0.015) & 0.848\\
            \textbf{PBSC~1}   & 2.420 (0.147) & -0.588 (0.033) & 0.933 (0.010) & 0.850\\
            \textbf{PBSC~2}  & 2.421 (0.143) & -0.540 (0.035) & 0.938 (0.007) & 0.846\\
            PBSC~3             & 2.426 (0.163) & -0.549 (0.037) & 0.840 (0.006) & 0.812\\
            \hline \hline
        \end{tabular}
        \caption{Average results for $\mu$ over 1000 independent test data sets for the OU process example with standard errors shown in brackets where appropriate.  The posteriors compared are the true posterior (True-post), the original approximate posterior (Approx-post), the adjusted posterior found using the location-scale transformation of \citet{Bon2025} (BSC), and the adjusted posteriors found using the polynomial transformation of order 1 (PBSC~1), order 2 (PBSC~2), and order 3 (PBSC~3).  \textbf{Highlighted} are the posterior approximations with metric values that are closest to True-post's metric values overall.}
        \label{tab:oumu}
    \end{table}

    \FloatBarrier
    
    \subsection{Susceptible-Infected-Recovered Model}\label{subsec:SIR}
    We consider a stochastic version of the SIR model adapted from \citet{Whitehouse2023}.  This SIR model has a fixed population size $P$, time steps~$t~\in~\{1, 2, \ldots, T\}$, and four random latent states~$Y_t~=~[S_t, A_t, I_t, R_t]$ that correspond to susceptible, asymptomatic, infected, and recovered individuals. These latent states have an initial condition $Y_0$, and their evolution follows an incremental update at each discrete time step $t$ given by
    \begin{align*}
        S_{t+1} &= S_{t} - a_t + s_t, & A_{t+1} &= a_t, \\
        I_{t+1} &= I_{t} + A_{t} - r_t, & R_{t+1} &= R_{t} + r_t - s_t,
    \end{align*}
    where $a_t \sim \text{Binom}(S_t, 1-e^{(-\theta_0(A_t + I_t)/P)})$ is the number of new infections, \mbox{$r_t \sim \text{Binom}(I_t, 1 - e^{-\theta_1})$} is the number of new recovered individuals, and $s_t \sim \text{Binom}(R_t, 0.1)$ is the number of new susceptible individuals.  The observation model is given by the observed number of infected at time $t$ with 
    \begin{equation}\label{eq:sirdgp}
        X_t \sim \text{Binom}(I_t, \theta_2).
    \end{equation}
    We assume that $P=763$, $T=20$, and~$Y_0~=~[762, 0, 1, 0]$ are known, and we attempt to infer $\theta_0, \theta_1$ and $\theta_2$.  We use independent priors for all three parameters with~$\theta_0 \sim \mathcal{N}(0, 10^2), \theta_1\sim \mathcal{N}(0, 10^2)$ and $\theta_2 \sim \mathcal{N}(0, 10^2)$.

    For this example, we do not have access to a likelihood function that is computationally tractable.  Since a particle filter can unbiasedly estimate the likelihood of state-space models, we use pMCMC with $10\ 000$ particles and a Metropolis-within-Gibbs algorithm \citep{Robert2004} to sample from the true posterior.  For the approximate model, we use the Poisson approximate likelihood \citep[PAL,][]{Whitehouse2023}, which approximates the state transitions and the observation model with independent Poisson processes.  We estimate the PAL approximate posterior using a Metropolis-within-Gibbs algorithm.  Despite using a highly optimised implementation of pMCMC, the computational cost of a single evaluation of the true likelihood (0.199 seconds) is almost 400 times slower than a single evaluation of the approximate likelihood (0.0005 seconds).  To ensure that PAL produces an approximate posterior that is significantly different from the true posterior, we adjusted the true DGP underpinning the model from the example in \citet{Whitehouse2023}.  That is, we include an asymptomatic state and consider the possibility that recovered individuals can become susceptible again.  These two changes provide an approximate model which needs calibration while still allowing us to use the optimised code from \citet{Whitehouse2023} with only minor modifications; see Appendix~\ref{app:code}.
    
\begin{figure}[h!]
        \centering{
        \resizebox{0.95\textwidth}{!}{
            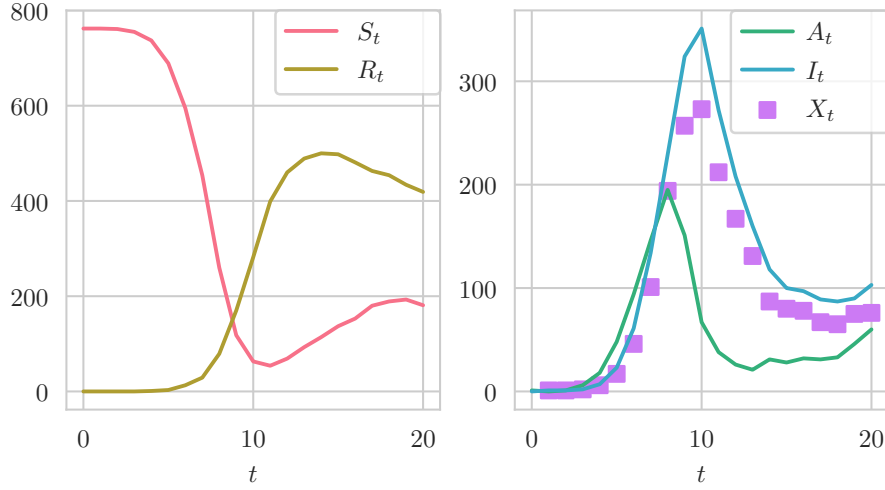
            }
        }
        \caption{Number of observed infected individuals $X_t$ for the observed data set for $t \in \{1, 2, \ldots, T\}$, where $T=20$ with initial condition $Y_0 = [762, 0, 1, 0]$ and true parameter values $\theta_0=1.5, \theta_1=0.5$ and $\theta_2=0.8$.  Also shown is the evolution of the four latent states over the same period with susceptible $S_t,$ and recovered $R_t$ on the left and asymptomatic $A_t,$ and infected $I_t$ on the right.}
        \label{fig:obssir}
    \end{figure}

    To generate the observed data, we simulate a single realisation of the true model where observations are generated according to Eq. \eqref{eq:sirdgp} with true parameter values $\theta_0 = 1.5, \theta_1 = 0.5$ and~$\theta_2~= 0.8$.  Figure~\ref{fig:obssir} displays the observed data and all four latent states.  For the test data, we take $1000$ independent realisations of the true model using the same true parameter values.  For the calibration distribution in round $0$, we use the approximate posterior based on the PAL and generate $M=100$ independent calibration samples at each round.

    For brevity, we detail the results up to round $4$ of our sequential transformation (SBSC 4), corresponding to a simulation budget of $500$ calibration data sets across the five rounds.  Appendix~\ref{app:resultsSIR} details results for higher rounds, including plots of the $89\%$ credible interval for the posterior distribution of each test data set.  The results for the MSE, bias, SD, and FCP$_{0.89}$ for $\theta_0, \theta_1$, and $\theta_2$ are presented in \mbox{Tables~\ref{tab:sirt0}--\ref{tab:sirt2},} respectively.  Approx-post performs poorly with significant undercoverage and high MSE and bias across all three parameters.  BSC substantially improves on Approx-post across all metrics.  Even so, BSC still results in undercoverage and slightly high MSE and bias.  Each successive round continues to improve most metrics in the adjusted posterior until SBSC 3.  The adjusted posteriors produced by SBSC 3 and SBSC 4 have comparable MSE, bias, SD, and coverage.  This trend continues for higher rounds, with SBSC 5 to SBSC 9 also exhibiting similar results to SBSC 3 and SBSC 4; see Appendix~\ref{app:resultsSIR}.
    
\begin{figure}[h]
        \centering{
        \resizebox{0.95\textwidth}{!}{
            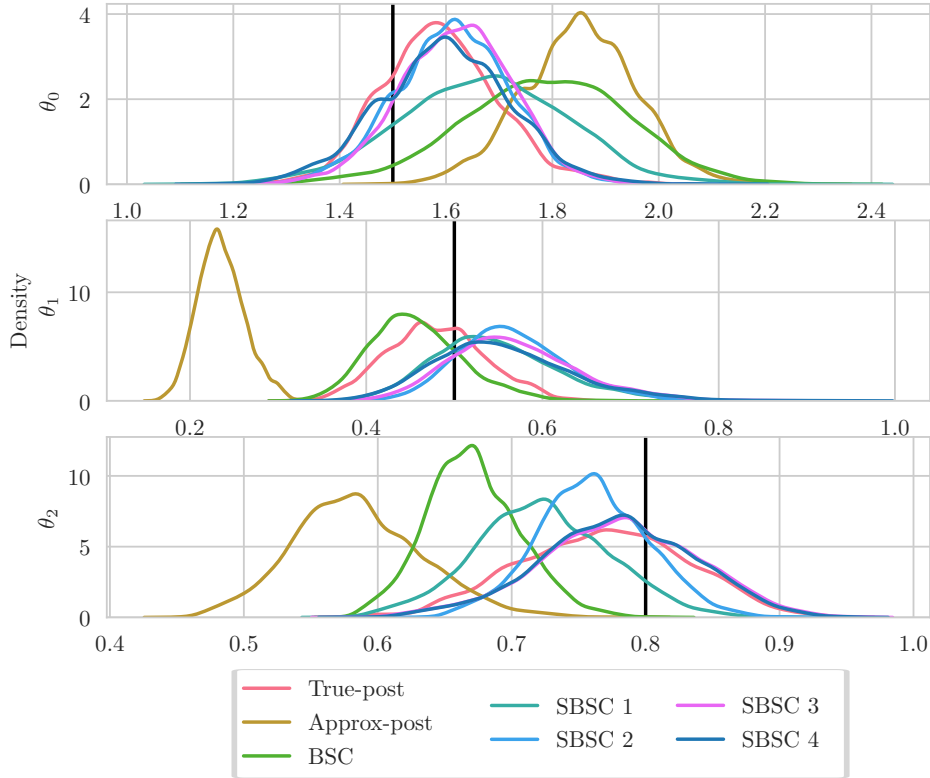
            }
        }
        \caption{Univariate densities estimates of approximations to the SIR model posterior distribution conditioned on the observed data for $\theta_0$ (top), $\theta_1$ (middle), and~$\theta_2$ (bottom).  The posteriors shown are the true posterior (True-post), the original approximate posterior (\mbox{Approx-post}), the adjusted posterior found using the location-scale transformation of \citet{Bon2025} (BSC), and the adjusted posteriors found using the sequential transformation over rounds $1-4$ (SBSC 1--4).  The parameter values used to generate the observed data are indicated with a black vertical line.}
        \label{fig:PALdensT2}
    \end{figure}

    Figure~\ref{fig:PALmiscovT0} displays the miscoverage of $\theta_0, \theta_1$ and $\theta_2$ for $\alpha \in (0, 1)$.  The pattern of miscoverage across all $\alpha$ levels is consistent with the FCP$_{0.89}$ results for all three parameters, showing that these results extend to all coverage levels.  All three marginal posterior density estimates for each of the competing methods, when conditioned on the observed data, are shown in Figure~\ref{fig:PALdensT2}.  This figure highlights how after only a few rounds of the sequential application of BSC, the adjusted posteriors reflect the density of True-post, which further rounds maintain.

    \begin{table}[!hbtp]
        \normalsize
        \setlength{\tabcolsep}{12pt}
        \begin{tabular}{ lcccc }
            \hline \hline
            Method            & MSE             & Bias             & SD             & FCP$_{0.89}$ \\
            \hline
            True-post         & 0.0310 (6.2e-4) & -0.0002 (2.6e-3) & 0.1370 (4.9e-4) & 0.959\\
            \hline
            Approx-post       & 0.0672 (1.8e-3) & -0.2094 (3.6e-3) & 0.0994 (2.7e-4) & 0.327\\
            BSC               & 0.0589 (1.4e-3) & -0.1512 (3.5e-3) & 0.1489 (1.2e-3) & 0.759\\
            SBSC 1            & 0.0337 (0.7e-3) & -0.0435 (3.3e-3) & 0.1438 (4.7e-4) & 0.966\\
            SBSC 2            & 0.0202 (0.5e-3) & ~0.0110 (3.2e-3) & 0.0992 (2.8e-4) & 0.878\\
            \textbf{SBSC 3}   & 0.0194 (0.5e-3) & ~0.0046 (3.2e-3) & 0.0952 (3.5e-4) & 0.864\\
            SBSC 4            & 0.0237 (0.5e-3) & ~0.0242 (3.2e-3) & 0.1143 (3.3e-4) & 0.909\\
            \hline
            \end{tabular}
            \caption{Average results for $\theta_0$ over 1000 independent test data sets for the SIR example with standard errors shown in brackets where appropriate.  The posteriors compared are the true posterior (True-post), the original approximate posterior (Approx-post), the adjusted posterior found using the location-scale transformation of \citet{Bon2025} (BSC), and the adjusted posteriors found using the sequential transformation over rounds $1-4$ (SBSC 1--4).  \textbf{Highlighted} is the posterior approximation with metric values that are closest to True-post's metric values overall.}
            \label{tab:sirt0}
    \end{table}  
    
    \begin{table}[!hbtp]
        \normalsize
        \setlength{\tabcolsep}{12pt}
        \begin{tabular}{ lcccc }
            \hline \hline
            Method            & MSE             & Bias            & SD              & FCP$_{0.89}$ \\
            \hline
            True-post         & 0.0010 (3.8e-5) & -0.0070 (7.7e-4) & 0.0290 (1.7e-4) & 0.945\\
            \hline
            Approx-post       & 0.0748 (5.6e-4) & ~0.2695 (1.1e-3) & 0.0294 (0.2e-3) & 0.000\\
            BSC               & 0.0118 (2.8e-4) & ~0.0528 (2.2e-3) & 0.0618 (0.7e-3) & 0.614\\
            \textbf{SBSC 1}   & 0.0172 (7.5e-4) & -0.0364 (2.6e-3) & 0.0877 (1.2e-3) & 0.910\\
            SBSC 2            & 0.0152 (6.4e-4) & -0.0537 (2.7e-3) & 0.0696 (0.6e-3) & 0.785\\
            SBSC 3            & 0.0183 (7.5e-4) & -0.0556 (2.7e-3) & 0.0843 (0.9e-3) & 0.875\\
            SBSC 4            & 0.0191 (7.7e-4) & -0.0448 (2.7e-3) & 0.0937 (1.1e-3) & 0.932\\
            \hline
            \end{tabular}
            \caption{Average results for $\theta_1$ over 1000 independent test data sets for the SIR example with standard errors shown in brackets where appropriate.  The posteriors compared are the true posterior (True-post), the original approximate posterior (Approx-post), the adjusted posterior found using the location-scale transformation of \citet{Bon2025} (BSC), and the adjusted posteriors found using the sequential transformation over rounds $1-4$ (SBSC 1--4).  \textbf{Highlighted} is the posterior approximation with metric values that are closest to True-post's metric values overall.}
            \label{tab:sirt1}
    \end{table}
    
    \begin{figure}[!h]
        \centering{
            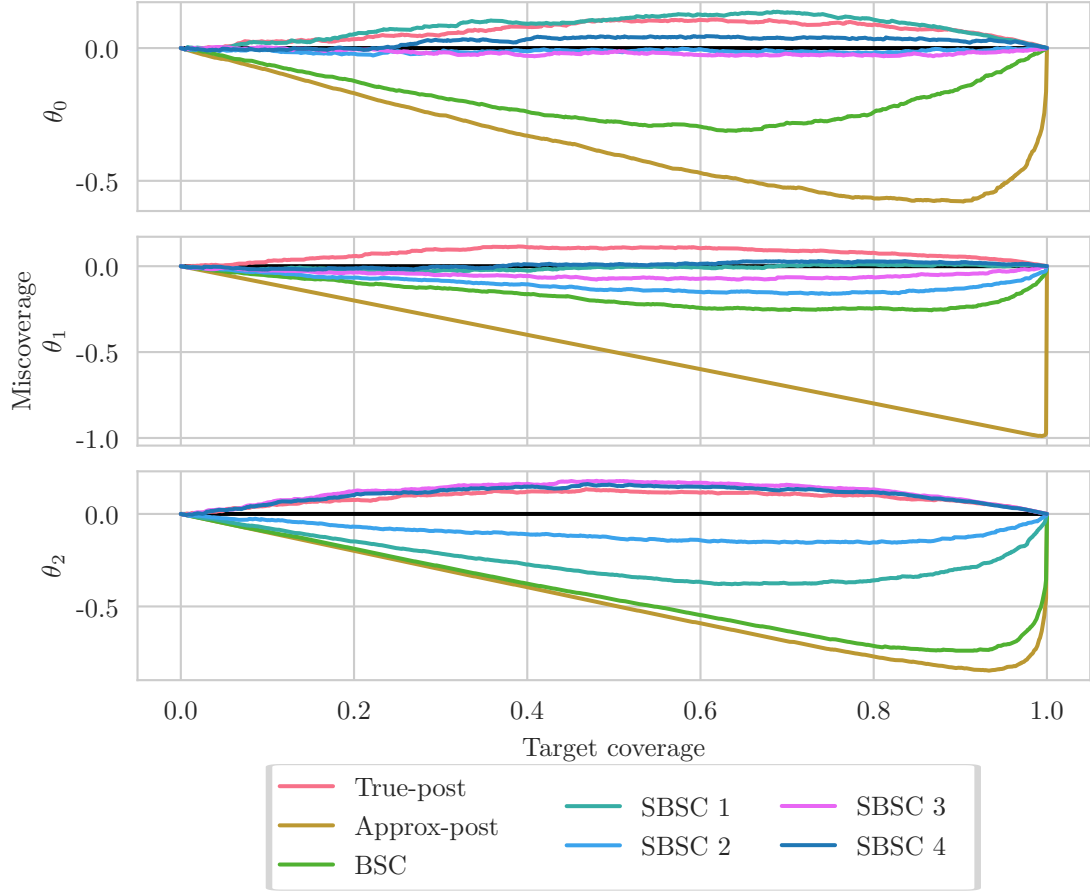
        } 
        \caption{Miscoverage for $\theta_0$ (top), $\theta_1$ (middle), and $\theta_2$ (bottom) across coverage levels $\alpha \in (0, 1)$ over the 1000 independent test data sets for the SIR example.  The posteriors compared are the true posterior (True-post), the original approximate posterior (Approx-post), the adjusted posterior found using the location-scale transformation of \citet{Bon2025} (BSC), and the adjusted posteriors found using the sequential transformation over rounds $1-4$ (SBSC 1--4).}
        \label{fig:PALmiscovT0}
    \end{figure}

    \begin{table}[!hbtp]
        \normalsize
        \setlength{\tabcolsep}{12pt}
        \begin{tabular}{ lcccc }
            \hline \hline
            Method            & MSE             & Bias             & SD              & FCP$_{0.89}$ \\
            \hline
            True-post         & 0.0037 (8.3e-5) & -0.0118 (1.1e-3) & 0.0466 (2.4e-4) & 0.963\\
            \hline
            Approx-post       & 0.0479 (7.1e-4) & ~0.2027 (1.9e-3) & 0.0560 (3.6e-4) & 0.070\\
            BSC               & 0.0187 (3.9e-4) & ~0.1187 (1.7e-3) & 0.0398 (3.9e-4) & 0.165\\
            SBSC 1            & 0.0102 (2.0e-4) & ~0.0701 (1.5e-3) & 0.0557 (3.9e-4) & 0.620\\
            SBSC 2            & 0.0051 (1.1e-4) & ~0.0334 (1.3e-3) & 0.0458 (3.9e-4) & 0.778\\
            \textbf{SBSC 3}   & 0.0059 (0.7e-4) & ~0.0165 (1.2e-3) & 0.0647 (3.6e-4) & 0.982\\
            \textbf{SBSC 4}   & 0.0058 (0.7e-4) & ~0.0175 (1.2e-3) & 0.0632 (3.5e-4) & 0.978\\
            \hline
            \end{tabular}
            \caption{Average results for $\theta_2$ over 1000 independent test data sets for the SIR example with standard errors shown in brackets where appropriate.  The posteriors compared are the true posterior (True-post), the original approximate posterior (Approx-post), the adjusted posterior found using the location-scale transformation of \citet{Bon2025} (BSC), and the adjusted posteriors found using the sequential transformation over rounds $1-4$ (SBSC 1--4).  \textbf{Highlighted} are the posterior approximations with metric values that are closest to True-post's metric values overall.}
            \label{tab:sirt2}
    \end{table}

\FloatBarrier 

\section{Discussion}\label{sec::Dis}
    In this paper, we explored a richer class of transformations for BSC, enabling more accurate adjustment of approximate distributions across a wider range of problems.  We proposed two flexible transformations.  The first, PBSC, is a polynomial extension that provides a step towards BSC being capable of achieving amortised inference.  The second, SBSC, sequentially applies BSC by updating the calibration distribution over several rounds.  This enables calibration of approximate models with posteriors that have low support for the true parameter values.  In Section~\ref{sec:SS}, our simulation studies demonstrate that PBSC and SBSC provide increased flexibility over BSC without significantly increasing computational cost.  We also explored a new diagnostic for the BSC framework, complementing the miscoverage diagnostic of \citet{Bon2025}.  We now discuss several limitations of our approach and directions for future research.


    The simulation study in Section~\ref{subsec:OU} showcased PBSC on an approximate model that introduces error differently across the parameter space.  Comparing standard BSC with PBSC on this example highlighted the advantage of incorporating polynomial terms into the transformation.  To account for the changing error, BSC could only inflate the variance of the adjusted posterior, while PBSC applied a more appropriate correction.  Despite these gains, the polynomial transformation assumes that the bias introduced by the approximate model can be captured by a smooth function.  Additionally, while we allow $A$ to capture all dependence between parameters, PBSC could still suffer from the curse of dimensionality.  We did not test dimensions beyond $d=3$, and scaling to higher-dimensional problems remains an open question.  Future work could consider alternative transformations that either target a richer class of functions or are specifically designed for higher-dimensional problems.

    In Section~\ref{subsec:SIR}, we explored an example where the approximate posterior exhibited substantial undercoverage and bias, as shown in Tables~\ref{tab:sirt0}--\ref{tab:sirt2}.  These results indicate that the true parameter values had negligible posterior support under the approximate distribution.  After only a few rounds, SBSC produced a well-specified adjusted posterior.  However, SBSC assumes that the error introduced by the approximate model does not change significantly across the parameter space.  A natural extension would be to incorporate PBSC into SBSC, allowing the initial calibration distribution to be the prior.  Such an approach could correct approximate models for which the true parameters lie in a low-probability region.  It could also handle models whose error changes significantly across the parameter space.  An alternative approach could be to utilise a predictively oriented posterior \citep{Shen2025, McLatchie2025} as the calibration distribution.  This would allow calibration to focus on parameter values that generate data that are similar to the observed data, potentially avoiding extrapolation. 

    One limitation neither PBSC nor SBSC addresses is calibrating the approximate posterior when the DGP itself is misspecified.  However, posterior predictive checks can help identify when the data generated using samples from the adjusted posterior are different from the observed data.  Future work could investigate whether incorporating the posterior predictive distribution into the optimisation problem improves BSC's robustness to model misspecification. 

    We considered only unit weighting in our simulation studies.  This approach is exact when PBSC uses the prior as the calibration distribution.  However, for SBSC, unit weighting is only an asymptotically consistent estimator of the importance weights.  Future work could therefore explore alternative weighting strategies to further improve performance.

    Finally, we did not explore alternative choices for the scoring rule beyond the energy score with~$\beta=1$.  Additionally, we utilised a black-box optimisation algorithm in our simulation studies.  The choice of scoring rule or optimisation algorithm may therefore affect the performance of PBSC and SBSC, and adjusting either could further improve the BSC framework.
    
    Taken together, PBSC and SBSC substantially broaden the class of models the BSC framework can calibrate. Both achieve this by allowing calibration samples to be drawn from less informative distributions, including the prior, without a large increase in the number of model simulations.  This flexibility enables the BSC framework to target computationally intensive models with location-varying error or poor coverage of the true parameter values.  We hope this work encourages wider adoption of score-based calibration as an alternative to conventional simulation-based inference.

        
 
\backmatter


\bmhead{Acknowledgements}
CD, DJW, and DJN were supported by an Australian Research Council Discovery Project (DP260101525). \\
AB and CD were supported by an Australian Research Council Future Fellowship (FT210100260). \\
DJW is supported by an Australian Research Council Early Career Researcher Award (DE250100396). \\




\noindent

\begin{appendices}





\section{Appendix A. Additional Results from Ornstein-Uhlenbeck Process}\label{app:resultsOU}
\begin{figure}[h]
        \centering{
        \resizebox{0.95\textwidth}{!}{
            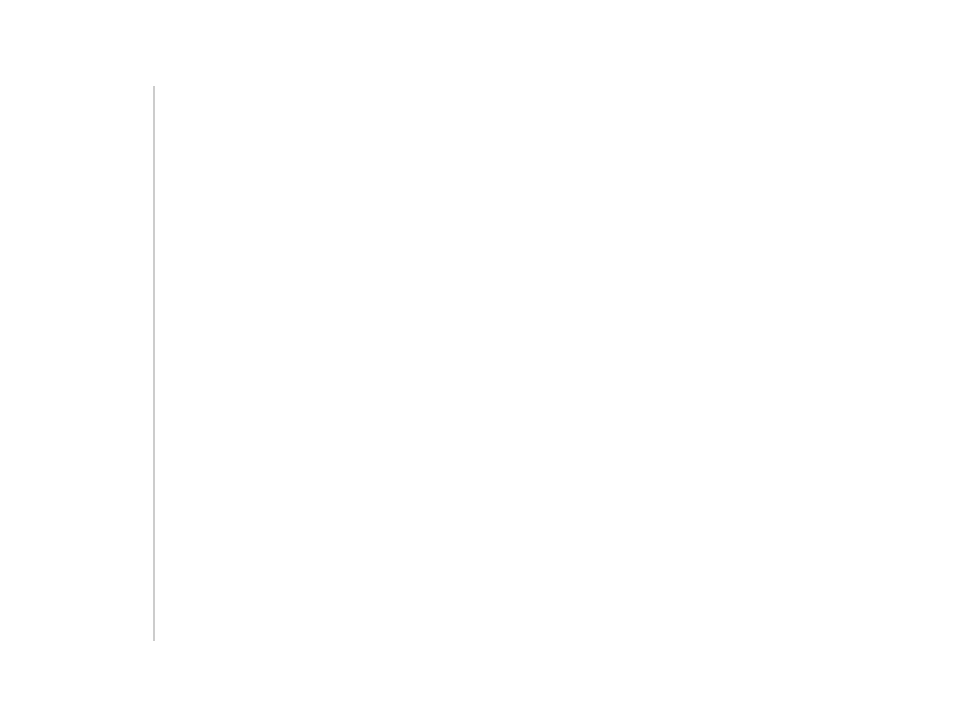
            }
        }
        \caption{89\% highest posterior credible intervals of each test data set for $\gamma$ for the OU example.  The posteriors compared are the true posterior (True-post), the original approximate posterior (Approx-post), the adjusted posterior found using the location-scale transformation of \citet{Bon2025}, and the adjusted posteriors found using the polynomial transformation of order 1 (PBSC~1), order 2 (PBSC~2) and order 3 (PBSC~3).}
        \label{fig:ougaac}
    \end{figure}

\begin{figure}[h]
        \centering{
        \resizebox{\textwidth}{!}{
            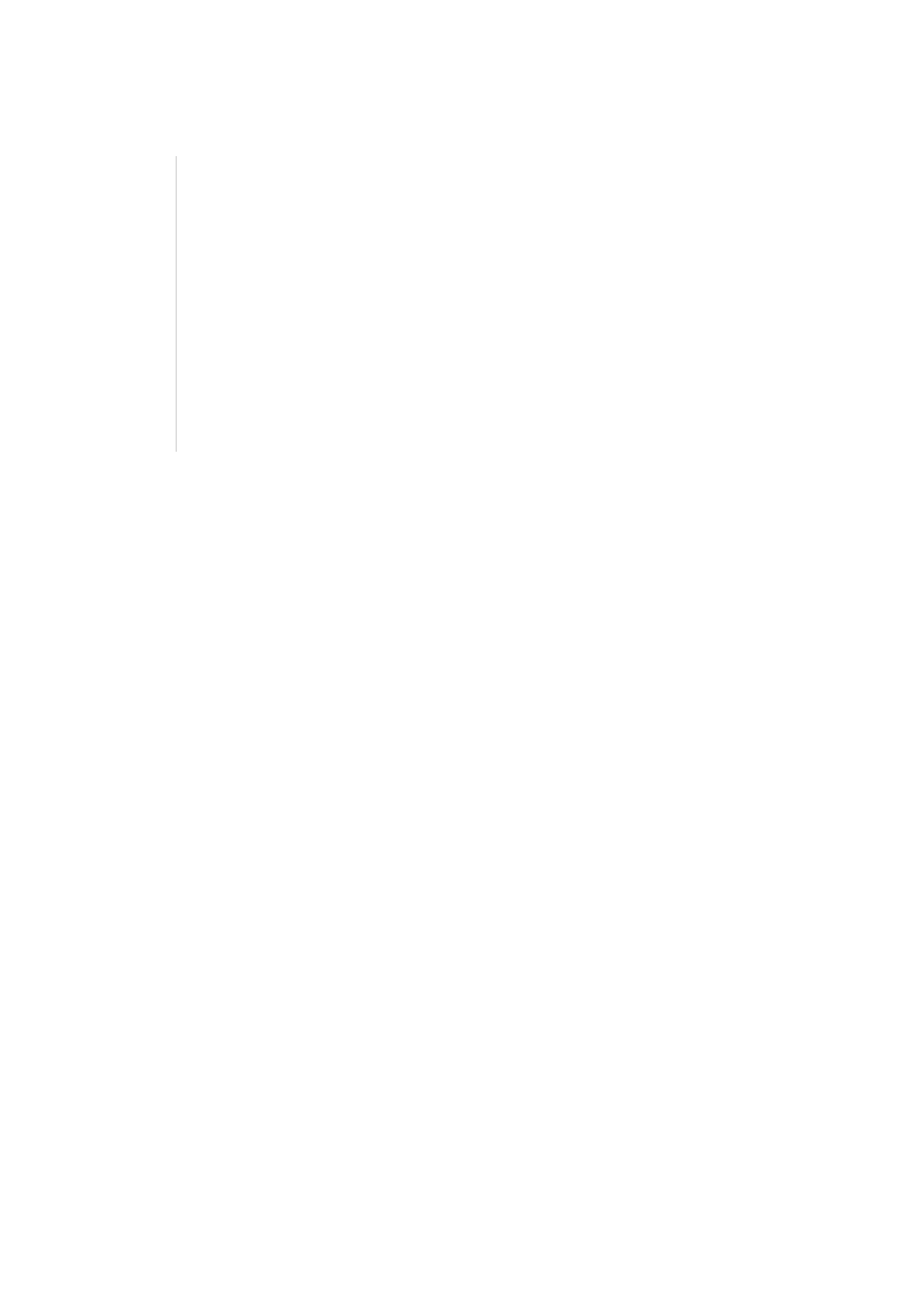
            }
        }
        \caption{Miscoverage for $\gamma$ across coverage levels $\alpha \in (0, 1)$ over the 1000 independent test data sets for the OU example for the true posterior (True-post), the original approximate posterior (Approx-post), the adjusted posterior found using the location-scale transformation of \citet{Bon2025}, and the adjusted posteriors found using the polynomial transformation of order 1 (PBSC~1), order 2 (PBSC~2) and order 3 (PBSC~3).}
        \label{fig:ougacr}
    \end{figure}

\begin{figure}[h]
        \centering{
        \resizebox{0.95\textwidth}{!}{
            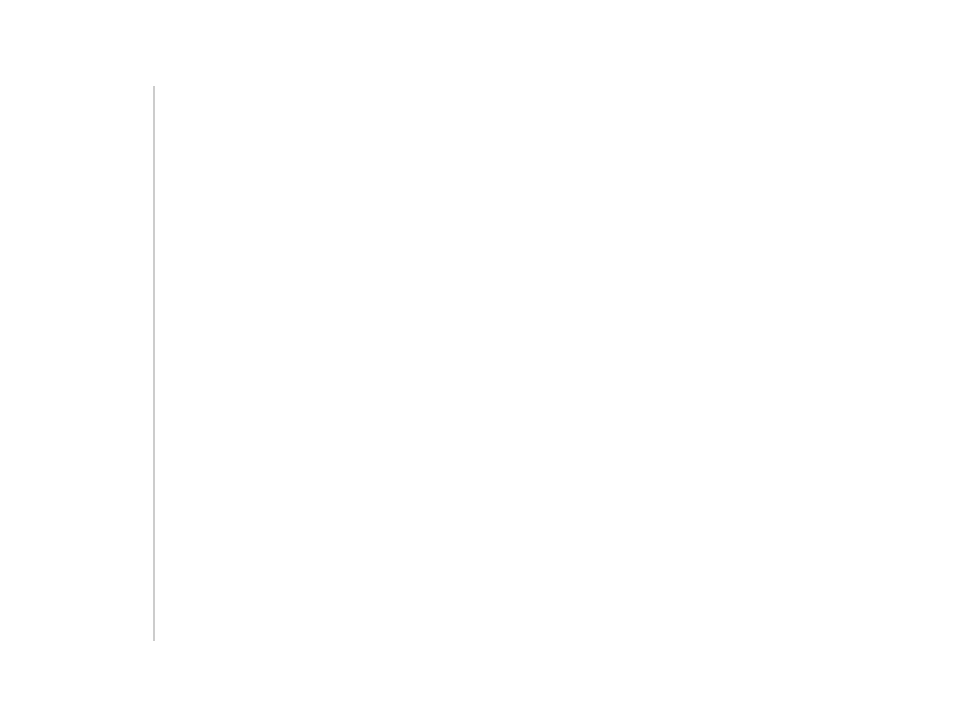
            }
        }
        \caption{Miscoverage for $D$ across coverage levels $\alpha \in (0, 1)$ over the 1000 independent test data sets for the OU example for the true posterior (True-post), the original approximate posterior (Approx-post), the adjusted posterior found using the location-scale transformation of \citet{Bon2025}, and the adjusted posteriors found using the polynomial transformation of order 1 (PBSC~1), order 2 (PBSC~2) and order 3 (PBSC~3).}
        \label{fig:ouDac}
    \end{figure}

\begin{figure}[h]
        \centering{
        \resizebox{\textwidth}{!}{
            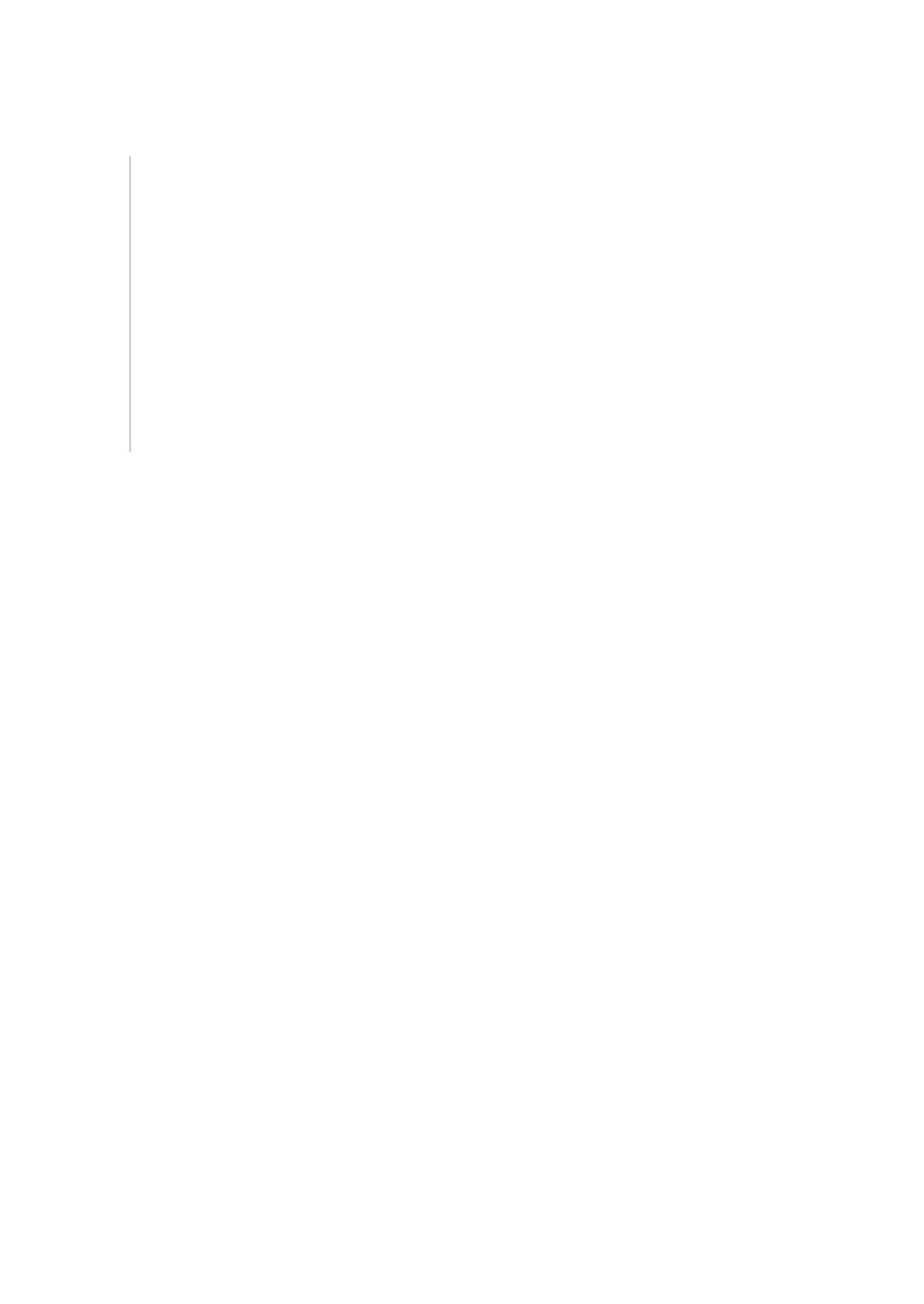
            }
        }
        \caption{89\% highest posterior credible intervals of each test data set for $D$ for the OU example.  The posteriors compared are the true posterior (True-post), the original approximate posterior (Approx-post), the adjusted posterior found using the location-scale transformation of \citet{Bon2025}, and the adjusted posteriors found using the polynomial transformation of order 1 (PBSC~1), order 2 (PBSC~2) and order 3 (PBSC~3).} 
        \label{fig:ouDcr}
    \end{figure}

    In this section, we detail the additional results for parameters $\gamma$ and $D$ from Section~\ref{subsec:OU}.  Table~\ref{tab:ouga} and Table~\ref{tab:ouD} present the results for the MSE, bias, SD, and FCP$_{0.89}$ for $\gamma$ and $D$, respectively.  Figure~\ref{fig:ougaac} and Figure~\ref{fig:ouDac} display the miscoverage across coverage levels $\alpha \in (0, 1)$ for $\gamma$ and $D$, respectively.  Finally, Figure~\ref{fig:ougacr} and Figure~\ref{fig:ouDcr} display the $89\%$ credible interval plots for $\gamma$ and $D$, respectively.
    \begin{table}[!hbtp]
        \normalsize
        \setlength{\tabcolsep}{12pt}
        \begin{tabular}{ lcccc }
            \hline \hline
            Method            & MSE           & Bias           & SD           & FCP$_{0.89}$ \\
            \hline
            True-post         & 0.285 (0.006) & -0.010 (0.012) & 0.372 (0.010) & 0.891\\
            \hline
            Approx-post       & 0.308 (0.006) & -0.004 (0.012) & 0.403 (0.002) & 0.915\\
            BSC               & 0.252 (0.006) & -0.029 (0.012) & 0.312 (0.027) & 0.789\\
            PBSC~1            & 0.297 (0.006) & -0.068 (0.012) & 0.382 (0.004) & 0.891\\
            PBSC~2           & 0.263 (0.006) & -0.064 (0.012) & 0.335 (0.001) & 0.847\\
            PBSC~3             & 0.285 (0.008) & -0.040 (0.012) & 0.326 (0.056) & 0.721\\
            \hline \hline
        \end{tabular}
        \caption{Average results for $\gamma$ over 1000 independent test data sets for the OU process example with standard errors shown in brackets where appropriate.  The posteriors compared are the true posterior (True-post), the original approximate posterior (Approx-post), the adjusted posterior found using the location-scale transformation of \citet{Bon2025} (BSC), and the adjusted posteriors found using the polynomial transformation of order 1 (PBSC~1), order 2 (PBSC~2), and order 3 (PBSC~3).  \textbf{Highlighted} is the posterior approximation with metric values that are closest to True-post's metric values overall.}
        \label{tab:ouga}
    \end{table}

    \begin{table}[!hbtp]
        \normalsize
        \setlength{\tabcolsep}{12pt}
        \begin{tabular}{ lcccc }
            \hline \hline
            Method            & MSE           & Bias           & SD           & FCP$_{0.89}$ \\
            \hline
            True-post         & 17.43 (1.186) &  0.040 (0.094) & 2.141 (0.063) & 0.893\\
            \hline
            Approx-post       & 19.21 (1.311) &  0.289 (0.096) & 2.282 (0.068) & 0.921\\
            BSC               & 17.76 (1.234) &  0.198 (0.096) & 2.159 (0.061) & 0.917\\
            PBSC~1            & 17.25 (1.151) & -0.268 (0.095) & 2.079 (0.061) & 0.859\\
            PBSC~2           & 16.95 (1.195) & -0.216 (0.096) & 1.975 (0.061) & 0.873\\
            PBSC~3             & 17.17 (1.168) & -0.201 (0.096) & 2.111 (0.058) & 0.909\\
            \hline \hline
        \end{tabular}
        \caption{Average results for $D$ over 1000 independent test data sets for the OU process example with standard errors shown in brackets where appropriate.  The posteriors compared are the true posterior (True-post), the original approximate posterior (Approx-post), the adjusted posterior found using the location-scale transformation of \citet{Bon2025} (BSC), and the adjusted posteriors found using the polynomial transformation of order 1 (PBSC~1), order 2 (PBSC~2), and order 3 (PBSC~3).  \textbf{Highlighted} is the posterior approximation with metric values that are closest to True-post's metric values overall.}
        
        \label{tab:ouD}
    \end{table}

\FloatBarrier
\section{Appendix B. Additional Results from Susceptible-Infected-Recovered Model}\label{app:resultsSIR}
\begin{figure}[h]
        \centering{
        \resizebox{0.95\textwidth}{!}{
            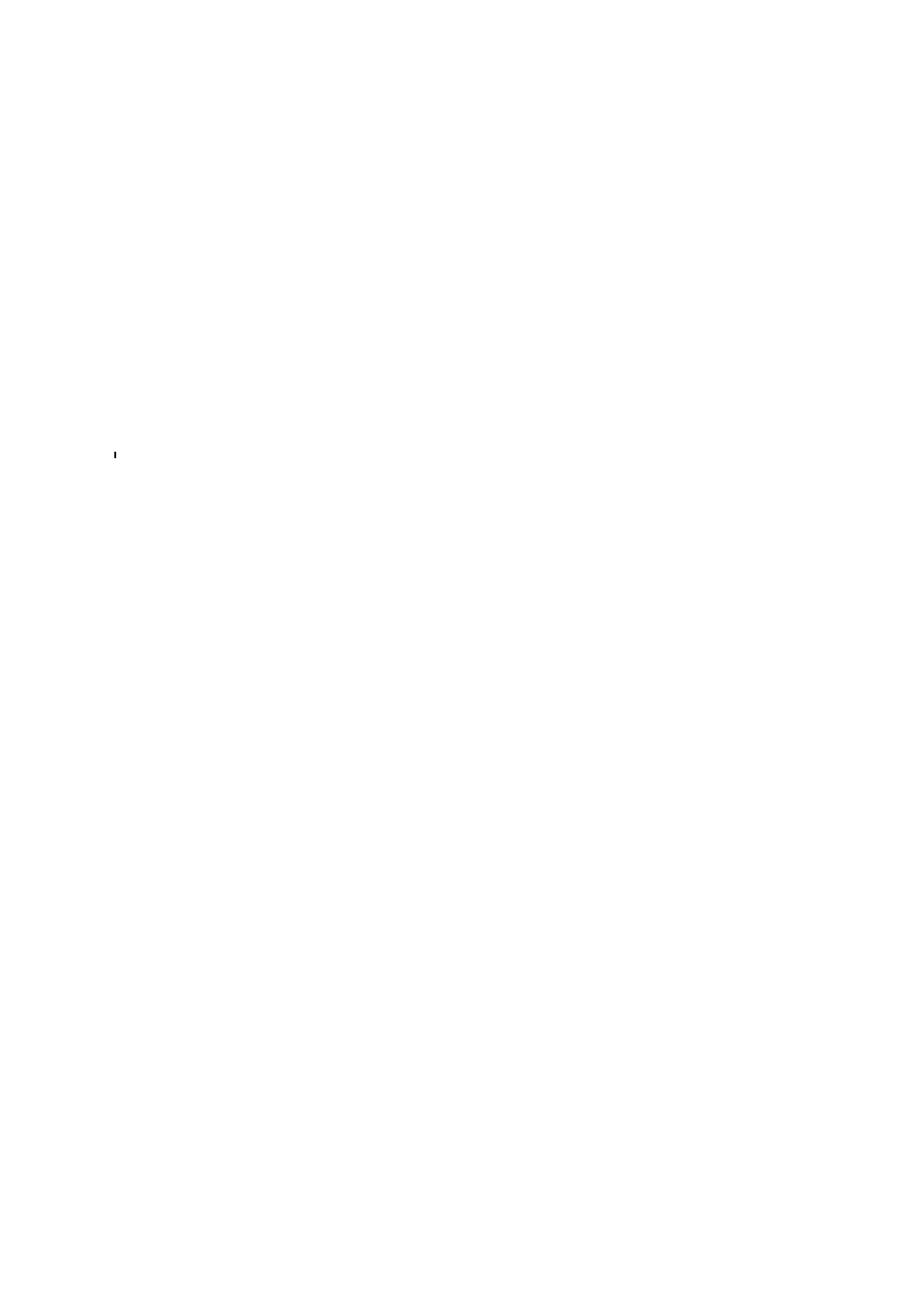
            }
        } 
        \caption{89\% highest posterior credible intervals of each test data set for $\theta_0$ for the SIR example in Section~\ref{subsec:SIR}.  The posteriors compared are the true posterior (True-post), the original approximate posterior (Approx-post), the adjusted posterior found using the location-scale transformation of \citet{Bon2025} (BSC), and the adjusted posteriors found using the sequential application of BSC for round 1 (SBSC~1), round 2 (SBSC~2) and round 3 (SBSC~3).}
        \label{fig:PALBFT0}
    \end{figure}

\begin{figure}[h]
        \centering{
        \resizebox{0.95\textwidth}{!}{
            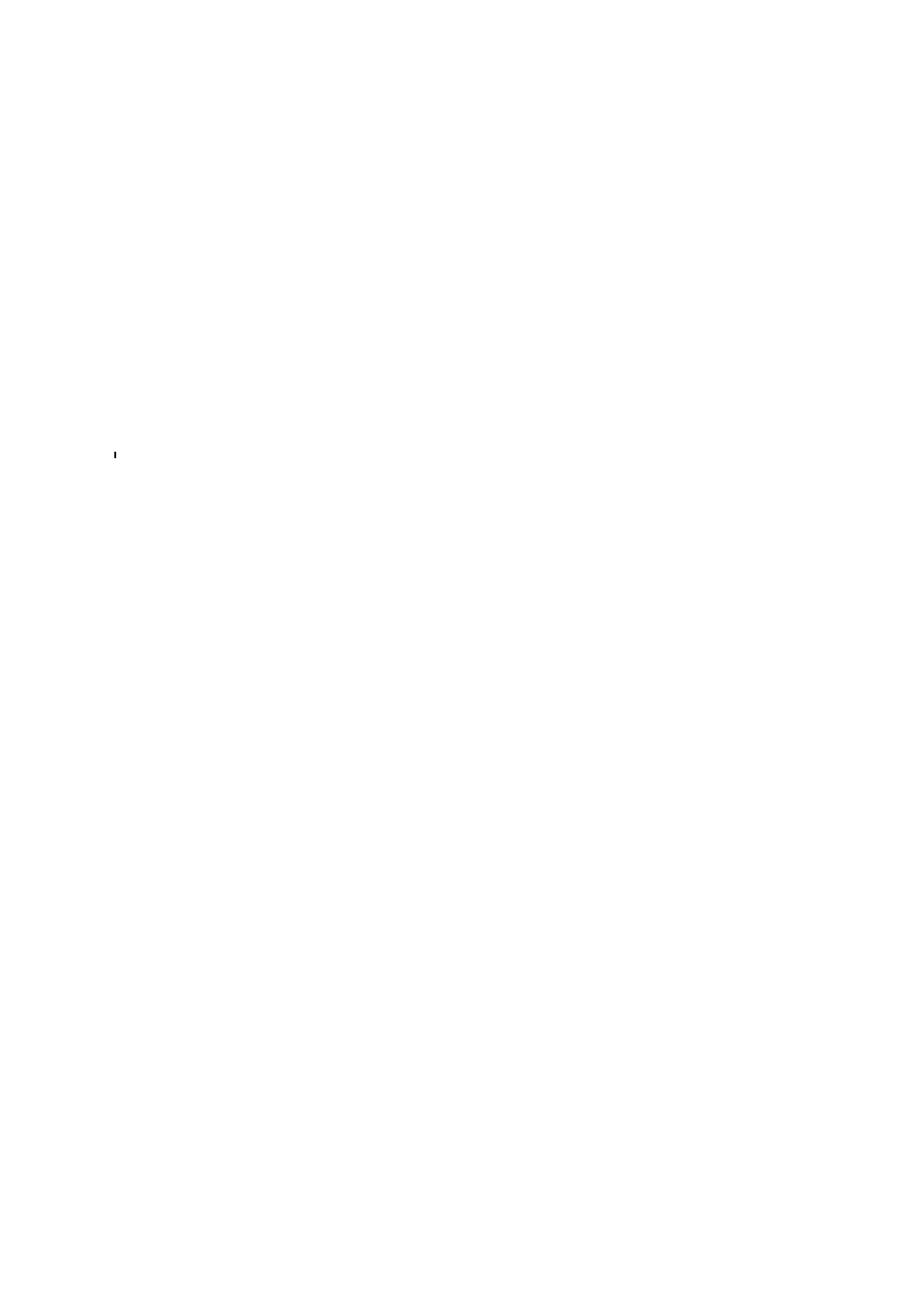
            }
        } 
        \caption{89\% highest posterior credible intervals of each test data set for $\theta_1$ for the SIR example in Section~\ref{subsec:SIR}.  The posteriors compared are the true posterior (True-post), the original approximate posterior (Approx-post), the adjusted posterior found using the location-scale transformation of \citet{Bon2025} (BSC), and the adjusted posteriors found using the sequential application of BSC for round 1 (SBSC~1), round 2 (SBSC~2) and round 3 (SBSC~3).}
        \label{fig:PALBFT1}
    \end{figure}

\begin{figure}[h]
        \centering{
        \resizebox{0.95\textwidth}{!}{
            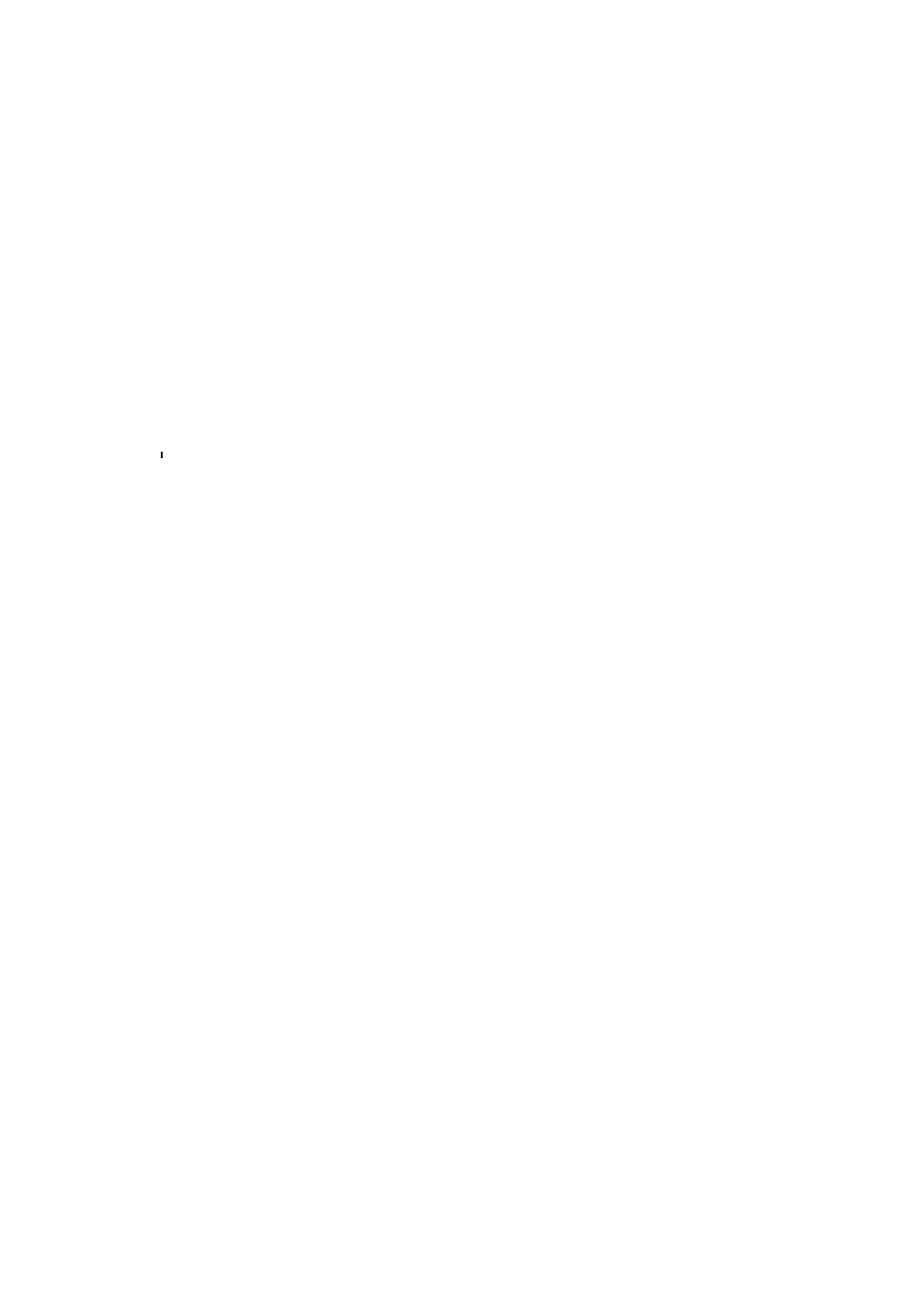
            }
        } 
        \caption{89\% highest posterior credible intervals of each test data set for $\theta_2$ for the SIR example in Section~\ref{subsec:SIR}.  The posteriors compared are the true posterior (True-post), the original approximate posterior (Approx-post), the adjusted posterior found using the location-scale transformation of \citet{Bon2025} (BSC), and the adjusted posteriors found using the sequential application of BSC for round 1 (SBSC~1), round 2 (SBSC~2) and round 3 (SBSC~3).}
        \label{fig:PALBFT2}
    \end{figure}

    In this section, we detail the additional results from Section~\ref{subsec:SIR}.  Table~\ref{tab:sirextt0}, Table~\ref{tab:sirextt1} and Table~\ref{tab:sirextt2} present the results for SBSC~$5-9$ for $\theta_0, \theta_1$ and $\theta_2$, respectively.  Figure~\ref{fig:PALBFT0}, Figure~\ref{fig:PALBFT1}, and Figure~\ref{fig:PALBFT2} display the $89\%$ credible interval plots for SBSC~$5-9$ for $\theta_0, \theta_1$ and $\theta_2$, respectively.
    \begin{table}[!hbtp]
        \normalsize
        \setlength{\tabcolsep}{12pt}
        \begin{tabular}{ lcccc }
            \hline \hline
            Method            & MSE             & Bias             & SD              & FCP$_{0.89}$ \\
            \hline
            True-post         & 0.0310 (6.2e-4) & -0.0002 (2.6e-3) & 0.1370 (4.9e-4) & 0.959\\
            \hline
            Approx-post       & 0.0672 (1.8e-3) & -0.2094 (3.6e-3) & 0.0994 (2.7e-4) & 0.327\\
            SBSC 5            & 0.0207 (0.5e-3) & ~0.0208 (3.2e-3) & 0.1006 (3.3e-3) & 0.875\\
            SBSC 6            & 0.0309 (0.5e-3) & ~0.0039 (3.2e-3) & 0.1430 (4.2e-4) & 0.978\\
            SBSC 7            & 0.0160 (0.5e-3) & ~0.0289 (3.1e-3) & 0.0720 (3.2e-4) & 0.721\\
            SBSC 8            & 0.0206 (0.5e-3) & ~0.0031 (3.2e-3) & 0.1010 (3.2e-4) & 0.891\\
            SBSC 9            & 0.0201 (0.5e-3) & ~0.0180 (3.2e-3) & 0.0981 (3.4e-4) & 0.870\\
            \hline
            \end{tabular}
            \caption{Average results for $\theta_0$ over 1000 independent test data sets for the SIR example with standard errors shown in brackets where appropriate.  The posteriors compared are the true posterior (True-post), the original approximate posterior (Approx-post), the adjusted posterior found using the location-scale transformation of \citet{Bon2025} (BSC), and the adjusted posteriors found using the sequential transformation over rounds $5-9$ (SBSC 5--9).}
            \label{tab:sirextt0}
    \end{table}  
    
    \begin{table}[!hbtp]
        \normalsize
        \setlength{\tabcolsep}{12pt}
        \begin{tabular}{ lcccc }
            \hline \hline
            Method            & MSE             & Bias             & SD              & FCP$_{0.89}$ \\
            \hline
            True-post         & 0.0010 (3.8e-5) & -0.0070 (7.7e-4) & 0.0290 (1.7e-4) & 0.945\\
            \hline
            Approx-post       & 0.0748 (5.6e-4) & ~0.2695 (1.1e-3) & 0.0294 (0.2e-3) & 0.000\\
            SBSC 5            & 0.0153 (6.1e-4) & -0.0413 (2.6e-3) & 0.0784 (0.8e-3) & 0.880\\
            SBSC 6            & 0.0190 (6.8e-4) & -0.0474 (2.7e-3) & 0.0949 (0.8e-3) & 0.939\\
            SBSC 7            & 0.0187 (7.7e-4) & -0.0458 (2.7e-3) & 0.0911 (1.1e-3) & 0.926\\
            SBSC 8            & 0.0179 (7.4e-4) & -0.0500 (2.7e-3) & 0.0854 (1.0e-3) & 0.894\\
            SBSC 9            & 0.0164 (6.3e-4) & -0.0419 (2.6e-3) & 0.0844 (0.8e-3) & 0.908\\
            \hline
            \end{tabular}
            \caption{Average results for $\theta_1$ over 1000 independent test data sets for the SIR example with standard errors shown in brackets where appropriate.  The posteriors compared are the true posterior (True-post), the original approximate posterior (Approx-post), the adjusted posterior found using the location-scale transformation of \citet{Bon2025} (BSC), and the adjusted posteriors found using the sequential transformation over rounds $5-9$ (SBSC 5--9).}
            \label{tab:sirextt1}
    \end{table}
    
    \begin{table}[!hbtp]
        \normalsize
        \setlength{\tabcolsep}{12pt}
        \begin{tabular}{ lcccc }
            \hline \hline
            Method            & MSE             & Bias             & SD              & FCP$_{0.89}$ \\
            \hline
            True-post         & 0.0037 (8.3e-5) & -0.0118 (1.1e-3) & 0.0466 (2.4e-4) & 0.963\\
            \hline
            Approx-post       & 0.0479 (7.1e-4) & ~0.2027 (1.9e-3) & 0.0560 (3.6e-4) & 0.070\\
            SBSC 5            & 0.0055 (0.8e-4) & ~0.0190 (1.2e-3) & 0.0603 (1.7e-4) & 0.965\\
            SBSC 6            & 0.0038 (0.7e-4) & ~0.0142 (1.3e-3) & 0.0445 (3.9e-4) & 0.890\\
            SBSC 7            & 0.0070 (0.7e-4) & ~0.0191 (1.1e-3) & 0.0728 (2.7e-4) & 0.990\\
            SBSC 8            & 0.0061 (0.7e-4) & ~0.0141 (1.1e-3) & 0.0676 (2.5e-4) & 0.988\\
            SBSC 9            & 0.0053 (0.7e-4) & ~0.0158 (1.2e-3) & 0.0601 (2.2e-4) & 0.973\\
            \hline
            \end{tabular}
            \caption{Average results for $\theta_2$ over 1000 independent test data sets for the SIR example with standard errors shown in brackets where appropriate.  The posteriors compared are the true posterior (True-post), the original approximate posterior (Approx-post), the adjusted posterior found using the location-scale transformation of \citet{Bon2025} (BSC), and the adjusted posteriors found using the sequential transformation over rounds $5-9$ (SBSC 5--9).}
            \label{tab:sirextt2}
    \end{table}
    
\FloatBarrier
\section{Appendix C. Package and Code Acknowledgments}\label{app:code}
We implement the BSC framework (including our transformation extensions), the DGP for both examples, and the No-U-Turn Hamiltonian Monte Carlo algorithm in Python using the JAX, NumPy, and SciPy packages with code available for our examples at \url{https://github.com/Lemiltock/PBSC-SBSC-Simstudy}.  We use the same Rcpp implementation of pMCMC as \citet{Whitehouse2023}, using R statistical software (v4.1.2).  We make minor changes to the PAL and pMCMC implementations, specifically we modify the priors and apply appropriate transformations to sample on the real number line for all three parameters.  The original implementation for pMCMC and PAL is available at \url{https://github.com/LorenzoRimella/PAL/blob/main/RealDataExperiments/BSFLU}. 
\end{appendices}


\bibliography{bibliography}

\end{document}

%% file: tmpplt/ouesoptim2.pdf_tex
\begingroup%
  \makeatletter%
  \providecommand\color[2][]{%
    \errmessage{(Inkscape) Color is used for the text in Inkscape, but the package 'color.sty' is not loaded}%
    \renewcommand\color[2][]{}%
  }%
  \providecommand\transparent[1]{%
    \errmessage{(Inkscape) Transparency is used (non-zero) for the text in Inkscape, but the package 'transparent.sty' is not loaded}%
    \renewcommand\transparent[1]{}%
  }%
  \providecommand\rotatebox[2]{#2}%
  \newcommand*\fsize{\dimexpr\f@size pt\relax}%
  \newcommand*\lineheight[1]{\fontsize{\fsize}{#1\fsize}\selectfont}%
  \ifx\svgwidth\undefined%
    \setlength{\unitlength}{415.44000244bp}%
    \ifx\svgscale\undefined%
      \relax%
    \else%
      \setlength{\unitlength}{\unitlength * \real{\svgscale}}%
    \fi%
  \else%
    \setlength{\unitlength}{\svgwidth}%
  \fi%
  \global\let\svgwidth\undefined%
  \global\let\svgscale\undefined%
  \makeatother%
  \begin{picture}(1,0.48873482)%
    \lineheight{1}%
    \setlength\tabcolsep{0pt}%
    \put(0,0){\includegraphics[width=\unitlength,page=1]{ouesoptim2.pdf}}%
    \put(0.16022727,0.04750617){\color[rgb]{0.14901961,0.14901961,0.14901961}\makebox(0,0)[t]{\lineheight{0}\smash{\begin{tabular}[t]{c}0\end{tabular}}}}%
    \put(0,0){\includegraphics[width=\unitlength,page=2]{ouesoptim2.pdf}}%
    \put(0.26087664,0.04750617){\color[rgb]{0.14901961,0.14901961,0.14901961}\makebox(0,0)[t]{\lineheight{0}\smash{\begin{tabular}[t]{c}1\end{tabular}}}}%
    \put(0,0){\includegraphics[width=\unitlength,page=3]{ouesoptim2.pdf}}%
    \put(0.36152598,0.04750617){\color[rgb]{0.14901961,0.14901961,0.14901961}\makebox(0,0)[t]{\lineheight{0}\smash{\begin{tabular}[t]{c}2\end{tabular}}}}%
    \put(0,0){\includegraphics[width=\unitlength,page=4]{ouesoptim2.pdf}}%
    \put(0.46217533,0.04750617){\color[rgb]{0.14901961,0.14901961,0.14901961}\makebox(0,0)[t]{\lineheight{0}\smash{\begin{tabular}[t]{c}3\end{tabular}}}}%
    \put(0,0){\includegraphics[width=\unitlength,page=5]{ouesoptim2.pdf}}%
    \put(0.56282468,0.04750617){\color[rgb]{0.14901961,0.14901961,0.14901961}\makebox(0,0)[t]{\lineheight{0}\smash{\begin{tabular}[t]{c}4\end{tabular}}}}%
    \put(0,0){\includegraphics[width=\unitlength,page=6]{ouesoptim2.pdf}}%
    \put(0.66347399,0.04750617){\color[rgb]{0.14901961,0.14901961,0.14901961}\makebox(0,0)[t]{\lineheight{0}\smash{\begin{tabular}[t]{c}5\end{tabular}}}}%
    \put(0,0){\includegraphics[width=\unitlength,page=7]{ouesoptim2.pdf}}%
    \put(0.76412338,0.04750617){\color[rgb]{0.14901961,0.14901961,0.14901961}\makebox(0,0)[t]{\lineheight{0}\smash{\begin{tabular}[t]{c}6\end{tabular}}}}%
    \put(0,0){\includegraphics[width=\unitlength,page=8]{ouesoptim2.pdf}}%
    \put(0.86477269,0.04750617){\color[rgb]{0.14901961,0.14901961,0.14901961}\makebox(0,0)[t]{\lineheight{0}\smash{\begin{tabular}[t]{c}7\end{tabular}}}}%
    \put(0.51249998,0.01458172){\color[rgb]{0.14901961,0.14901961,0.14901961}\makebox(0,0)[t]{\lineheight{0}\smash{\begin{tabular}[t]{c}$I$\end{tabular}}}}%
    \put(0,0){\includegraphics[width=\unitlength,page=9]{ouesoptim2.pdf}}%
    \put(0.10815039,0.09840002){\color[rgb]{0.14901961,0.14901961,0.14901961}\makebox(0,0)[rt]{\lineheight{0}\smash{\begin{tabular}[t]{r}1.5\end{tabular}}}}%
    \put(0,0){\includegraphics[width=\unitlength,page=10]{ouesoptim2.pdf}}%
    \put(0.10815039,0.17598125){\color[rgb]{0.14901961,0.14901961,0.14901961}\makebox(0,0)[rt]{\lineheight{0}\smash{\begin{tabular}[t]{r}1.6\end{tabular}}}}%
    \put(0,0){\includegraphics[width=\unitlength,page=11]{ouesoptim2.pdf}}%
    \put(0.10815039,0.25356248){\color[rgb]{0.14901961,0.14901961,0.14901961}\makebox(0,0)[rt]{\lineheight{0}\smash{\begin{tabular}[t]{r}1.7\end{tabular}}}}%
    \put(0,0){\includegraphics[width=\unitlength,page=12]{ouesoptim2.pdf}}%
    \put(0.10815039,0.33114369){\color[rgb]{0.14901961,0.14901961,0.14901961}\makebox(0,0)[rt]{\lineheight{0}\smash{\begin{tabular}[t]{r}1.8\end{tabular}}}}%
    \put(0,0){\includegraphics[width=\unitlength,page=13]{ouesoptim2.pdf}}%
    \put(0.10815039,0.4087249){\color[rgb]{0.14901961,0.14901961,0.14901961}\makebox(0,0)[rt]{\lineheight{0}\smash{\begin{tabular}[t]{r}1.9\end{tabular}}}}%
    \put(0.05523587,0.27080876){\color[rgb]{0.14901961,0.14901961,0.14901961}\rotatebox{90}{\makebox(0,0)[t]{\lineheight{0}\smash{\begin{tabular}[t]{c}Energy score\end{tabular}}}}}%
    \put(0,0){\includegraphics[width=\unitlength,page=14]{ouesoptim2.pdf}}%
    \put(0.73299712,0.4190178){\color[rgb]{0.14901961,0.14901961,0.14901961}\makebox(0,0)[lt]{\lineheight{0}\smash{\begin{tabular}[t]{l}Train Scores\end{tabular}}}}%
    \put(0,0){\includegraphics[width=\unitlength,page=15]{ouesoptim2.pdf}}%
    \put(0.73299712,0.38368629){\color[rgb]{0.14901961,0.14901961,0.14901961}\makebox(0,0)[lt]{\lineheight{0}\smash{\begin{tabular}[t]{l}Test Scores\end{tabular}}}}%
  \end{picture}%
\endgroup%

%% file: tmpplt/palesoptim2.pdf_tex
\begingroup%
  \makeatletter%
  \providecommand\color[2][]{%
    \errmessage{(Inkscape) Color is used for the text in Inkscape, but the package 'color.sty' is not loaded}%
    \renewcommand\color[2][]{}%
  }%
  \providecommand\transparent[1]{%
    \errmessage{(Inkscape) Transparency is used (non-zero) for the text in Inkscape, but the package 'transparent.sty' is not loaded}%
    \renewcommand\transparent[1]{}%
  }%
  \providecommand\rotatebox[2]{#2}%
  \newcommand*\fsize{\dimexpr\f@size pt\relax}%
  \newcommand*\lineheight[1]{\fontsize{\fsize}{#1\fsize}\selectfont}%
  \ifx\svgwidth\undefined%
    \setlength{\unitlength}{415.44000244bp}%
    \ifx\svgscale\undefined%
      \relax%
    \else%
      \setlength{\unitlength}{\unitlength * \real{\svgscale}}%
    \fi%
  \else%
    \setlength{\unitlength}{\svgwidth}%
  \fi%
  \global\let\svgwidth\undefined%
  \global\let\svgscale\undefined%
  \makeatother%
  \begin{picture}(1,0.48873482)%
    \lineheight{1}%
    \setlength\tabcolsep{0pt}%
    \put(0,0){\includegraphics[width=\unitlength,page=1]{palesoptim2.pdf}}%
    \put(0.16022727,0.05833806){\color[rgb]{0.14901961,0.14901961,0.14901961}\makebox(0,0)[t]{\lineheight{0}\smash{\begin{tabular}[t]{c}0\end{tabular}}}}%
    \put(0,0){\includegraphics[width=\unitlength,page=2]{palesoptim2.pdf}}%
    \put(0.23851011,0.05833806){\color[rgb]{0.14901961,0.14901961,0.14901961}\makebox(0,0)[t]{\lineheight{0}\smash{\begin{tabular}[t]{c}1\end{tabular}}}}%
    \put(0,0){\includegraphics[width=\unitlength,page=3]{palesoptim2.pdf}}%
    \put(0.31679294,0.05833806){\color[rgb]{0.14901961,0.14901961,0.14901961}\makebox(0,0)[t]{\lineheight{0}\smash{\begin{tabular}[t]{c}2\end{tabular}}}}%
    \put(0,0){\includegraphics[width=\unitlength,page=4]{palesoptim2.pdf}}%
    \put(0.39507577,0.05833806){\color[rgb]{0.14901961,0.14901961,0.14901961}\makebox(0,0)[t]{\lineheight{0}\smash{\begin{tabular}[t]{c}3\end{tabular}}}}%
    \put(0,0){\includegraphics[width=\unitlength,page=5]{palesoptim2.pdf}}%
    \put(0.47335857,0.05833806){\color[rgb]{0.14901961,0.14901961,0.14901961}\makebox(0,0)[t]{\lineheight{0}\smash{\begin{tabular}[t]{c}4\end{tabular}}}}%
    \put(0,0){\includegraphics[width=\unitlength,page=6]{palesoptim2.pdf}}%
    \put(0.5516414,0.05833806){\color[rgb]{0.14901961,0.14901961,0.14901961}\makebox(0,0)[t]{\lineheight{0}\smash{\begin{tabular}[t]{c}5\end{tabular}}}}%
    \put(0,0){\includegraphics[width=\unitlength,page=7]{palesoptim2.pdf}}%
    \put(0.62992427,0.05833806){\color[rgb]{0.14901961,0.14901961,0.14901961}\makebox(0,0)[t]{\lineheight{0}\smash{\begin{tabular}[t]{c}6\end{tabular}}}}%
    \put(0,0){\includegraphics[width=\unitlength,page=8]{palesoptim2.pdf}}%
    \put(0.7082071,0.05833806){\color[rgb]{0.14901961,0.14901961,0.14901961}\makebox(0,0)[t]{\lineheight{0}\smash{\begin{tabular}[t]{c}7\end{tabular}}}}%
    \put(0,0){\includegraphics[width=\unitlength,page=9]{palesoptim2.pdf}}%
    \put(0.78648986,0.05833806){\color[rgb]{0.14901961,0.14901961,0.14901961}\makebox(0,0)[t]{\lineheight{0}\smash{\begin{tabular}[t]{c}8\end{tabular}}}}%
    \put(0,0){\includegraphics[width=\unitlength,page=10]{palesoptim2.pdf}}%
    \put(0.86477269,0.05833806){\color[rgb]{0.14901961,0.14901961,0.14901961}\makebox(0,0)[t]{\lineheight{0}\smash{\begin{tabular}[t]{c}9\end{tabular}}}}%
    \put(0.51249998,0.02541361){\color[rgb]{0.14901961,0.14901961,0.14901961}\makebox(0,0)[t]{\lineheight{0}\smash{\begin{tabular}[t]{c}$r$\end{tabular}}}}%
    \put(0,0){\includegraphics[width=\unitlength,page=11]{palesoptim2.pdf}}%
    \put(0.10815039,0.10104166){\color[rgb]{0.14901961,0.14901961,0.14901961}\makebox(0,0)[rt]{\lineheight{0}\smash{\begin{tabular}[t]{r}0.0\end{tabular}}}}%
    \put(0,0){\includegraphics[width=\unitlength,page=12]{palesoptim2.pdf}}%
    \put(0.10815039,0.16774604){\color[rgb]{0.14901961,0.14901961,0.14901961}\makebox(0,0)[rt]{\lineheight{0}\smash{\begin{tabular}[t]{r}0.2\end{tabular}}}}%
    \put(0,0){\includegraphics[width=\unitlength,page=13]{palesoptim2.pdf}}%
    \put(0.10815039,0.23445039){\color[rgb]{0.14901961,0.14901961,0.14901961}\makebox(0,0)[rt]{\lineheight{0}\smash{\begin{tabular}[t]{r}0.4\end{tabular}}}}%
    \put(0,0){\includegraphics[width=\unitlength,page=14]{palesoptim2.pdf}}%
    \put(0.10815039,0.30115479){\color[rgb]{0.14901961,0.14901961,0.14901961}\makebox(0,0)[rt]{\lineheight{0}\smash{\begin{tabular}[t]{r}0.6\end{tabular}}}}%
    \put(0,0){\includegraphics[width=\unitlength,page=15]{palesoptim2.pdf}}%
    \put(0.10815039,0.36785918){\color[rgb]{0.14901961,0.14901961,0.14901961}\makebox(0,0)[rt]{\lineheight{0}\smash{\begin{tabular}[t]{r}0.8\end{tabular}}}}%
    \put(0,0){\includegraphics[width=\unitlength,page=16]{palesoptim2.pdf}}%
    \put(0.10815039,0.43456355){\color[rgb]{0.14901961,0.14901961,0.14901961}\makebox(0,0)[rt]{\lineheight{0}\smash{\begin{tabular}[t]{r}1.0\end{tabular}}}}%
    \put(0.05523587,0.28164065){\color[rgb]{0.14901961,0.14901961,0.14901961}\rotatebox{90}{\makebox(0,0)[t]{\lineheight{0}\smash{\begin{tabular}[t]{c}Energy distance\end{tabular}}}}}%
    \put(0,0){\includegraphics[width=\unitlength,page=17]{palesoptim2.pdf}}%
  \end{picture}%
\endgroup%

%% file: tmpplt/diag.pdf_tex
\begingroup%
  \makeatletter%
  \providecommand\color[2][]{%
    \errmessage{(Inkscape) Color is used for the text in Inkscape, but the package 'color.sty' is not loaded}%
    \renewcommand\color[2][]{}%
  }%
  \providecommand\transparent[1]{%
    \errmessage{(Inkscape) Transparency is used (non-zero) for the text in Inkscape, but the package 'transparent.sty' is not loaded}%
    \renewcommand\transparent[1]{}%
  }%
  \providecommand\rotatebox[2]{#2}%
  \newcommand*\fsize{\dimexpr\f@size pt\relax}%
  \newcommand*\lineheight[1]{\fontsize{\fsize}{#1\fsize}\selectfont}%
  \ifx\svgwidth\undefined%
    \setlength{\unitlength}{415.44000244bp}%
    \ifx\svgscale\undefined%
      \relax%
    \else%
      \setlength{\unitlength}{\unitlength * \real{\svgscale}}%
    \fi%
  \else%
    \setlength{\unitlength}{\svgwidth}%
  \fi%
  \global\let\svgwidth\undefined%
  \global\let\svgscale\undefined%
  \makeatother%
  \begin{picture}(1,0.65164645)%
    \lineheight{1}%
    \setlength\tabcolsep{0pt}%
    \put(0,0){\includegraphics[width=\unitlength,page=1]{diag.pdf}}%
    \put(0.14349119,0.41635094){\color[rgb]{0.14901961,0.14901961,0.14901961}\makebox(0,0)[t]{\lineheight{0}\smash{\begin{tabular}[t]{c}0.0\end{tabular}}}}%
    \put(0,0){\includegraphics[width=\unitlength,page=2]{diag.pdf}}%
    \put(0.3016144,0.41635094){\color[rgb]{0.14901961,0.14901961,0.14901961}\makebox(0,0)[t]{\lineheight{0}\smash{\begin{tabular}[t]{c}0.2\end{tabular}}}}%
    \put(0,0){\includegraphics[width=\unitlength,page=3]{diag.pdf}}%
    \put(0.45973761,0.41635094){\color[rgb]{0.14901961,0.14901961,0.14901961}\makebox(0,0)[t]{\lineheight{0}\smash{\begin{tabular}[t]{c}0.4\end{tabular}}}}%
    \put(0,0){\includegraphics[width=\unitlength,page=4]{diag.pdf}}%
    \put(0.61786085,0.41635094){\color[rgb]{0.14901961,0.14901961,0.14901961}\makebox(0,0)[t]{\lineheight{0}\smash{\begin{tabular}[t]{c}0.6\end{tabular}}}}%
    \put(0,0){\includegraphics[width=\unitlength,page=5]{diag.pdf}}%
    \put(0.77598405,0.41635094){\color[rgb]{0.14901961,0.14901961,0.14901961}\makebox(0,0)[t]{\lineheight{0}\smash{\begin{tabular}[t]{c}0.8\end{tabular}}}}%
    \put(0,0){\includegraphics[width=\unitlength,page=6]{diag.pdf}}%
    \put(0.93410726,0.41635094){\color[rgb]{0.14901961,0.14901961,0.14901961}\makebox(0,0)[t]{\lineheight{0}\smash{\begin{tabular}[t]{c}1.0\end{tabular}}}}%
    \put(0.53879921,0.38342651){\color[rgb]{0.14901961,0.14901961,0.14901961}\makebox(0,0)[t]{\lineheight{0}\smash{\begin{tabular}[t]{c}Target coverage\end{tabular}}}}%
    \put(0,0){\includegraphics[width=\unitlength,page=7]{diag.pdf}}%
    \put(0.08711077,0.49045079){\color[rgb]{0.14901961,0.14901961,0.14901961}\makebox(0,0)[rt]{\lineheight{0}\smash{\begin{tabular}[t]{r}-0.2\end{tabular}}}}%
    \put(0,0){\includegraphics[width=\unitlength,page=8]{diag.pdf}}%
    \put(0.08711077,0.56665707){\color[rgb]{0.14901961,0.14901961,0.14901961}\makebox(0,0)[rt]{\lineheight{0}\smash{\begin{tabular}[t]{r}0.0\end{tabular}}}}%
    \put(0.02551193,0.54055671){\color[rgb]{0.14901961,0.14901961,0.14901961}\rotatebox{90}{\makebox(0,0)[t]{\lineheight{0}\smash{\begin{tabular}[t]{c}Miscoverage\end{tabular}}}}}%
    \put(0,0){\includegraphics[width=\unitlength,page=9]{diag.pdf}}%
    \put(0.15426309,0.14155994){\color[rgb]{0.14901961,0.14901961,0.14901961}\makebox(0,0)[t]{\lineheight{0}\smash{\begin{tabular}[t]{c}1.5\end{tabular}}}}%
    \put(0,0){\includegraphics[width=\unitlength,page=10]{diag.pdf}}%
    \put(0.30277143,0.14155994){\color[rgb]{0.14901961,0.14901961,0.14901961}\makebox(0,0)[t]{\lineheight{0}\smash{\begin{tabular}[t]{c}2.0\end{tabular}}}}%
    \put(0,0){\includegraphics[width=\unitlength,page=11]{diag.pdf}}%
    \put(0.45127975,0.14155994){\color[rgb]{0.14901961,0.14901961,0.14901961}\makebox(0,0)[t]{\lineheight{0}\smash{\begin{tabular}[t]{c}2.5\end{tabular}}}}%
    \put(0,0){\includegraphics[width=\unitlength,page=12]{diag.pdf}}%
    \put(0.08711077,0.1781306){\color[rgb]{0.14901961,0.14901961,0.14901961}\makebox(0,0)[rt]{\lineheight{0}\smash{\begin{tabular}[t]{r}1.6\end{tabular}}}}%
    \put(0,0){\includegraphics[width=\unitlength,page=13]{diag.pdf}}%
    \put(0.08711077,0.24108273){\color[rgb]{0.14901961,0.14901961,0.14901961}\makebox(0,0)[rt]{\lineheight{0}\smash{\begin{tabular}[t]{r}1.8\end{tabular}}}}%
    \put(0,0){\includegraphics[width=\unitlength,page=14]{diag.pdf}}%
    \put(0.08711077,0.30403486){\color[rgb]{0.14901961,0.14901961,0.14901961}\makebox(0,0)[rt]{\lineheight{0}\smash{\begin{tabular}[t]{r}2.0\end{tabular}}}}%
    \put(0.03419624,0.26576572){\color[rgb]{0.14901961,0.14901961,0.14901961}\rotatebox{90}{\makebox(0,0)[t]{\lineheight{0}\smash{\begin{tabular}[t]{c}Calibration sample\end{tabular}}}}}%
    \put(0,0){\includegraphics[width=\unitlength,page=15]{diag.pdf}}%
    \put(0.64002135,0.14155994){\color[rgb]{0.14901961,0.14901961,0.14901961}\makebox(0,0)[t]{\lineheight{0}\smash{\begin{tabular}[t]{c}1.0\end{tabular}}}}%
    \put(0,0){\includegraphics[width=\unitlength,page=16]{diag.pdf}}%
    \put(0.76865569,0.14155994){\color[rgb]{0.14901961,0.14901961,0.14901961}\makebox(0,0)[t]{\lineheight{0}\smash{\begin{tabular}[t]{c}1.5\end{tabular}}}}%
    \put(0,0){\includegraphics[width=\unitlength,page=17]{diag.pdf}}%
    \put(0.89728996,0.14155994){\color[rgb]{0.14901961,0.14901961,0.14901961}\makebox(0,0)[t]{\lineheight{0}\smash{\begin{tabular}[t]{c}2.0\end{tabular}}}}%
    \put(0,0){\includegraphics[width=\unitlength,page=18]{diag.pdf}}%
    \put(0.57118002,0.18664161){\color[rgb]{0.14901961,0.14901961,0.14901961}\makebox(0,0)[rt]{\lineheight{0}\smash{\begin{tabular}[t]{r}1.25\end{tabular}}}}%
    \put(0,0){\includegraphics[width=\unitlength,page=19]{diag.pdf}}%
    \put(0.57118002,0.2489739){\color[rgb]{0.14901961,0.14901961,0.14901961}\makebox(0,0)[rt]{\lineheight{0}\smash{\begin{tabular}[t]{r}1.50\end{tabular}}}}%
    \put(0,0){\includegraphics[width=\unitlength,page=20]{diag.pdf}}%
    \put(0.57118002,0.31130622){\color[rgb]{0.14901961,0.14901961,0.14901961}\makebox(0,0)[rt]{\lineheight{0}\smash{\begin{tabular}[t]{r}1.75\end{tabular}}}}%
    \put(0,0){\includegraphics[width=\unitlength,page=21]{diag.pdf}}%
    \put(0.53879921,0.1086355){\color[rgb]{0.14901961,0.14901961,0.14901961}\makebox(0,0)[t]{\lineheight{0}\smash{\begin{tabular}[t]{c}89\% credible interval\end{tabular}}}}%
    \put(0,0){\includegraphics[width=\unitlength,page=22]{diag.pdf}}%
    \put(0.50499859,0.06388169){\color[rgb]{0.14901961,0.14901961,0.14901961}\makebox(0,0)[lt]{\lineheight{0}\smash{\begin{tabular}[t]{l}SBSC 1\end{tabular}}}}%
    \put(0,0){\includegraphics[width=\unitlength,page=23]{diag.pdf}}%
    \put(0.50499859,0.02855017){\color[rgb]{0.14901961,0.14901961,0.14901961}\makebox(0,0)[lt]{\lineheight{0}\smash{\begin{tabular}[t]{l}SBSC 2\end{tabular}}}}%
    \put(0,0){\includegraphics[width=\unitlength,page=24]{diag.pdf}}%
    \put(0.27303484,0.06393484){\color[rgb]{0.14901961,0.14901961,0.14901961}\makebox(0,0)[lt]{\lineheight{0}\smash{\begin{tabular}[t]{l}Approx-post\end{tabular}}}}%
    \put(0,0){\includegraphics[width=\unitlength,page=25]{diag.pdf}}%
    \put(0.27303484,0.02860332){\color[rgb]{0.14901961,0.14901961,0.14901961}\makebox(0,0)[lt]{\lineheight{0}\smash{\begin{tabular}[t]{l}SBSC 0\end{tabular}}}}%
    \put(0,0){\includegraphics[width=\unitlength,page=26]{diag.pdf}}%
    \put(0.73674962,0.06405789){\color[rgb]{0.14901961,0.14901961,0.14901961}\makebox(0,0)[lt]{\lineheight{0}\smash{\begin{tabular}[t]{l}SBSC 3\end{tabular}}}}%
    \put(0,0){\includegraphics[width=\unitlength,page=27]{diag.pdf}}%
    \put(0.73674962,0.02872638){\color[rgb]{0.14901961,0.14901961,0.14901961}\makebox(0,0)[lt]{\lineheight{0}\smash{\begin{tabular}[t]{l}SBSC 4\end{tabular}}}}%
  \end{picture}%
\endgroup%

%% file: tmpplt/realobsdataou.pdf_tex
\begingroup%
  \makeatletter%
  \providecommand\color[2][]{%
    \errmessage{(Inkscape) Color is used for the text in Inkscape, but the package 'color.sty' is not loaded}%
    \renewcommand\color[2][]{}%
  }%
  \providecommand\transparent[1]{%
    \errmessage{(Inkscape) Transparency is used (non-zero) for the text in Inkscape, but the package 'transparent.sty' is not loaded}%
    \renewcommand\transparent[1]{}%
  }%
  \providecommand\rotatebox[2]{#2}%
  \newcommand*\fsize{\dimexpr\f@size pt\relax}%
  \newcommand*\lineheight[1]{\fontsize{\fsize}{#1\fsize}\selectfont}%
  \ifx\svgwidth\undefined%
    \setlength{\unitlength}{415.44000244bp}%
    \ifx\svgscale\undefined%
      \relax%
    \else%
      \setlength{\unitlength}{\unitlength * \real{\svgscale}}%
    \fi%
  \else%
    \setlength{\unitlength}{\svgwidth}%
  \fi%
  \global\let\svgwidth\undefined%
  \global\let\svgscale\undefined%
  \makeatother%
  \begin{picture}(1,0.48873482)%
    \lineheight{1}%
    \setlength\tabcolsep{0pt}%
    \put(0,0){\includegraphics[width=\unitlength,page=1]{realobsdataou.pdf}}%
    \put(0.16022727,0.05472743){\makebox(0,0)[t]{\lineheight{0}\smash{\begin{tabular}[t]{c}-10\end{tabular}}}}%
    \put(0,0){\includegraphics[width=\unitlength,page=2]{realobsdataou.pdf}}%
    \put(0.30701369,0.05472743){\makebox(0,0)[t]{\lineheight{0}\smash{\begin{tabular}[t]{c}-5\end{tabular}}}}%
    \put(0,0){\includegraphics[width=\unitlength,page=3]{realobsdataou.pdf}}%
    \put(0.45380011,0.05472743){\makebox(0,0)[t]{\lineheight{0}\smash{\begin{tabular}[t]{c}0\end{tabular}}}}%
    \put(0,0){\includegraphics[width=\unitlength,page=4]{realobsdataou.pdf}}%
    \put(0.60058654,0.05472743){\makebox(0,0)[t]{\lineheight{0}\smash{\begin{tabular}[t]{c}5\end{tabular}}}}%
    \put(0,0){\includegraphics[width=\unitlength,page=5]{realobsdataou.pdf}}%
    \put(0.74737294,0.05472743){\makebox(0,0)[t]{\lineheight{0}\smash{\begin{tabular}[t]{c}10\end{tabular}}}}%
    \put(0,0){\includegraphics[width=\unitlength,page=6]{realobsdataou.pdf}}%
    \put(0.89415938,0.05472743){\makebox(0,0)[t]{\lineheight{0}\smash{\begin{tabular}[t]{c}15\end{tabular}}}}%
    \put(0.51249998,0.02180298){\makebox(0,0)[t]{\lineheight{0}\smash{\begin{tabular}[t]{c}$X_T$\end{tabular}}}}%
    \put(0,0){\includegraphics[width=\unitlength,page=7]{realobsdataou.pdf}}%
    \put(0.10815039,0.08072206){\makebox(0,0)[rt]{\lineheight{0}\smash{\begin{tabular}[t]{r}0.000\end{tabular}}}}%
    \put(0,0){\includegraphics[width=\unitlength,page=8]{realobsdataou.pdf}}%
    \put(0.10815039,0.14269036){\makebox(0,0)[rt]{\lineheight{0}\smash{\begin{tabular}[t]{r}0.025\end{tabular}}}}%
    \put(0,0){\includegraphics[width=\unitlength,page=9]{realobsdataou.pdf}}%
    \put(0.10815039,0.20465865){\makebox(0,0)[rt]{\lineheight{0}\smash{\begin{tabular}[t]{r}0.050\end{tabular}}}}%
    \put(0,0){\includegraphics[width=\unitlength,page=10]{realobsdataou.pdf}}%
    \put(0.10815039,0.26662695){\makebox(0,0)[rt]{\lineheight{0}\smash{\begin{tabular}[t]{r}0.075\end{tabular}}}}%
    \put(0,0){\includegraphics[width=\unitlength,page=11]{realobsdataou.pdf}}%
    \put(0.10815039,0.32859525){\makebox(0,0)[rt]{\lineheight{0}\smash{\begin{tabular}[t]{r}0.100\end{tabular}}}}%
    \put(0,0){\includegraphics[width=\unitlength,page=12]{realobsdataou.pdf}}%
    \put(0.10815039,0.39056356){\makebox(0,0)[rt]{\lineheight{0}\smash{\begin{tabular}[t]{r}0.125\end{tabular}}}}%
    \put(0,0){\includegraphics[width=\unitlength,page=13]{realobsdataou.pdf}}%
    \put(0.10815039,0.45253186){\makebox(0,0)[rt]{\lineheight{0}\smash{\begin{tabular}[t]{r}0.150\end{tabular}}}}%
    \put(0.02393622,0.27803002){\rotatebox{90}{\makebox(0,0)[t]{\lineheight{0}\smash{\begin{tabular}[t]{c}Pr$(X_T)$\end{tabular}}}}}%
    \put(0,0){\includegraphics[width=\unitlength,page=14]{realobsdataou.pdf}}%
    \put(0.20675669,0.43294656){\makebox(0,0)[lt]{\lineheight{0}\smash{\begin{tabular}[t]{l}Simulations\end{tabular}}}}%
    \put(0,0){\includegraphics[width=\unitlength,page=15]{realobsdataou.pdf}}%
    \put(0.20675669,0.39039379){\makebox(0,0)[lt]{\lineheight{0}\smash{\begin{tabular}[t]{l}Approximate\\Distribution\end{tabular}}}}%
    \put(0,0){\includegraphics[width=\unitlength,page=16]{realobsdataou.pdf}}%
    \put(0.20675669,0.32617723){\makebox(0,0)[lt]{\lineheight{0}\smash{\begin{tabular}[t]{l}True\\Distribution\end{tabular}}}}%
  \end{picture}%
\endgroup%

%% file: tmpplt/miscovplot1k.pdf_tex
\begingroup%
  \makeatletter%
  \providecommand\color[2][]{%
    \errmessage{(Inkscape) Color is used for the text in Inkscape, but the package 'color.sty' is not loaded}%
    \renewcommand\color[2][]{}%
  }%
  \providecommand\transparent[1]{%
    \errmessage{(Inkscape) Transparency is used (non-zero) for the text in Inkscape, but the package 'transparent.sty' is not loaded}%
    \renewcommand\transparent[1]{}%
  }%
  \providecommand\rotatebox[2]{#2}%
  \newcommand*\fsize{\dimexpr\f@size pt\relax}%
  \newcommand*\lineheight[1]{\fontsize{\fsize}{#1\fsize}\selectfont}%
  \ifx\svgwidth\undefined%
    \setlength{\unitlength}{460.79998779bp}%
    \ifx\svgscale\undefined%
      \relax%
    \else%
      \setlength{\unitlength}{\unitlength * \real{\svgscale}}%
    \fi%
  \else%
    \setlength{\unitlength}{\svgwidth}%
  \fi%
  \global\let\svgwidth\undefined%
  \global\let\svgscale\undefined%
  \makeatother%
  \begin{picture}(1,0.75000003)%
    \lineheight{1}%
    \setlength\tabcolsep{0pt}%
    \put(0,0){\includegraphics[width=\unitlength,page=1]{miscovplot1k.pdf}}%
    \put(0.16022727,0.05081939){\color[rgb]{0.14901961,0.14901961,0.14901961}\makebox(0,0)[t]{\lineheight{0}\smash{\begin{tabular}[t]{c}0.0\end{tabular}}}}%
    \put(0,0){\includegraphics[width=\unitlength,page=2]{miscovplot1k.pdf}}%
    \put(0.30113638,0.05081939){\color[rgb]{0.14901961,0.14901961,0.14901961}\makebox(0,0)[t]{\lineheight{0}\smash{\begin{tabular}[t]{c}0.2\end{tabular}}}}%
    \put(0,0){\includegraphics[width=\unitlength,page=3]{miscovplot1k.pdf}}%
    \put(0.44204548,0.05081939){\color[rgb]{0.14901961,0.14901961,0.14901961}\makebox(0,0)[t]{\lineheight{0}\smash{\begin{tabular}[t]{c}0.4\end{tabular}}}}%
    \put(0,0){\includegraphics[width=\unitlength,page=4]{miscovplot1k.pdf}}%
    \put(0.58295457,0.05081939){\color[rgb]{0.14901961,0.14901961,0.14901961}\makebox(0,0)[t]{\lineheight{0}\smash{\begin{tabular}[t]{c}0.6\end{tabular}}}}%
    \put(0,0){\includegraphics[width=\unitlength,page=5]{miscovplot1k.pdf}}%
    \put(0.72386366,0.05081939){\color[rgb]{0.14901961,0.14901961,0.14901961}\makebox(0,0)[t]{\lineheight{0}\smash{\begin{tabular}[t]{c}0.8\end{tabular}}}}%
    \put(0,0){\includegraphics[width=\unitlength,page=6]{miscovplot1k.pdf}}%
    \put(0.86477275,0.05081939){\color[rgb]{0.14901961,0.14901961,0.14901961}\makebox(0,0)[t]{\lineheight{0}\smash{\begin{tabular}[t]{c}1.0\end{tabular}}}}%
    \put(0.51250002,0.02113594){\color[rgb]{0.14901961,0.14901961,0.14901961}\makebox(0,0)[t]{\lineheight{0}\smash{\begin{tabular}[t]{c}Target coverage\end{tabular}}}}%
    \put(0,0){\includegraphics[width=\unitlength,page=7]{miscovplot1k.pdf}}%
    \put(0.10980903,0.12123625){\color[rgb]{0.14901961,0.14901961,0.14901961}\makebox(0,0)[rt]{\lineheight{0}\smash{\begin{tabular}[t]{r}-0.8\end{tabular}}}}%
    \put(0,0){\includegraphics[width=\unitlength,page=8]{miscovplot1k.pdf}}%
    \put(0.10980903,0.233296){\color[rgb]{0.14901961,0.14901961,0.14901961}\makebox(0,0)[rt]{\lineheight{0}\smash{\begin{tabular}[t]{r}-0.6\end{tabular}}}}%
    \put(0,0){\includegraphics[width=\unitlength,page=9]{miscovplot1k.pdf}}%
    \put(0.10980903,0.34535578){\color[rgb]{0.14901961,0.14901961,0.14901961}\makebox(0,0)[rt]{\lineheight{0}\smash{\begin{tabular}[t]{r}-0.4\end{tabular}}}}%
    \put(0,0){\includegraphics[width=\unitlength,page=10]{miscovplot1k.pdf}}%
    \put(0.10980903,0.45741556){\color[rgb]{0.14901961,0.14901961,0.14901961}\makebox(0,0)[rt]{\lineheight{0}\smash{\begin{tabular}[t]{r}-0.2\end{tabular}}}}%
    \put(0,0){\includegraphics[width=\unitlength,page=11]{miscovplot1k.pdf}}%
    \put(0.10980903,0.56947531){\color[rgb]{0.14901961,0.14901961,0.14901961}\makebox(0,0)[rt]{\lineheight{0}\smash{\begin{tabular}[t]{r}0.0\end{tabular}}}}%
    \put(0.05427382,0.37125002){\color[rgb]{0.14901961,0.14901961,0.14901961}\rotatebox{90}{\makebox(0,0)[t]{\lineheight{0}\smash{\begin{tabular}[t]{c}Miscoverage\end{tabular}}}}}%
    \put(0,0){\includegraphics[width=\unitlength,page=12]{miscovplot1k.pdf}}%
    \put(0.20529514,0.26581232){\color[rgb]{0.14901961,0.14901961,0.14901961}\makebox(0,0)[lt]{\lineheight{0}\smash{\begin{tabular}[t]{l}True-post\end{tabular}}}}%
    \put(0,0){\includegraphics[width=\unitlength,page=13]{miscovplot1k.pdf}}%
    \put(0.20529514,0.23395877){\color[rgb]{0.14901961,0.14901961,0.14901961}\makebox(0,0)[lt]{\lineheight{0}\smash{\begin{tabular}[t]{l}Approx-post\end{tabular}}}}%
    \put(0,0){\includegraphics[width=\unitlength,page=14]{miscovplot1k.pdf}}%
    \put(0.20529514,0.20210518){\color[rgb]{0.14901961,0.14901961,0.14901961}\makebox(0,0)[lt]{\lineheight{0}\smash{\begin{tabular}[t]{l}BSC\end{tabular}}}}%
    \put(0,0){\includegraphics[width=\unitlength,page=15]{miscovplot1k.pdf}}%
    \put(0.20529514,0.1702516){\color[rgb]{0.14901961,0.14901961,0.14901961}\makebox(0,0)[lt]{\lineheight{0}\smash{\begin{tabular}[t]{l}PBSC 1\end{tabular}}}}%
    \put(0,0){\includegraphics[width=\unitlength,page=16]{miscovplot1k.pdf}}%
    \put(0.20529514,0.13839802){\color[rgb]{0.14901961,0.14901961,0.14901961}\makebox(0,0)[lt]{\lineheight{0}\smash{\begin{tabular}[t]{l}PBSC 2\end{tabular}}}}%
    \put(0,0){\includegraphics[width=\unitlength,page=17]{miscovplot1k.pdf}}%
    \put(0.20529514,0.1065445){\color[rgb]{0.14901961,0.14901961,0.14901961}\makebox(0,0)[lt]{\lineheight{0}\smash{\begin{tabular}[t]{l}PBSC 3\end{tabular}}}}%
  \end{picture}%
\endgroup%

%% file: tmpplt/OUmu1kbf.pdf_tex
\begingroup%
  \makeatletter%
  \providecommand\color[2][]{%
    \errmessage{(Inkscape) Color is used for the text in Inkscape, but the package 'color.sty' is not loaded}%
    \renewcommand\color[2][]{}%
  }%
  \providecommand\transparent[1]{%
    \errmessage{(Inkscape) Transparency is used (non-zero) for the text in Inkscape, but the package 'transparent.sty' is not loaded}%
    \renewcommand\transparent[1]{}%
  }%
  \providecommand\rotatebox[2]{#2}%
  \newcommand*\fsize{\dimexpr\f@size pt\relax}%
  \newcommand*\lineheight[1]{\fontsize{\fsize}{#1\fsize}\selectfont}%
  \ifx\svgwidth\undefined%
    \setlength{\unitlength}{476.64001465bp}%
    \ifx\svgscale\undefined%
      \relax%
    \else%
      \setlength{\unitlength}{\unitlength * \real{\svgscale}}%
    \fi%
  \else%
    \setlength{\unitlength}{\svgwidth}%
  \fi%
  \global\let\svgwidth\undefined%
  \global\let\svgscale\undefined%
  \makeatother%
  \begin{picture}(1,1.41238669)%
    \lineheight{1}%
    \setlength\tabcolsep{0pt}%
    \put(0,0){\includegraphics[width=\unitlength,page=1]{OUmu1kbf.pdf}}%
    \put(0.138396,0.89240845){\color[rgb]{0.14901961,0.14901961,0.14901961}\makebox(0,0)[t]{\lineheight{0}\smash{\begin{tabular}[t]{c}-40\end{tabular}}}}%
    \put(0,0){\includegraphics[width=\unitlength,page=2]{OUmu1kbf.pdf}}%
    \put(0.22464025,0.89240845){\color[rgb]{0.14901961,0.14901961,0.14901961}\makebox(0,0)[t]{\lineheight{0}\smash{\begin{tabular}[t]{c}-20\end{tabular}}}}%
    \put(0,0){\includegraphics[width=\unitlength,page=3]{OUmu1kbf.pdf}}%
    \put(0.3108845,0.89240845){\color[rgb]{0.14901961,0.14901961,0.14901961}\makebox(0,0)[t]{\lineheight{0}\smash{\begin{tabular}[t]{c}0\end{tabular}}}}%
    \put(0,0){\includegraphics[width=\unitlength,page=4]{OUmu1kbf.pdf}}%
    \put(0.39712873,0.89240845){\color[rgb]{0.14901961,0.14901961,0.14901961}\makebox(0,0)[t]{\lineheight{0}\smash{\begin{tabular}[t]{c}20\end{tabular}}}}%
    \put(0,0){\includegraphics[width=\unitlength,page=5]{OUmu1kbf.pdf}}%
    \put(0.11031386,0.9955994){\color[rgb]{0.14901961,0.14901961,0.14901961}\makebox(0,0)[rt]{\lineheight{0}\smash{\begin{tabular}[t]{r}-20\end{tabular}}}}%
    \put(0,0){\includegraphics[width=\unitlength,page=6]{OUmu1kbf.pdf}}%
    \put(0.11031386,1.08066148){\color[rgb]{0.14901961,0.14901961,0.14901961}\makebox(0,0)[rt]{\lineheight{0}\smash{\begin{tabular}[t]{r}0\end{tabular}}}}%
    \put(0,0){\includegraphics[width=\unitlength,page=7]{OUmu1kbf.pdf}}%
    \put(0.11031386,1.16572357){\color[rgb]{0.14901961,0.14901961,0.14901961}\makebox(0,0)[rt]{\lineheight{0}\smash{\begin{tabular}[t]{r}20\end{tabular}}}}%
    \put(0,0){\includegraphics[width=\unitlength,page=8]{OUmu1kbf.pdf}}%
    \put(0.36096047,0.94628166){\color[rgb]{0.14901961,0.14901961,0.14901961}\makebox(0,0)[lt]{\lineheight{0}\smash{\begin{tabular}[t]{l}True-post\end{tabular}}}}%
    \put(0,0){\includegraphics[width=\unitlength,page=9]{OUmu1kbf.pdf}}%
    \put(0.63642095,0.89240845){\color[rgb]{0.14901961,0.14901961,0.14901961}\makebox(0,0)[t]{\lineheight{0}\smash{\begin{tabular}[t]{c}-20\end{tabular}}}}%
    \put(0,0){\includegraphics[width=\unitlength,page=10]{OUmu1kbf.pdf}}%
    \put(0.73010153,0.89240845){\color[rgb]{0.14901961,0.14901961,0.14901961}\makebox(0,0)[t]{\lineheight{0}\smash{\begin{tabular}[t]{c}0\end{tabular}}}}%
    \put(0,0){\includegraphics[width=\unitlength,page=11]{OUmu1kbf.pdf}}%
    \put(0.82378212,0.89240845){\color[rgb]{0.14901961,0.14901961,0.14901961}\makebox(0,0)[t]{\lineheight{0}\smash{\begin{tabular}[t]{c}20\end{tabular}}}}%
    \put(0,0){\includegraphics[width=\unitlength,page=12]{OUmu1kbf.pdf}}%
    \put(0.53304112,0.9955994){\color[rgb]{0.14901961,0.14901961,0.14901961}\makebox(0,0)[rt]{\lineheight{0}\smash{\begin{tabular}[t]{r}-20\end{tabular}}}}%
    \put(0,0){\includegraphics[width=\unitlength,page=13]{OUmu1kbf.pdf}}%
    \put(0.53304112,1.08066148){\color[rgb]{0.14901961,0.14901961,0.14901961}\makebox(0,0)[rt]{\lineheight{0}\smash{\begin{tabular}[t]{r}0\end{tabular}}}}%
    \put(0,0){\includegraphics[width=\unitlength,page=14]{OUmu1kbf.pdf}}%
    \put(0.53304112,1.16572357){\color[rgb]{0.14901961,0.14901961,0.14901961}\makebox(0,0)[rt]{\lineheight{0}\smash{\begin{tabular}[t]{r}20\end{tabular}}}}%
    \put(0,0){\includegraphics[width=\unitlength,page=15]{OUmu1kbf.pdf}}%
    \put(0.75449575,0.94628166){\color[rgb]{0.14901961,0.14901961,0.14901961}\makebox(0,0)[lt]{\lineheight{0}\smash{\begin{tabular}[t]{l}Approx-post\end{tabular}}}}%
    \put(0,0){\includegraphics[width=\unitlength,page=16]{OUmu1kbf.pdf}}%
    \put(0.21369371,0.50857161){\color[rgb]{0.14901961,0.14901961,0.14901961}\makebox(0,0)[t]{\lineheight{0}\smash{\begin{tabular}[t]{c}-20\end{tabular}}}}%
    \put(0,0){\includegraphics[width=\unitlength,page=17]{OUmu1kbf.pdf}}%
    \put(0.30737428,0.50857161){\color[rgb]{0.14901961,0.14901961,0.14901961}\makebox(0,0)[t]{\lineheight{0}\smash{\begin{tabular}[t]{c}0\end{tabular}}}}%
    \put(0,0){\includegraphics[width=\unitlength,page=18]{OUmu1kbf.pdf}}%
    \put(0.40105487,0.50857161){\color[rgb]{0.14901961,0.14901961,0.14901961}\makebox(0,0)[t]{\lineheight{0}\smash{\begin{tabular}[t]{c}20\end{tabular}}}}%
    \put(0,0){\includegraphics[width=\unitlength,page=19]{OUmu1kbf.pdf}}%
    \put(0.11031386,0.61176256){\color[rgb]{0.14901961,0.14901961,0.14901961}\makebox(0,0)[rt]{\lineheight{0}\smash{\begin{tabular}[t]{r}-20\end{tabular}}}}%
    \put(0,0){\includegraphics[width=\unitlength,page=20]{OUmu1kbf.pdf}}%
    \put(0.11031386,0.69682461){\color[rgb]{0.14901961,0.14901961,0.14901961}\makebox(0,0)[rt]{\lineheight{0}\smash{\begin{tabular}[t]{r}0\end{tabular}}}}%
    \put(0,0){\includegraphics[width=\unitlength,page=21]{OUmu1kbf.pdf}}%
    \put(0.11031386,0.78188672){\color[rgb]{0.14901961,0.14901961,0.14901961}\makebox(0,0)[rt]{\lineheight{0}\smash{\begin{tabular}[t]{r}20\end{tabular}}}}%
    \put(0,0){\includegraphics[width=\unitlength,page=22]{OUmu1kbf.pdf}}%
    \put(0.41187352,0.56244483){\color[rgb]{0.14901961,0.14901961,0.14901961}\makebox(0,0)[lt]{\lineheight{0}\smash{\begin{tabular}[t]{l}BSC\end{tabular}}}}%
    \put(0,0){\includegraphics[width=\unitlength,page=23]{OUmu1kbf.pdf}}%
    \put(0.5564118,0.50857161){\color[rgb]{0.14901961,0.14901961,0.14901961}\makebox(0,0)[t]{\lineheight{0}\smash{\begin{tabular}[t]{c}-40\end{tabular}}}}%
    \put(0,0){\includegraphics[width=\unitlength,page=24]{OUmu1kbf.pdf}}%
    \put(0.64437752,0.50857161){\color[rgb]{0.14901961,0.14901961,0.14901961}\makebox(0,0)[t]{\lineheight{0}\smash{\begin{tabular}[t]{c}-20\end{tabular}}}}%
    \put(0,0){\includegraphics[width=\unitlength,page=25]{OUmu1kbf.pdf}}%
    \put(0.73234323,0.50857161){\color[rgb]{0.14901961,0.14901961,0.14901961}\makebox(0,0)[t]{\lineheight{0}\smash{\begin{tabular}[t]{c}0\end{tabular}}}}%
    \put(0,0){\includegraphics[width=\unitlength,page=26]{OUmu1kbf.pdf}}%
    \put(0.82030894,0.50857161){\color[rgb]{0.14901961,0.14901961,0.14901961}\makebox(0,0)[t]{\lineheight{0}\smash{\begin{tabular}[t]{c}20\end{tabular}}}}%
    \put(0,0){\includegraphics[width=\unitlength,page=27]{OUmu1kbf.pdf}}%
    \put(0.53304112,0.61176256){\color[rgb]{0.14901961,0.14901961,0.14901961}\makebox(0,0)[rt]{\lineheight{0}\smash{\begin{tabular}[t]{r}-20\end{tabular}}}}%
    \put(0,0){\includegraphics[width=\unitlength,page=28]{OUmu1kbf.pdf}}%
    \put(0.53304112,0.69682461){\color[rgb]{0.14901961,0.14901961,0.14901961}\makebox(0,0)[rt]{\lineheight{0}\smash{\begin{tabular}[t]{r}0\end{tabular}}}}%
    \put(0,0){\includegraphics[width=\unitlength,page=29]{OUmu1kbf.pdf}}%
    \put(0.53304112,0.78188672){\color[rgb]{0.14901961,0.14901961,0.14901961}\makebox(0,0)[rt]{\lineheight{0}\smash{\begin{tabular}[t]{r}20\end{tabular}}}}%
    \put(0,0){\includegraphics[width=\unitlength,page=30]{OUmu1kbf.pdf}}%
    \put(0.8064709,0.56244483){\color[rgb]{0.14901961,0.14901961,0.14901961}\makebox(0,0)[lt]{\lineheight{0}\smash{\begin{tabular}[t]{l}PBSC 1\end{tabular}}}}%
    \put(0,0){\includegraphics[width=\unitlength,page=31]{OUmu1kbf.pdf}}%
    \put(0.1412499,0.12473471){\color[rgb]{0.14901961,0.14901961,0.14901961}\makebox(0,0)[t]{\lineheight{0}\smash{\begin{tabular}[t]{c}-40\end{tabular}}}}%
    \put(0,0){\includegraphics[width=\unitlength,page=32]{OUmu1kbf.pdf}}%
    \put(0.22871249,0.12473471){\color[rgb]{0.14901961,0.14901961,0.14901961}\makebox(0,0)[t]{\lineheight{0}\smash{\begin{tabular}[t]{c}-20\end{tabular}}}}%
    \put(0,0){\includegraphics[width=\unitlength,page=33]{OUmu1kbf.pdf}}%
    \put(0.31617507,0.12473471){\color[rgb]{0.14901961,0.14901961,0.14901961}\makebox(0,0)[t]{\lineheight{0}\smash{\begin{tabular}[t]{c}0\end{tabular}}}}%
    \put(0,0){\includegraphics[width=\unitlength,page=34]{OUmu1kbf.pdf}}%
    \put(0.40363766,0.12473471){\color[rgb]{0.14901961,0.14901961,0.14901961}\makebox(0,0)[t]{\lineheight{0}\smash{\begin{tabular}[t]{c}20\end{tabular}}}}%
    \put(0,0){\includegraphics[width=\unitlength,page=35]{OUmu1kbf.pdf}}%
    \put(0.11031386,0.22792566){\color[rgb]{0.14901961,0.14901961,0.14901961}\makebox(0,0)[rt]{\lineheight{0}\smash{\begin{tabular}[t]{r}-20\end{tabular}}}}%
    \put(0,0){\includegraphics[width=\unitlength,page=36]{OUmu1kbf.pdf}}%
    \put(0.11031386,0.31298777){\color[rgb]{0.14901961,0.14901961,0.14901961}\makebox(0,0)[rt]{\lineheight{0}\smash{\begin{tabular}[t]{r}0\end{tabular}}}}%
    \put(0,0){\includegraphics[width=\unitlength,page=37]{OUmu1kbf.pdf}}%
    \put(0.11031386,0.39804988){\color[rgb]{0.14901961,0.14901961,0.14901961}\makebox(0,0)[rt]{\lineheight{0}\smash{\begin{tabular}[t]{r}20\end{tabular}}}}%
    \put(0,0){\includegraphics[width=\unitlength,page=38]{OUmu1kbf.pdf}}%
    \put(0.38380594,0.17860799){\color[rgb]{0.14901961,0.14901961,0.14901961}\makebox(0,0)[lt]{\lineheight{0}\smash{\begin{tabular}[t]{l}PBSC 2\end{tabular}}}}%
    \put(0,0){\includegraphics[width=\unitlength,page=39]{OUmu1kbf.pdf}}%
    \put(0.58205434,0.12473471){\color[rgb]{0.14901961,0.14901961,0.14901961}\makebox(0,0)[t]{\lineheight{0}\smash{\begin{tabular}[t]{c}-40\end{tabular}}}}%
    \put(0,0){\includegraphics[width=\unitlength,page=40]{OUmu1kbf.pdf}}%
    \put(0.6610883,0.12473471){\color[rgb]{0.14901961,0.14901961,0.14901961}\makebox(0,0)[t]{\lineheight{0}\smash{\begin{tabular}[t]{c}-20\end{tabular}}}}%
    \put(0,0){\includegraphics[width=\unitlength,page=41]{OUmu1kbf.pdf}}%
    \put(0.74012232,0.12473471){\color[rgb]{0.14901961,0.14901961,0.14901961}\makebox(0,0)[t]{\lineheight{0}\smash{\begin{tabular}[t]{c}0\end{tabular}}}}%
    \put(0,0){\includegraphics[width=\unitlength,page=42]{OUmu1kbf.pdf}}%
    \put(0.81915627,0.12473471){\color[rgb]{0.14901961,0.14901961,0.14901961}\makebox(0,0)[t]{\lineheight{0}\smash{\begin{tabular}[t]{c}20\end{tabular}}}}%
    \put(0,0){\includegraphics[width=\unitlength,page=43]{OUmu1kbf.pdf}}%
    \put(0.89819023,0.12473471){\color[rgb]{0.14901961,0.14901961,0.14901961}\makebox(0,0)[t]{\lineheight{0}\smash{\begin{tabular}[t]{c}40\end{tabular}}}}%
    \put(0,0){\includegraphics[width=\unitlength,page=44]{OUmu1kbf.pdf}}%
    \put(0.53304112,0.22792566){\color[rgb]{0.14901961,0.14901961,0.14901961}\makebox(0,0)[rt]{\lineheight{0}\smash{\begin{tabular}[t]{r}-20\end{tabular}}}}%
    \put(0,0){\includegraphics[width=\unitlength,page=45]{OUmu1kbf.pdf}}%
    \put(0.53304112,0.31298777){\color[rgb]{0.14901961,0.14901961,0.14901961}\makebox(0,0)[rt]{\lineheight{0}\smash{\begin{tabular}[t]{r}0\end{tabular}}}}%
    \put(0,0){\includegraphics[width=\unitlength,page=46]{OUmu1kbf.pdf}}%
    \put(0.53304112,0.39804988){\color[rgb]{0.14901961,0.14901961,0.14901961}\makebox(0,0)[rt]{\lineheight{0}\smash{\begin{tabular}[t]{r}20\end{tabular}}}}%
    \put(0,0){\includegraphics[width=\unitlength,page=47]{OUmu1kbf.pdf}}%
    \put(0.80675287,0.17860799){\color[rgb]{0.14901961,0.14901961,0.14901961}\makebox(0,0)[lt]{\lineheight{0}\smash{\begin{tabular}[t]{l}PBSC 3\end{tabular}}}}%
    \put(0.5,0.01935981){\color[rgb]{0.14901961,0.14901961,0.14901961}\makebox(0,0)[t]{\lineheight{0}\smash{\begin{tabular}[t]{c}89\% credible interval\end{tabular}}}}%
    \put(0.03913001,0.58793389){\color[rgb]{0.14901961,0.14901961,0.14901961}\rotatebox{90}{\makebox(0,0)[lt]{\lineheight{0}\smash{\begin{tabular}[t]{l}Calibration sample\end{tabular}}}}}%
  \end{picture}%
\endgroup%

%% file: tmpplt/obsdata.pdf_tex
\begingroup%
  \makeatletter%
  \providecommand\color[2][]{%
    \errmessage{(Inkscape) Color is used for the text in Inkscape, but the package 'color.sty' is not loaded}%
    \renewcommand\color[2][]{}%
  }%
  \providecommand\transparent[1]{%
    \errmessage{(Inkscape) Transparency is used (non-zero) for the text in Inkscape, but the package 'transparent.sty' is not loaded}%
    \renewcommand\transparent[1]{}%
  }%
  \providecommand\rotatebox[2]{#2}%
  \newcommand*\fsize{\dimexpr\f@size pt\relax}%
  \newcommand*\lineheight[1]{\fontsize{\fsize}{#1\fsize}\selectfont}%
  \ifx\svgwidth\undefined%
    \setlength{\unitlength}{415.44000244bp}%
    \ifx\svgscale\undefined%
      \relax%
    \else%
      \setlength{\unitlength}{\unitlength * \real{\svgscale}}%
    \fi%
  \else%
    \setlength{\unitlength}{\svgwidth}%
  \fi%
  \global\let\svgwidth\undefined%
  \global\let\svgscale\undefined%
  \makeatother%
  \begin{picture}(1,0.48873482)%
    \lineheight{1}%
    \setlength\tabcolsep{0pt}%
    \put(0,0){\includegraphics[width=\unitlength,page=1]{obsdata.pdf}}%
    \put(0.1410124,0.04750617){\color[rgb]{0.14901961,0.14901961,0.14901961}\makebox(0,0)[t]{\lineheight{0}\smash{\begin{tabular}[t]{c}0\end{tabular}}}}%
    \put(0,0){\includegraphics[width=\unitlength,page=2]{obsdata.pdf}}%
    \put(0.30113635,0.04750617){\color[rgb]{0.14901961,0.14901961,0.14901961}\makebox(0,0)[t]{\lineheight{0}\smash{\begin{tabular}[t]{c}10\end{tabular}}}}%
    \put(0,0){\includegraphics[width=\unitlength,page=3]{obsdata.pdf}}%
    \put(0.46126033,0.04750617){\color[rgb]{0.14901961,0.14901961,0.14901961}\makebox(0,0)[t]{\lineheight{0}\smash{\begin{tabular}[t]{c}20\end{tabular}}}}%
    \put(0.30113635,0.01458172){\color[rgb]{0.14901961,0.14901961,0.14901961}\makebox(0,0)[t]{\lineheight{0}\smash{\begin{tabular}[t]{c}$t$\end{tabular}}}}%
    \put(0,0){\includegraphics[width=\unitlength,page=4]{obsdata.pdf}}%
    \put(0.10815039,0.09060652){\color[rgb]{0.14901961,0.14901961,0.14901961}\makebox(0,0)[rt]{\lineheight{0}\smash{\begin{tabular}[t]{r}0\end{tabular}}}}%
    \put(0,0){\includegraphics[width=\unitlength,page=5]{obsdata.pdf}}%
    \put(0.10815039,0.18040032){\color[rgb]{0.14901961,0.14901961,0.14901961}\makebox(0,0)[rt]{\lineheight{0}\smash{\begin{tabular}[t]{r}200\end{tabular}}}}%
    \put(0,0){\includegraphics[width=\unitlength,page=6]{obsdata.pdf}}%
    \put(0.10815039,0.27019413){\color[rgb]{0.14901961,0.14901961,0.14901961}\makebox(0,0)[rt]{\lineheight{0}\smash{\begin{tabular}[t]{r}400\end{tabular}}}}%
    \put(0,0){\includegraphics[width=\unitlength,page=7]{obsdata.pdf}}%
    \put(0.10815039,0.35998792){\color[rgb]{0.14901961,0.14901961,0.14901961}\makebox(0,0)[rt]{\lineheight{0}\smash{\begin{tabular}[t]{r}600\end{tabular}}}}%
    \put(0,0){\includegraphics[width=\unitlength,page=8]{obsdata.pdf}}%
    \put(0.10815039,0.44978173){\color[rgb]{0.14901961,0.14901961,0.14901961}\makebox(0,0)[rt]{\lineheight{0}\smash{\begin{tabular}[t]{r}800\end{tabular}}}}%
    \put(0,0){\includegraphics[width=\unitlength,page=9]{obsdata.pdf}}%
    \put(0.39909463,0.43119884){\color[rgb]{0.14901961,0.14901961,0.14901961}\makebox(0,0)[lt]{\lineheight{0}\smash{\begin{tabular}[t]{l}$S_t$\end{tabular}}}}%
    \put(0,0){\includegraphics[width=\unitlength,page=10]{obsdata.pdf}}%
    \put(0.39909463,0.39519785){\color[rgb]{0.14901961,0.14901961,0.14901961}\makebox(0,0)[lt]{\lineheight{0}\smash{\begin{tabular}[t]{l}$R_t$\end{tabular}}}}%
    \put(0,0){\includegraphics[width=\unitlength,page=11]{obsdata.pdf}}%
    \put(0.56373968,0.04750617){\color[rgb]{0.14901961,0.14901961,0.14901961}\makebox(0,0)[t]{\lineheight{0}\smash{\begin{tabular}[t]{c}0\end{tabular}}}}%
    \put(0,0){\includegraphics[width=\unitlength,page=12]{obsdata.pdf}}%
    \put(0.72386367,0.04750617){\color[rgb]{0.14901961,0.14901961,0.14901961}\makebox(0,0)[t]{\lineheight{0}\smash{\begin{tabular}[t]{c}10\end{tabular}}}}%
    \put(0,0){\includegraphics[width=\unitlength,page=13]{obsdata.pdf}}%
    \put(0.88398759,0.04750617){\color[rgb]{0.14901961,0.14901961,0.14901961}\makebox(0,0)[t]{\lineheight{0}\smash{\begin{tabular}[t]{c}20\end{tabular}}}}%
    \put(0.72386367,0.01458172){\color[rgb]{0.14901961,0.14901961,0.14901961}\makebox(0,0)[t]{\lineheight{0}\smash{\begin{tabular}[t]{c}$t$\end{tabular}}}}%
    \put(0,0){\includegraphics[width=\unitlength,page=14]{obsdata.pdf}}%
    \put(0.53087767,0.09060652){\color[rgb]{0.14901961,0.14901961,0.14901961}\makebox(0,0)[rt]{\lineheight{0}\smash{\begin{tabular}[t]{r}0\end{tabular}}}}%
    \put(0,0){\includegraphics[width=\unitlength,page=15]{obsdata.pdf}}%
    \put(0.53087767,0.18807501){\color[rgb]{0.14901961,0.14901961,0.14901961}\makebox(0,0)[rt]{\lineheight{0}\smash{\begin{tabular}[t]{r}100\end{tabular}}}}%
    \put(0,0){\includegraphics[width=\unitlength,page=16]{obsdata.pdf}}%
    \put(0.53087767,0.28554349){\color[rgb]{0.14901961,0.14901961,0.14901961}\makebox(0,0)[rt]{\lineheight{0}\smash{\begin{tabular}[t]{r}200\end{tabular}}}}%
    \put(0,0){\includegraphics[width=\unitlength,page=17]{obsdata.pdf}}%
    \put(0.53087767,0.38301198){\color[rgb]{0.14901961,0.14901961,0.14901961}\makebox(0,0)[rt]{\lineheight{0}\smash{\begin{tabular}[t]{r}300\end{tabular}}}}%
    \put(0,0){\includegraphics[width=\unitlength,page=18]{obsdata.pdf}}%
    \put(0.82202022,0.43076587){\color[rgb]{0.14901961,0.14901961,0.14901961}\makebox(0,0)[lt]{\lineheight{0}\smash{\begin{tabular}[t]{l}$A_t$\end{tabular}}}}%
    \put(0,0){\includegraphics[width=\unitlength,page=19]{obsdata.pdf}}%
    \put(0.82202022,0.39476488){\color[rgb]{0.14901961,0.14901961,0.14901961}\makebox(0,0)[lt]{\lineheight{0}\smash{\begin{tabular}[t]{l}$I_t$\end{tabular}}}}%
    \put(0,0){\includegraphics[width=\unitlength,page=20]{obsdata.pdf}}%
    \put(0.82202022,0.35876389){\color[rgb]{0.14901961,0.14901961,0.14901961}\makebox(0,0)[lt]{\lineheight{0}\smash{\begin{tabular}[t]{l}$X_t$\end{tabular}}}}%
  \end{picture}%
\endgroup%

%% file: tmpplt/miscovplott21k10.pdf_tex
\begingroup%
  \makeatletter%
  \providecommand\color[2][]{%
    \errmessage{(Inkscape) Color is used for the text in Inkscape, but the package 'color.sty' is not loaded}%
    \renewcommand\color[2][]{}%
  }%
  \providecommand\transparent[1]{%
    \errmessage{(Inkscape) Transparency is used (non-zero) for the text in Inkscape, but the package 'transparent.sty' is not loaded}%
    \renewcommand\transparent[1]{}%
  }%
  \providecommand\rotatebox[2]{#2}%
  \newcommand*\fsize{\dimexpr\f@size pt\relax}%
  \newcommand*\lineheight[1]{\fontsize{\fsize}{#1\fsize}\selectfont}%
  \ifx\svgwidth\undefined%
    \setlength{\unitlength}{460.79998779bp}%
    \ifx\svgscale\undefined%
      \relax%
    \else%
      \setlength{\unitlength}{\unitlength * \real{\svgscale}}%
    \fi%
  \else%
    \setlength{\unitlength}{\svgwidth}%
  \fi%
  \global\let\svgwidth\undefined%
  \global\let\svgscale\undefined%
  \makeatother%
  \begin{picture}(1,0.75000003)%
    \lineheight{1}%
    \setlength\tabcolsep{0pt}%
    \put(0,0){\includegraphics[width=\unitlength,page=1]{miscovplott21k10.pdf}}%
    \put(0.10980903,0.58133754){\color[rgb]{0.14901961,0.14901961,0.14901961}\makebox(0,0)[rt]{\lineheight{0}\smash{\begin{tabular}[t]{r}-0.5\end{tabular}}}}%
    \put(0,0){\includegraphics[width=\unitlength,page=2]{miscovplott21k10.pdf}}%
    \put(0.10980903,0.68931782){\color[rgb]{0.14901961,0.14901961,0.14901961}\makebox(0,0)[rt]{\lineheight{0}\smash{\begin{tabular}[t]{r}0.0\end{tabular}}}}%
    \put(0,0){\includegraphics[width=\unitlength,page=3]{miscovplott21k10.pdf}}%
    \put(0.10980903,0.3721528){\color[rgb]{0.14901961,0.14901961,0.14901961}\makebox(0,0)[rt]{\lineheight{0}\smash{\begin{tabular}[t]{r}-1.0\end{tabular}}}}%
    \put(0,0){\includegraphics[width=\unitlength,page=4]{miscovplott21k10.pdf}}%
    \put(0.10980903,0.44202239){\color[rgb]{0.14901961,0.14901961,0.14901961}\makebox(0,0)[rt]{\lineheight{0}\smash{\begin{tabular}[t]{r}-0.5\end{tabular}}}}%
    \put(0,0){\includegraphics[width=\unitlength,page=5]{miscovplott21k10.pdf}}%
    \put(0.10980903,0.51189195){\color[rgb]{0.14901961,0.14901961,0.14901961}\makebox(0,0)[rt]{\lineheight{0}\smash{\begin{tabular}[t]{r}0.0\end{tabular}}}}%
    \put(0.03799778,0.45914065){\color[rgb]{0.14901961,0.14901961,0.14901961}\rotatebox{90}{\makebox(0,0)[t]{\lineheight{0}\smash{\begin{tabular}[t]{c}Miscoverage\end{tabular}}}}}%
    \put(0,0){\includegraphics[width=\unitlength,page=6]{miscovplott21k10.pdf}}%
    \put(0.16022727,0.15173085){\color[rgb]{0.14901961,0.14901961,0.14901961}\makebox(0,0)[t]{\lineheight{0}\smash{\begin{tabular}[t]{c}0.0\end{tabular}}}}%
    \put(0,0){\includegraphics[width=\unitlength,page=7]{miscovplott21k10.pdf}}%
    \put(0.30113638,0.15173085){\color[rgb]{0.14901961,0.14901961,0.14901961}\makebox(0,0)[t]{\lineheight{0}\smash{\begin{tabular}[t]{c}0.2\end{tabular}}}}%
    \put(0,0){\includegraphics[width=\unitlength,page=8]{miscovplott21k10.pdf}}%
    \put(0.44204548,0.15173085){\color[rgb]{0.14901961,0.14901961,0.14901961}\makebox(0,0)[t]{\lineheight{0}\smash{\begin{tabular}[t]{c}0.4\end{tabular}}}}%
    \put(0,0){\includegraphics[width=\unitlength,page=9]{miscovplott21k10.pdf}}%
    \put(0.58295457,0.15173085){\color[rgb]{0.14901961,0.14901961,0.14901961}\makebox(0,0)[t]{\lineheight{0}\smash{\begin{tabular}[t]{c}0.6\end{tabular}}}}%
    \put(0,0){\includegraphics[width=\unitlength,page=10]{miscovplott21k10.pdf}}%
    \put(0.72386366,0.15173085){\color[rgb]{0.14901961,0.14901961,0.14901961}\makebox(0,0)[t]{\lineheight{0}\smash{\begin{tabular}[t]{c}0.8\end{tabular}}}}%
    \put(0,0){\includegraphics[width=\unitlength,page=11]{miscovplott21k10.pdf}}%
    \put(0.86477275,0.15173085){\color[rgb]{0.14901961,0.14901961,0.14901961}\makebox(0,0)[t]{\lineheight{0}\smash{\begin{tabular}[t]{c}1.0\end{tabular}}}}%
    \put(0.51250002,0.1220474){\color[rgb]{0.14901961,0.14901961,0.14901961}\makebox(0,0)[t]{\lineheight{0}\smash{\begin{tabular}[t]{c}Target coverage\end{tabular}}}}%
    \put(0,0){\includegraphics[width=\unitlength,page=12]{miscovplott21k10.pdf}}%
    \put(0.10980903,0.23511033){\color[rgb]{0.14901961,0.14901961,0.14901961}\makebox(0,0)[rt]{\lineheight{0}\smash{\begin{tabular}[t]{r}-0.5\end{tabular}}}}%
    \put(0,0){\includegraphics[width=\unitlength,page=13]{miscovplott21k10.pdf}}%
    \put(0.10980903,0.3103597){\color[rgb]{0.14901961,0.14901961,0.14901961}\makebox(0,0)[rt]{\lineheight{0}\smash{\begin{tabular}[t]{r}0.0\end{tabular}}}}%
    \put(0,0){\includegraphics[width=\unitlength,page=14]{miscovplott21k10.pdf}}%
    \put(0.30140657,0.08955727){\color[rgb]{0.14901961,0.14901961,0.14901961}\makebox(0,0)[lt]{\lineheight{0}\smash{\begin{tabular}[t]{l}True-post\end{tabular}}}}%
    \put(0,0){\includegraphics[width=\unitlength,page=15]{miscovplott21k10.pdf}}%
    \put(0.30140657,0.05770368){\color[rgb]{0.14901961,0.14901961,0.14901961}\makebox(0,0)[lt]{\lineheight{0}\smash{\begin{tabular}[t]{l}Approx-post\end{tabular}}}}%
    \put(0,0){\includegraphics[width=\unitlength,page=16]{miscovplott21k10.pdf}}%
    \put(0.30140657,0.02585013){\color[rgb]{0.14901961,0.14901961,0.14901961}\makebox(0,0)[lt]{\lineheight{0}\smash{\begin{tabular}[t]{l}BSC\end{tabular}}}}%
    \put(0,0){\includegraphics[width=\unitlength,page=17]{miscovplott21k10.pdf}}%
    \put(0.53483126,0.07332151){\color[rgb]{0.14901961,0.14901961,0.14901961}\makebox(0,0)[lt]{\lineheight{0}\smash{\begin{tabular}[t]{l}SBSC 1\end{tabular}}}}%
    \put(0,0){\includegraphics[width=\unitlength,page=18]{miscovplott21k10.pdf}}%
    \put(0.53483126,0.04146793){\color[rgb]{0.14901961,0.14901961,0.14901961}\makebox(0,0)[lt]{\lineheight{0}\smash{\begin{tabular}[t]{l}SBSC 2\end{tabular}}}}%
    \put(0,0){\includegraphics[width=\unitlength,page=19]{miscovplott21k10.pdf}}%
    \put(0.70978699,0.07345519){\color[rgb]{0.14901961,0.14901961,0.14901961}\makebox(0,0)[lt]{\lineheight{0}\smash{\begin{tabular}[t]{l}SBSC 3\end{tabular}}}}%
    \put(0,0){\includegraphics[width=\unitlength,page=20]{miscovplott21k10.pdf}}%
    \put(0.70978699,0.04160167){\color[rgb]{0.14901961,0.14901961,0.14901961}\makebox(0,0)[lt]{\lineheight{0}\smash{\begin{tabular}[t]{l}SBSC 4\end{tabular}}}}%
    \put(0.06361689,0.64632393){\color[rgb]{0.14901961,0.14901961,0.14901961}\rotatebox{90}{\makebox(0,0)[t]{\lineheight{0}\smash{\begin{tabular}[t]{c}$\theta_0$\end{tabular}}}}}%
    \put(0.0630042,0.45963946){\color[rgb]{0.14901961,0.14901961,0.14901961}\rotatebox{90}{\makebox(0,0)[t]{\lineheight{0}\smash{\begin{tabular}[t]{c}$\theta_1$\end{tabular}}}}}%
    \put(0.06232509,0.27657103){\color[rgb]{0.14901961,0.14901961,0.14901961}\rotatebox{90}{\makebox(0,0)[t]{\lineheight{0}\smash{\begin{tabular}[t]{c}$\theta_2$\end{tabular}}}}}%
  \end{picture}%
\endgroup%

%% file: tmpplt/densityPAL1k.pdf_tex
\begingroup%
  \makeatletter%
  \providecommand\color[2][]{%
    \errmessage{(Inkscape) Color is used for the text in Inkscape, but the package 'color.sty' is not loaded}%
    \renewcommand\color[2][]{}%
  }%
  \providecommand\transparent[1]{%
    \errmessage{(Inkscape) Transparency is used (non-zero) for the text in Inkscape, but the package 'transparent.sty' is not loaded}%
    \renewcommand\transparent[1]{}%
  }%
  \providecommand\rotatebox[2]{#2}%
  \newcommand*\fsize{\dimexpr\f@size pt\relax}%
  \newcommand*\lineheight[1]{\fontsize{\fsize}{#1\fsize}\selectfont}%
  \ifx\svgwidth\undefined%
    \setlength{\unitlength}{460.79998779bp}%
    \ifx\svgscale\undefined%
      \relax%
    \else%
      \setlength{\unitlength}{\unitlength * \real{\svgscale}}%
    \fi%
  \else%
    \setlength{\unitlength}{\svgwidth}%
  \fi%
  \global\let\svgwidth\undefined%
  \global\let\svgscale\undefined%
  \makeatother%
  \begin{picture}(1,0.75000003)%
    \lineheight{1}%
    \setlength\tabcolsep{0pt}%
    \put(0,0){\includegraphics[width=\unitlength,page=1]{densityPAL1k.pdf}}%
    \put(0.14397896,0.53333625){\color[rgb]{0.14901961,0.14901961,0.14901961}\makebox(0,0)[t]{\lineheight{0}\smash{\begin{tabular}[t]{c}1.0\end{tabular}}}}%
    \put(0,0){\includegraphics[width=\unitlength,page=2]{densityPAL1k.pdf}}%
    \put(0.24409594,0.53333625){\color[rgb]{0.14901961,0.14901961,0.14901961}\makebox(0,0)[t]{\lineheight{0}\smash{\begin{tabular}[t]{c}1.2\end{tabular}}}}%
    \put(0,0){\includegraphics[width=\unitlength,page=3]{densityPAL1k.pdf}}%
    \put(0.34421293,0.53333625){\color[rgb]{0.14901961,0.14901961,0.14901961}\makebox(0,0)[t]{\lineheight{0}\smash{\begin{tabular}[t]{c}1.4\end{tabular}}}}%
    \put(0,0){\includegraphics[width=\unitlength,page=4]{densityPAL1k.pdf}}%
    \put(0.44432992,0.53333625){\color[rgb]{0.14901961,0.14901961,0.14901961}\makebox(0,0)[t]{\lineheight{0}\smash{\begin{tabular}[t]{c}1.6\end{tabular}}}}%
    \put(0,0){\includegraphics[width=\unitlength,page=5]{densityPAL1k.pdf}}%
    \put(0.54444689,0.53333625){\color[rgb]{0.14901961,0.14901961,0.14901961}\makebox(0,0)[t]{\lineheight{0}\smash{\begin{tabular}[t]{c}1.8\end{tabular}}}}%
    \put(0,0){\includegraphics[width=\unitlength,page=6]{densityPAL1k.pdf}}%
    \put(0.64456385,0.53333625){\color[rgb]{0.14901961,0.14901961,0.14901961}\makebox(0,0)[t]{\lineheight{0}\smash{\begin{tabular}[t]{c}2.0\end{tabular}}}}%
    \put(0,0){\includegraphics[width=\unitlength,page=7]{densityPAL1k.pdf}}%
    \put(0.74468085,0.53333625){\color[rgb]{0.14901961,0.14901961,0.14901961}\makebox(0,0)[t]{\lineheight{0}\smash{\begin{tabular}[t]{c}2.2\end{tabular}}}}%
    \put(0,0){\includegraphics[width=\unitlength,page=8]{densityPAL1k.pdf}}%
    \put(0.84479785,0.53333625){\color[rgb]{0.14901961,0.14901961,0.14901961}\makebox(0,0)[t]{\lineheight{0}\smash{\begin{tabular}[t]{c}2.4\end{tabular}}}}%
    \put(0,0){\includegraphics[width=\unitlength,page=9]{densityPAL1k.pdf}}%
    \put(0.10980903,0.55677205){\color[rgb]{0.14901961,0.14901961,0.14901961}\makebox(0,0)[rt]{\lineheight{0}\smash{\begin{tabular}[t]{r}0\end{tabular}}}}%
    \put(0,0){\includegraphics[width=\unitlength,page=10]{densityPAL1k.pdf}}%
    \put(0.10980903,0.63691787){\color[rgb]{0.14901961,0.14901961,0.14901961}\makebox(0,0)[rt]{\lineheight{0}\smash{\begin{tabular}[t]{r}2\end{tabular}}}}%
    \put(0,0){\includegraphics[width=\unitlength,page=11]{densityPAL1k.pdf}}%
    \put(0.10980903,0.71706367){\color[rgb]{0.14901961,0.14901961,0.14901961}\makebox(0,0)[rt]{\lineheight{0}\smash{\begin{tabular}[t]{r}4\end{tabular}}}}%
    \put(0,0){\includegraphics[width=\unitlength,page=12]{densityPAL1k.pdf}}%
    \put(0.2033343,0.3295127){\color[rgb]{0.14901961,0.14901961,0.14901961}\makebox(0,0)[t]{\lineheight{0}\smash{\begin{tabular}[t]{c}0.2\end{tabular}}}}%
    \put(0,0){\includegraphics[width=\unitlength,page=13]{densityPAL1k.pdf}}%
    \put(0.36922837,0.3295127){\color[rgb]{0.14901961,0.14901961,0.14901961}\makebox(0,0)[t]{\lineheight{0}\smash{\begin{tabular}[t]{c}0.4\end{tabular}}}}%
    \put(0,0){\includegraphics[width=\unitlength,page=14]{densityPAL1k.pdf}}%
    \put(0.53512247,0.3295127){\color[rgb]{0.14901961,0.14901961,0.14901961}\makebox(0,0)[t]{\lineheight{0}\smash{\begin{tabular}[t]{c}0.6\end{tabular}}}}%
    \put(0,0){\includegraphics[width=\unitlength,page=15]{densityPAL1k.pdf}}%
    \put(0.70101654,0.3295127){\color[rgb]{0.14901961,0.14901961,0.14901961}\makebox(0,0)[t]{\lineheight{0}\smash{\begin{tabular}[t]{c}0.8\end{tabular}}}}%
    \put(0,0){\includegraphics[width=\unitlength,page=16]{densityPAL1k.pdf}}%
    \put(0.86691057,0.3295127){\color[rgb]{0.14901961,0.14901961,0.14901961}\makebox(0,0)[t]{\lineheight{0}\smash{\begin{tabular}[t]{c}1.0\end{tabular}}}}%
    \put(0,0){\includegraphics[width=\unitlength,page=17]{densityPAL1k.pdf}}%
    \put(0.10980903,0.35294852){\color[rgb]{0.14901961,0.14901961,0.14901961}\makebox(0,0)[rt]{\lineheight{0}\smash{\begin{tabular}[t]{r}0\end{tabular}}}}%
    \put(0,0){\includegraphics[width=\unitlength,page=18]{densityPAL1k.pdf}}%
    \put(0.10980903,0.4551638){\color[rgb]{0.14901961,0.14901961,0.14901961}\makebox(0,0)[rt]{\lineheight{0}\smash{\begin{tabular}[t]{r}10\end{tabular}}}}%
    \put(0.04946899,0.44611982){\color[rgb]{0.14901961,0.14901961,0.14901961}\rotatebox{90}{\makebox(0,0)[t]{\lineheight{0}\smash{\begin{tabular}[t]{c}Density\end{tabular}}}}}%
    \put(0,0){\includegraphics[width=\unitlength,page=19]{densityPAL1k.pdf}}%
    \put(0.12793431,0.12568918){\color[rgb]{0.14901961,0.14901961,0.14901961}\makebox(0,0)[t]{\lineheight{0}\smash{\begin{tabular}[t]{c}0.4\end{tabular}}}}%
    \put(0,0){\includegraphics[width=\unitlength,page=20]{densityPAL1k.pdf}}%
    \put(0.2540072,0.12568918){\color[rgb]{0.14901961,0.14901961,0.14901961}\makebox(0,0)[t]{\lineheight{0}\smash{\begin{tabular}[t]{c}0.5\end{tabular}}}}%
    \put(0,0){\includegraphics[width=\unitlength,page=21]{densityPAL1k.pdf}}%
    \put(0.38008012,0.12568918){\color[rgb]{0.14901961,0.14901961,0.14901961}\makebox(0,0)[t]{\lineheight{0}\smash{\begin{tabular}[t]{c}0.6\end{tabular}}}}%
    \put(0,0){\includegraphics[width=\unitlength,page=22]{densityPAL1k.pdf}}%
    \put(0.50615302,0.12568918){\color[rgb]{0.14901961,0.14901961,0.14901961}\makebox(0,0)[t]{\lineheight{0}\smash{\begin{tabular}[t]{c}0.7\end{tabular}}}}%
    \put(0,0){\includegraphics[width=\unitlength,page=23]{densityPAL1k.pdf}}%
    \put(0.63222589,0.12568918){\color[rgb]{0.14901961,0.14901961,0.14901961}\makebox(0,0)[t]{\lineheight{0}\smash{\begin{tabular}[t]{c}0.8\end{tabular}}}}%
    \put(0,0){\includegraphics[width=\unitlength,page=24]{densityPAL1k.pdf}}%
    \put(0.75829879,0.12568918){\color[rgb]{0.14901961,0.14901961,0.14901961}\makebox(0,0)[t]{\lineheight{0}\smash{\begin{tabular}[t]{c}0.9\end{tabular}}}}%
    \put(0,0){\includegraphics[width=\unitlength,page=25]{densityPAL1k.pdf}}%
    \put(0.88437169,0.12568918){\color[rgb]{0.14901961,0.14901961,0.14901961}\makebox(0,0)[t]{\lineheight{0}\smash{\begin{tabular}[t]{c}1.0\end{tabular}}}}%
    \put(0,0){\includegraphics[width=\unitlength,page=26]{densityPAL1k.pdf}}%
    \put(0.10980903,0.149125){\color[rgb]{0.14901961,0.14901961,0.14901961}\makebox(0,0)[rt]{\lineheight{0}\smash{\begin{tabular}[t]{r}0\end{tabular}}}}%
    \put(0,0){\includegraphics[width=\unitlength,page=27]{densityPAL1k.pdf}}%
    \put(0.10980903,0.21573193){\color[rgb]{0.14901961,0.14901961,0.14901961}\makebox(0,0)[rt]{\lineheight{0}\smash{\begin{tabular}[t]{r}5\end{tabular}}}}%
    \put(0,0){\includegraphics[width=\unitlength,page=28]{densityPAL1k.pdf}}%
    \put(0.10980903,0.28233887){\color[rgb]{0.14901961,0.14901961,0.14901961}\makebox(0,0)[rt]{\lineheight{0}\smash{\begin{tabular}[t]{r}10\end{tabular}}}}%
    \put(0,0){\includegraphics[width=\unitlength,page=29]{densityPAL1k.pdf}}%
    \put(0.31451227,0.08310883){\color[rgb]{0.14901961,0.14901961,0.14901961}\makebox(0,0)[lt]{\lineheight{0}\smash{\begin{tabular}[t]{l}True-post\end{tabular}}}}%
    \put(0,0){\includegraphics[width=\unitlength,page=30]{densityPAL1k.pdf}}%
    \put(0.31451227,0.05125525){\color[rgb]{0.14901961,0.14901961,0.14901961}\makebox(0,0)[lt]{\lineheight{0}\smash{\begin{tabular}[t]{l}Approx-post\end{tabular}}}}%
    \put(0,0){\includegraphics[width=\unitlength,page=31]{densityPAL1k.pdf}}%
    \put(0.31451227,0.0194017){\color[rgb]{0.14901961,0.14901961,0.14901961}\makebox(0,0)[lt]{\lineheight{0}\smash{\begin{tabular}[t]{l}BSC\end{tabular}}}}%
    \put(0,0){\includegraphics[width=\unitlength,page=32]{densityPAL1k.pdf}}%
    \put(0.54793696,0.06687308){\color[rgb]{0.14901961,0.14901961,0.14901961}\makebox(0,0)[lt]{\lineheight{0}\smash{\begin{tabular}[t]{l}SBSC 1\end{tabular}}}}%
    \put(0,0){\includegraphics[width=\unitlength,page=33]{densityPAL1k.pdf}}%
    \put(0.54793696,0.03501949){\color[rgb]{0.14901961,0.14901961,0.14901961}\makebox(0,0)[lt]{\lineheight{0}\smash{\begin{tabular}[t]{l}SBSC 2\end{tabular}}}}%
    \put(0,0){\includegraphics[width=\unitlength,page=34]{densityPAL1k.pdf}}%
    \put(0.72289269,0.06700675){\color[rgb]{0.14901961,0.14901961,0.14901961}\makebox(0,0)[lt]{\lineheight{0}\smash{\begin{tabular}[t]{l}SBSC 3\end{tabular}}}}%
    \put(0,0){\includegraphics[width=\unitlength,page=35]{densityPAL1k.pdf}}%
    \put(0.72289269,0.03515323){\color[rgb]{0.14901961,0.14901961,0.14901961}\makebox(0,0)[lt]{\lineheight{0}\smash{\begin{tabular}[t]{l}SBSC 4\end{tabular}}}}%
    \put(0.07672258,0.64313071){\color[rgb]{0.14901961,0.14901961,0.14901961}\rotatebox{90}{\makebox(0,0)[t]{\lineheight{0}\smash{\begin{tabular}[t]{c}$\theta_0$\end{tabular}}}}}%
    \put(0.07610991,0.44993582){\color[rgb]{0.14901961,0.14901961,0.14901961}\rotatebox{90}{\makebox(0,0)[t]{\lineheight{0}\smash{\begin{tabular}[t]{c}$\theta_1$\end{tabular}}}}}%
    \put(0.07543078,0.25059138){\color[rgb]{0.14901961,0.14901961,0.14901961}\rotatebox{90}{\makebox(0,0)[t]{\lineheight{0}\smash{\begin{tabular}[t]{c}$\theta_2$\end{tabular}}}}}%
  \end{picture}%
\endgroup%

%% file: tmpplt/miscovplot1kga.pdf_tex
\begingroup%
  \makeatletter%
  \providecommand\color[2][]{%
    \errmessage{(Inkscape) Color is used for the text in Inkscape, but the package 'color.sty' is not loaded}%
    \renewcommand\color[2][]{}%
  }%
  \providecommand\transparent[1]{%
    \errmessage{(Inkscape) Transparency is used (non-zero) for the text in Inkscape, but the package 'transparent.sty' is not loaded}%
    \renewcommand\transparent[1]{}%
  }%
  \providecommand\rotatebox[2]{#2}%
  \newcommand*\fsize{\dimexpr\f@size pt\relax}%
  \newcommand*\lineheight[1]{\fontsize{\fsize}{#1\fsize}\selectfont}%
  \ifx\svgwidth\undefined%
    \setlength{\unitlength}{460.79998779bp}%
    \ifx\svgscale\undefined%
      \relax%
    \else%
      \setlength{\unitlength}{\unitlength * \real{\svgscale}}%
    \fi%
  \else%
    \setlength{\unitlength}{\svgwidth}%
  \fi%
  \global\let\svgwidth\undefined%
  \global\let\svgscale\undefined%
  \makeatother%
  \begin{picture}(1,0.75000003)%
    \lineheight{1}%
    \setlength\tabcolsep{0pt}%
    \put(0,0){\includegraphics[width=\unitlength,page=1]{miscovplot1kga.pdf}}%
    \put(0.16022727,0.05081939){\color[rgb]{0.14901961,0.14901961,0.14901961}\makebox(0,0)[t]{\lineheight{0}\smash{\begin{tabular}[t]{c}0.0\end{tabular}}}}%
    \put(0,0){\includegraphics[width=\unitlength,page=2]{miscovplot1kga.pdf}}%
    \put(0.30113638,0.05081939){\color[rgb]{0.14901961,0.14901961,0.14901961}\makebox(0,0)[t]{\lineheight{0}\smash{\begin{tabular}[t]{c}0.2\end{tabular}}}}%
    \put(0,0){\includegraphics[width=\unitlength,page=3]{miscovplot1kga.pdf}}%
    \put(0.44204548,0.05081939){\color[rgb]{0.14901961,0.14901961,0.14901961}\makebox(0,0)[t]{\lineheight{0}\smash{\begin{tabular}[t]{c}0.4\end{tabular}}}}%
    \put(0,0){\includegraphics[width=\unitlength,page=4]{miscovplot1kga.pdf}}%
    \put(0.58295457,0.05081939){\color[rgb]{0.14901961,0.14901961,0.14901961}\makebox(0,0)[t]{\lineheight{0}\smash{\begin{tabular}[t]{c}0.6\end{tabular}}}}%
    \put(0,0){\includegraphics[width=\unitlength,page=5]{miscovplot1kga.pdf}}%
    \put(0.72386366,0.05081939){\color[rgb]{0.14901961,0.14901961,0.14901961}\makebox(0,0)[t]{\lineheight{0}\smash{\begin{tabular}[t]{c}0.8\end{tabular}}}}%
    \put(0,0){\includegraphics[width=\unitlength,page=6]{miscovplot1kga.pdf}}%
    \put(0.86477275,0.05081939){\color[rgb]{0.14901961,0.14901961,0.14901961}\makebox(0,0)[t]{\lineheight{0}\smash{\begin{tabular}[t]{c}1.0\end{tabular}}}}%
    \put(0.51250002,0.02113594){\color[rgb]{0.14901961,0.14901961,0.14901961}\makebox(0,0)[t]{\lineheight{0}\smash{\begin{tabular}[t]{c}Target coverage\end{tabular}}}}%
    \put(0,0){\includegraphics[width=\unitlength,page=7]{miscovplot1kga.pdf}}%
    \put(0.10980903,0.14296843){\color[rgb]{0.14901961,0.14901961,0.14901961}\makebox(0,0)[rt]{\lineheight{0}\smash{\begin{tabular}[t]{r}-0.15\end{tabular}}}}%
    \put(0,0){\includegraphics[width=\unitlength,page=8]{miscovplot1kga.pdf}}%
    \put(0.10980903,0.23947577){\color[rgb]{0.14901961,0.14901961,0.14901961}\makebox(0,0)[rt]{\lineheight{0}\smash{\begin{tabular}[t]{r}-0.10\end{tabular}}}}%
    \put(0,0){\includegraphics[width=\unitlength,page=9]{miscovplot1kga.pdf}}%
    \put(0.10980903,0.33598315){\color[rgb]{0.14901961,0.14901961,0.14901961}\makebox(0,0)[rt]{\lineheight{0}\smash{\begin{tabular}[t]{r}-0.05\end{tabular}}}}%
    \put(0,0){\includegraphics[width=\unitlength,page=10]{miscovplot1kga.pdf}}%
    \put(0.10980903,0.43249049){\color[rgb]{0.14901961,0.14901961,0.14901961}\makebox(0,0)[rt]{\lineheight{0}\smash{\begin{tabular}[t]{r}0.00\end{tabular}}}}%
    \put(0,0){\includegraphics[width=\unitlength,page=11]{miscovplot1kga.pdf}}%
    \put(0.10980903,0.52899784){\color[rgb]{0.14901961,0.14901961,0.14901961}\makebox(0,0)[rt]{\lineheight{0}\smash{\begin{tabular}[t]{r}0.05\end{tabular}}}}%
    \put(0,0){\includegraphics[width=\unitlength,page=12]{miscovplot1kga.pdf}}%
    \put(0.10980903,0.6255052){\color[rgb]{0.14901961,0.14901961,0.14901961}\makebox(0,0)[rt]{\lineheight{0}\smash{\begin{tabular}[t]{r}0.10\end{tabular}}}}%
    \put(0.04046631,0.37125002){\color[rgb]{0.14901961,0.14901961,0.14901961}\rotatebox{90}{\makebox(0,0)[t]{\lineheight{0}\smash{\begin{tabular}[t]{c}Miscoverage\end{tabular}}}}}%
    \put(0,0){\includegraphics[width=\unitlength,page=13]{miscovplot1kga.pdf}}%
    \put(0.20529514,0.26581232){\color[rgb]{0.14901961,0.14901961,0.14901961}\makebox(0,0)[lt]{\lineheight{0}\smash{\begin{tabular}[t]{l}True-post\end{tabular}}}}%
    \put(0,0){\includegraphics[width=\unitlength,page=14]{miscovplot1kga.pdf}}%
    \put(0.20529514,0.23395877){\color[rgb]{0.14901961,0.14901961,0.14901961}\makebox(0,0)[lt]{\lineheight{0}\smash{\begin{tabular}[t]{l}Approx-post\end{tabular}}}}%
    \put(0,0){\includegraphics[width=\unitlength,page=15]{miscovplot1kga.pdf}}%
    \put(0.20529514,0.20210518){\color[rgb]{0.14901961,0.14901961,0.14901961}\makebox(0,0)[lt]{\lineheight{0}\smash{\begin{tabular}[t]{l}BSC\end{tabular}}}}%
    \put(0,0){\includegraphics[width=\unitlength,page=16]{miscovplot1kga.pdf}}%
    \put(0.20529514,0.1702516){\color[rgb]{0.14901961,0.14901961,0.14901961}\makebox(0,0)[lt]{\lineheight{0}\smash{\begin{tabular}[t]{l}PBSC 1\end{tabular}}}}%
    \put(0,0){\includegraphics[width=\unitlength,page=17]{miscovplot1kga.pdf}}%
    \put(0.20529514,0.13839802){\color[rgb]{0.14901961,0.14901961,0.14901961}\makebox(0,0)[lt]{\lineheight{0}\smash{\begin{tabular}[t]{l}PBSC 2\end{tabular}}}}%
    \put(0,0){\includegraphics[width=\unitlength,page=18]{miscovplot1kga.pdf}}%
    \put(0.20529514,0.1065445){\color[rgb]{0.14901961,0.14901961,0.14901961}\makebox(0,0)[lt]{\lineheight{0}\smash{\begin{tabular}[t]{l}PBSC 3\end{tabular}}}}%
  \end{picture}%
\endgroup%

%% file: tmpplt/OUga1kbf.pdf_tex
\begingroup%
  \makeatletter%
  \providecommand\color[2][]{%
    \errmessage{(Inkscape) Color is used for the text in Inkscape, but the package 'color.sty' is not loaded}%
    \renewcommand\color[2][]{}%
  }%
  \providecommand\transparent[1]{%
    \errmessage{(Inkscape) Transparency is used (non-zero) for the text in Inkscape, but the package 'transparent.sty' is not loaded}%
    \renewcommand\transparent[1]{}%
  }%
  \providecommand\rotatebox[2]{#2}%
  \newcommand*\fsize{\dimexpr\f@size pt\relax}%
  \newcommand*\lineheight[1]{\fontsize{\fsize}{#1\fsize}\selectfont}%
  \ifx\svgwidth\undefined%
    \setlength{\unitlength}{476.64001465bp}%
    \ifx\svgscale\undefined%
      \relax%
    \else%
      \setlength{\unitlength}{\unitlength * \real{\svgscale}}%
    \fi%
  \else%
    \setlength{\unitlength}{\svgwidth}%
  \fi%
  \global\let\svgwidth\undefined%
  \global\let\svgscale\undefined%
  \makeatother%
  \begin{picture}(1,1.41238669)%
    \lineheight{1}%
    \setlength\tabcolsep{0pt}%
    \put(0,0){\includegraphics[width=\unitlength,page=1]{OUga1kbf.pdf}}%
    \put(0.19095956,0.89240845){\color[rgb]{0.14901961,0.14901961,0.14901961}\makebox(0,0)[t]{\lineheight{0}\smash{\begin{tabular}[t]{c}1\end{tabular}}}}%
    \put(0,0){\includegraphics[width=\unitlength,page=2]{OUga1kbf.pdf}}%
    \put(0.31991902,0.89240845){\color[rgb]{0.14901961,0.14901961,0.14901961}\makebox(0,0)[t]{\lineheight{0}\smash{\begin{tabular}[t]{c}2\end{tabular}}}}%
    \put(0,0){\includegraphics[width=\unitlength,page=3]{OUga1kbf.pdf}}%
    \put(0.44887848,0.89240845){\color[rgb]{0.14901961,0.14901961,0.14901961}\makebox(0,0)[t]{\lineheight{0}\smash{\begin{tabular}[t]{c}3\end{tabular}}}}%
    \put(0,0){\includegraphics[width=\unitlength,page=4]{OUga1kbf.pdf}}%
    \put(0.11031386,0.93347531){\color[rgb]{0.14901961,0.14901961,0.14901961}\makebox(0,0)[rt]{\lineheight{0}\smash{\begin{tabular}[t]{r}1.0\end{tabular}}}}%
    \put(0,0){\includegraphics[width=\unitlength,page=5]{OUga1kbf.pdf}}%
    \put(0.11031386,1.0019183){\color[rgb]{0.14901961,0.14901961,0.14901961}\makebox(0,0)[rt]{\lineheight{0}\smash{\begin{tabular}[t]{r}1.5\end{tabular}}}}%
    \put(0,0){\includegraphics[width=\unitlength,page=6]{OUga1kbf.pdf}}%
    \put(0.11031386,1.07036131){\color[rgb]{0.14901961,0.14901961,0.14901961}\makebox(0,0)[rt]{\lineheight{0}\smash{\begin{tabular}[t]{r}2.0\end{tabular}}}}%
    \put(0,0){\includegraphics[width=\unitlength,page=7]{OUga1kbf.pdf}}%
    \put(0.11031386,1.1388043){\color[rgb]{0.14901961,0.14901961,0.14901961}\makebox(0,0)[rt]{\lineheight{0}\smash{\begin{tabular}[t]{r}2.5\end{tabular}}}}%
    \put(0,0){\includegraphics[width=\unitlength,page=8]{OUga1kbf.pdf}}%
    \put(0.11031386,1.2072473){\color[rgb]{0.14901961,0.14901961,0.14901961}\makebox(0,0)[rt]{\lineheight{0}\smash{\begin{tabular}[t]{r}3.0\end{tabular}}}}%
    \put(0,0){\includegraphics[width=\unitlength,page=9]{OUga1kbf.pdf}}%
    \put(0.36096047,0.94628166){\color[rgb]{0.14901961,0.14901961,0.14901961}\makebox(0,0)[lt]{\lineheight{0}\smash{\begin{tabular}[t]{l}True-post\end{tabular}}}}%
    \put(0,0){\includegraphics[width=\unitlength,page=10]{OUga1kbf.pdf}}%
    \put(0.56800245,0.89240845){\color[rgb]{0.14901961,0.14901961,0.14901961}\makebox(0,0)[t]{\lineheight{0}\smash{\begin{tabular}[t]{c}1.0\end{tabular}}}}%
    \put(0,0){\includegraphics[width=\unitlength,page=11]{OUga1kbf.pdf}}%
    \put(0.64338011,0.89240845){\color[rgb]{0.14901961,0.14901961,0.14901961}\makebox(0,0)[t]{\lineheight{0}\smash{\begin{tabular}[t]{c}1.5\end{tabular}}}}%
    \put(0,0){\includegraphics[width=\unitlength,page=12]{OUga1kbf.pdf}}%
    \put(0.71875777,0.89240845){\color[rgb]{0.14901961,0.14901961,0.14901961}\makebox(0,0)[t]{\lineheight{0}\smash{\begin{tabular}[t]{c}2.0\end{tabular}}}}%
    \put(0,0){\includegraphics[width=\unitlength,page=13]{OUga1kbf.pdf}}%
    \put(0.79413543,0.89240845){\color[rgb]{0.14901961,0.14901961,0.14901961}\makebox(0,0)[t]{\lineheight{0}\smash{\begin{tabular}[t]{c}2.5\end{tabular}}}}%
    \put(0,0){\includegraphics[width=\unitlength,page=14]{OUga1kbf.pdf}}%
    \put(0.86951303,0.89240845){\color[rgb]{0.14901961,0.14901961,0.14901961}\makebox(0,0)[t]{\lineheight{0}\smash{\begin{tabular}[t]{c}3.0\end{tabular}}}}%
    \put(0,0){\includegraphics[width=\unitlength,page=15]{OUga1kbf.pdf}}%
    \put(0.53304112,0.93347531){\color[rgb]{0.14901961,0.14901961,0.14901961}\makebox(0,0)[rt]{\lineheight{0}\smash{\begin{tabular}[t]{r}1.0\end{tabular}}}}%
    \put(0,0){\includegraphics[width=\unitlength,page=16]{OUga1kbf.pdf}}%
    \put(0.53304112,1.0019183){\color[rgb]{0.14901961,0.14901961,0.14901961}\makebox(0,0)[rt]{\lineheight{0}\smash{\begin{tabular}[t]{r}1.5\end{tabular}}}}%
    \put(0,0){\includegraphics[width=\unitlength,page=17]{OUga1kbf.pdf}}%
    \put(0.53304112,1.07036131){\color[rgb]{0.14901961,0.14901961,0.14901961}\makebox(0,0)[rt]{\lineheight{0}\smash{\begin{tabular}[t]{r}2.0\end{tabular}}}}%
    \put(0,0){\includegraphics[width=\unitlength,page=18]{OUga1kbf.pdf}}%
    \put(0.53304112,1.1388043){\color[rgb]{0.14901961,0.14901961,0.14901961}\makebox(0,0)[rt]{\lineheight{0}\smash{\begin{tabular}[t]{r}2.5\end{tabular}}}}%
    \put(0,0){\includegraphics[width=\unitlength,page=19]{OUga1kbf.pdf}}%
    \put(0.53304112,1.2072473){\color[rgb]{0.14901961,0.14901961,0.14901961}\makebox(0,0)[rt]{\lineheight{0}\smash{\begin{tabular}[t]{r}3.0\end{tabular}}}}%
    \put(0,0){\includegraphics[width=\unitlength,page=20]{OUga1kbf.pdf}}%
    \put(0.75449575,0.94628166){\color[rgb]{0.14901961,0.14901961,0.14901961}\makebox(0,0)[lt]{\lineheight{0}\smash{\begin{tabular}[t]{l}Approx-post\end{tabular}}}}%
    \put(0,0){\includegraphics[width=\unitlength,page=21]{OUga1kbf.pdf}}%
    \put(0.19462182,0.50857161){\color[rgb]{0.14901961,0.14901961,0.14901961}\makebox(0,0)[t]{\lineheight{0}\smash{\begin{tabular}[t]{c}1\end{tabular}}}}%
    \put(0,0){\includegraphics[width=\unitlength,page=22]{OUga1kbf.pdf}}%
    \put(0.32183402,0.50857161){\color[rgb]{0.14901961,0.14901961,0.14901961}\makebox(0,0)[t]{\lineheight{0}\smash{\begin{tabular}[t]{c}2\end{tabular}}}}%
    \put(0,0){\includegraphics[width=\unitlength,page=23]{OUga1kbf.pdf}}%
    \put(0.44904623,0.50857161){\color[rgb]{0.14901961,0.14901961,0.14901961}\makebox(0,0)[t]{\lineheight{0}\smash{\begin{tabular}[t]{c}3\end{tabular}}}}%
    \put(0,0){\includegraphics[width=\unitlength,page=24]{OUga1kbf.pdf}}%
    \put(0.11031386,0.54963844){\color[rgb]{0.14901961,0.14901961,0.14901961}\makebox(0,0)[rt]{\lineheight{0}\smash{\begin{tabular}[t]{r}1.0\end{tabular}}}}%
    \put(0,0){\includegraphics[width=\unitlength,page=25]{OUga1kbf.pdf}}%
    \put(0.11031386,0.61808146){\color[rgb]{0.14901961,0.14901961,0.14901961}\makebox(0,0)[rt]{\lineheight{0}\smash{\begin{tabular}[t]{r}1.5\end{tabular}}}}%
    \put(0,0){\includegraphics[width=\unitlength,page=26]{OUga1kbf.pdf}}%
    \put(0.11031386,0.68652448){\color[rgb]{0.14901961,0.14901961,0.14901961}\makebox(0,0)[rt]{\lineheight{0}\smash{\begin{tabular}[t]{r}2.0\end{tabular}}}}%
    \put(0,0){\includegraphics[width=\unitlength,page=27]{OUga1kbf.pdf}}%
    \put(0.11031386,0.75496743){\color[rgb]{0.14901961,0.14901961,0.14901961}\makebox(0,0)[rt]{\lineheight{0}\smash{\begin{tabular}[t]{r}2.5\end{tabular}}}}%
    \put(0,0){\includegraphics[width=\unitlength,page=28]{OUga1kbf.pdf}}%
    \put(0.11031386,0.82341045){\color[rgb]{0.14901961,0.14901961,0.14901961}\makebox(0,0)[rt]{\lineheight{0}\smash{\begin{tabular}[t]{r}3.0\end{tabular}}}}%
    \put(0,0){\includegraphics[width=\unitlength,page=29]{OUga1kbf.pdf}}%
    \put(0.56800245,0.50857161){\color[rgb]{0.14901961,0.14901961,0.14901961}\makebox(0,0)[t]{\lineheight{0}\smash{\begin{tabular}[t]{c}1.0\end{tabular}}}}%
    \put(0,0){\includegraphics[width=\unitlength,page=30]{OUga1kbf.pdf}}%
    \put(0.64338011,0.50857161){\color[rgb]{0.14901961,0.14901961,0.14901961}\makebox(0,0)[t]{\lineheight{0}\smash{\begin{tabular}[t]{c}1.5\end{tabular}}}}%
    \put(0,0){\includegraphics[width=\unitlength,page=31]{OUga1kbf.pdf}}%
    \put(0.71875777,0.50857161){\color[rgb]{0.14901961,0.14901961,0.14901961}\makebox(0,0)[t]{\lineheight{0}\smash{\begin{tabular}[t]{c}2.0\end{tabular}}}}%
    \put(0,0){\includegraphics[width=\unitlength,page=32]{OUga1kbf.pdf}}%
    \put(0.79413543,0.50857161){\color[rgb]{0.14901961,0.14901961,0.14901961}\makebox(0,0)[t]{\lineheight{0}\smash{\begin{tabular}[t]{c}2.5\end{tabular}}}}%
    \put(0,0){\includegraphics[width=\unitlength,page=33]{OUga1kbf.pdf}}%
    \put(0.86951303,0.50857161){\color[rgb]{0.14901961,0.14901961,0.14901961}\makebox(0,0)[t]{\lineheight{0}\smash{\begin{tabular}[t]{c}3.0\end{tabular}}}}%
    \put(0,0){\includegraphics[width=\unitlength,page=34]{OUga1kbf.pdf}}%
    \put(0.53304112,0.54963844){\color[rgb]{0.14901961,0.14901961,0.14901961}\makebox(0,0)[rt]{\lineheight{0}\smash{\begin{tabular}[t]{r}1.0\end{tabular}}}}%
    \put(0,0){\includegraphics[width=\unitlength,page=35]{OUga1kbf.pdf}}%
    \put(0.53304112,0.61808146){\color[rgb]{0.14901961,0.14901961,0.14901961}\makebox(0,0)[rt]{\lineheight{0}\smash{\begin{tabular}[t]{r}1.5\end{tabular}}}}%
    \put(0,0){\includegraphics[width=\unitlength,page=36]{OUga1kbf.pdf}}%
    \put(0.53304112,0.68652448){\color[rgb]{0.14901961,0.14901961,0.14901961}\makebox(0,0)[rt]{\lineheight{0}\smash{\begin{tabular}[t]{r}2.0\end{tabular}}}}%
    \put(0,0){\includegraphics[width=\unitlength,page=37]{OUga1kbf.pdf}}%
    \put(0.53304112,0.75496743){\color[rgb]{0.14901961,0.14901961,0.14901961}\makebox(0,0)[rt]{\lineheight{0}\smash{\begin{tabular}[t]{r}2.5\end{tabular}}}}%
    \put(0,0){\includegraphics[width=\unitlength,page=38]{OUga1kbf.pdf}}%
    \put(0.53304112,0.82341045){\color[rgb]{0.14901961,0.14901961,0.14901961}\makebox(0,0)[rt]{\lineheight{0}\smash{\begin{tabular}[t]{r}3.0\end{tabular}}}}%
    \put(0,0){\includegraphics[width=\unitlength,page=39]{OUga1kbf.pdf}}%
    \put(0.14527518,0.12473471){\color[rgb]{0.14901961,0.14901961,0.14901961}\makebox(0,0)[t]{\lineheight{0}\smash{\begin{tabular}[t]{c}1.0\end{tabular}}}}%
    \put(0,0){\includegraphics[width=\unitlength,page=40]{OUga1kbf.pdf}}%
    \put(0.22065283,0.12473471){\color[rgb]{0.14901961,0.14901961,0.14901961}\makebox(0,0)[t]{\lineheight{0}\smash{\begin{tabular}[t]{c}1.5\end{tabular}}}}%
    \put(0,0){\includegraphics[width=\unitlength,page=41]{OUga1kbf.pdf}}%
    \put(0.29603049,0.12473471){\color[rgb]{0.14901961,0.14901961,0.14901961}\makebox(0,0)[t]{\lineheight{0}\smash{\begin{tabular}[t]{c}2.0\end{tabular}}}}%
    \put(0,0){\includegraphics[width=\unitlength,page=42]{OUga1kbf.pdf}}%
    \put(0.37140815,0.12473471){\color[rgb]{0.14901961,0.14901961,0.14901961}\makebox(0,0)[t]{\lineheight{0}\smash{\begin{tabular}[t]{c}2.5\end{tabular}}}}%
    \put(0,0){\includegraphics[width=\unitlength,page=43]{OUga1kbf.pdf}}%
    \put(0.44678581,0.12473471){\color[rgb]{0.14901961,0.14901961,0.14901961}\makebox(0,0)[t]{\lineheight{0}\smash{\begin{tabular}[t]{c}3.0\end{tabular}}}}%
    \put(0,0){\includegraphics[width=\unitlength,page=44]{OUga1kbf.pdf}}%
    \put(0.11031386,0.16580167){\color[rgb]{0.14901961,0.14901961,0.14901961}\makebox(0,0)[rt]{\lineheight{0}\smash{\begin{tabular}[t]{r}1.0\end{tabular}}}}%
    \put(0,0){\includegraphics[width=\unitlength,page=45]{OUga1kbf.pdf}}%
    \put(0.11031386,0.23424456){\color[rgb]{0.14901961,0.14901961,0.14901961}\makebox(0,0)[rt]{\lineheight{0}\smash{\begin{tabular}[t]{r}1.5\end{tabular}}}}%
    \put(0,0){\includegraphics[width=\unitlength,page=46]{OUga1kbf.pdf}}%
    \put(0.11031386,0.30268758){\color[rgb]{0.14901961,0.14901961,0.14901961}\makebox(0,0)[rt]{\lineheight{0}\smash{\begin{tabular}[t]{r}2.0\end{tabular}}}}%
    \put(0,0){\includegraphics[width=\unitlength,page=47]{OUga1kbf.pdf}}%
    \put(0.11031386,0.37113059){\color[rgb]{0.14901961,0.14901961,0.14901961}\makebox(0,0)[rt]{\lineheight{0}\smash{\begin{tabular}[t]{r}2.5\end{tabular}}}}%
    \put(0,0){\includegraphics[width=\unitlength,page=48]{OUga1kbf.pdf}}%
    \put(0.11031386,0.43957361){\color[rgb]{0.14901961,0.14901961,0.14901961}\makebox(0,0)[rt]{\lineheight{0}\smash{\begin{tabular}[t]{r}3.0\end{tabular}}}}%
    \put(0,0){\includegraphics[width=\unitlength,page=49]{OUga1kbf.pdf}}%
    \put(0.65122412,0.12473471){\color[rgb]{0.14901961,0.14901961,0.14901961}\makebox(0,0)[t]{\lineheight{0}\smash{\begin{tabular}[t]{c}1\end{tabular}}}}%
    \put(0,0){\includegraphics[width=\unitlength,page=50]{OUga1kbf.pdf}}%
    \put(0.75505361,0.12473471){\color[rgb]{0.14901961,0.14901961,0.14901961}\makebox(0,0)[t]{\lineheight{0}\smash{\begin{tabular}[t]{c}2\end{tabular}}}}%
    \put(0,0){\includegraphics[width=\unitlength,page=51]{OUga1kbf.pdf}}%
    \put(0.8588831,0.12473471){\color[rgb]{0.14901961,0.14901961,0.14901961}\makebox(0,0)[t]{\lineheight{0}\smash{\begin{tabular}[t]{c}3\end{tabular}}}}%
    \put(0,0){\includegraphics[width=\unitlength,page=52]{OUga1kbf.pdf}}%
    \put(0.53304112,0.16580167){\color[rgb]{0.14901961,0.14901961,0.14901961}\makebox(0,0)[rt]{\lineheight{0}\smash{\begin{tabular}[t]{r}1.0\end{tabular}}}}%
    \put(0,0){\includegraphics[width=\unitlength,page=53]{OUga1kbf.pdf}}%
    \put(0.53304112,0.23424456){\color[rgb]{0.14901961,0.14901961,0.14901961}\makebox(0,0)[rt]{\lineheight{0}\smash{\begin{tabular}[t]{r}1.5\end{tabular}}}}%
    \put(0,0){\includegraphics[width=\unitlength,page=54]{OUga1kbf.pdf}}%
    \put(0.53304112,0.30268758){\color[rgb]{0.14901961,0.14901961,0.14901961}\makebox(0,0)[rt]{\lineheight{0}\smash{\begin{tabular}[t]{r}2.0\end{tabular}}}}%
    \put(0,0){\includegraphics[width=\unitlength,page=55]{OUga1kbf.pdf}}%
    \put(0.53304112,0.37113059){\color[rgb]{0.14901961,0.14901961,0.14901961}\makebox(0,0)[rt]{\lineheight{0}\smash{\begin{tabular}[t]{r}2.5\end{tabular}}}}%
    \put(0,0){\includegraphics[width=\unitlength,page=56]{OUga1kbf.pdf}}%
    \put(0.53304112,0.43957361){\color[rgb]{0.14901961,0.14901961,0.14901961}\makebox(0,0)[rt]{\lineheight{0}\smash{\begin{tabular}[t]{r}3.0\end{tabular}}}}%
    \put(0,0){\includegraphics[width=\unitlength,page=57]{OUga1kbf.pdf}}%
    \put(0.5,0.01935981){\color[rgb]{0.14901961,0.14901961,0.14901961}\makebox(0,0)[t]{\lineheight{0}\smash{\begin{tabular}[t]{c}89\% credible interval\end{tabular}}}}%
    \put(0.03913001,0.58793389){\color[rgb]{0.14901961,0.14901961,0.14901961}\rotatebox{90}{\makebox(0,0)[lt]{\lineheight{0}\smash{\begin{tabular}[t]{l}Calibration sample\end{tabular}}}}}%
    \put(0,0){\includegraphics[width=\unitlength,page=58]{OUga1kbf.pdf}}%
    \put(0.41565037,0.55925846){\color[rgb]{0.14901961,0.14901961,0.14901961}\makebox(0,0)[lt]{\lineheight{0}\smash{\begin{tabular}[t]{l}BSC\end{tabular}}}}%
    \put(0,0){\includegraphics[width=\unitlength,page=59]{OUga1kbf.pdf}}%
    \put(0.81024775,0.55925846){\color[rgb]{0.14901961,0.14901961,0.14901961}\makebox(0,0)[lt]{\lineheight{0}\smash{\begin{tabular}[t]{l}PBSC 1\end{tabular}}}}%
    \put(0,0){\includegraphics[width=\unitlength,page=60]{OUga1kbf.pdf}}%
    \put(0.3875828,0.17542162){\color[rgb]{0.14901961,0.14901961,0.14901961}\makebox(0,0)[lt]{\lineheight{0}\smash{\begin{tabular}[t]{l}PBSC 2\end{tabular}}}}%
    \put(0,0){\includegraphics[width=\unitlength,page=61]{OUga1kbf.pdf}}%
    \put(0.81052973,0.17542162){\color[rgb]{0.14901961,0.14901961,0.14901961}\makebox(0,0)[lt]{\lineheight{0}\smash{\begin{tabular}[t]{l}PBSC 3\end{tabular}}}}%
  \end{picture}%
\endgroup%

%% file: tmpplt/miscovplot1kD.pdf_tex
\begingroup%
  \makeatletter%
  \providecommand\color[2][]{%
    \errmessage{(Inkscape) Color is used for the text in Inkscape, but the package 'color.sty' is not loaded}%
    \renewcommand\color[2][]{}%
  }%
  \providecommand\transparent[1]{%
    \errmessage{(Inkscape) Transparency is used (non-zero) for the text in Inkscape, but the package 'transparent.sty' is not loaded}%
    \renewcommand\transparent[1]{}%
  }%
  \providecommand\rotatebox[2]{#2}%
  \newcommand*\fsize{\dimexpr\f@size pt\relax}%
  \newcommand*\lineheight[1]{\fontsize{\fsize}{#1\fsize}\selectfont}%
  \ifx\svgwidth\undefined%
    \setlength{\unitlength}{460.79998779bp}%
    \ifx\svgscale\undefined%
      \relax%
    \else%
      \setlength{\unitlength}{\unitlength * \real{\svgscale}}%
    \fi%
  \else%
    \setlength{\unitlength}{\svgwidth}%
  \fi%
  \global\let\svgwidth\undefined%
  \global\let\svgscale\undefined%
  \makeatother%
  \begin{picture}(1,0.75000003)%
    \lineheight{1}%
    \setlength\tabcolsep{0pt}%
    \put(0,0){\includegraphics[width=\unitlength,page=1]{miscovplot1kD.pdf}}%
    \put(0.16022727,0.05081939){\color[rgb]{0.14901961,0.14901961,0.14901961}\makebox(0,0)[t]{\lineheight{0}\smash{\begin{tabular}[t]{c}0.0\end{tabular}}}}%
    \put(0,0){\includegraphics[width=\unitlength,page=2]{miscovplot1kD.pdf}}%
    \put(0.30113638,0.05081939){\color[rgb]{0.14901961,0.14901961,0.14901961}\makebox(0,0)[t]{\lineheight{0}\smash{\begin{tabular}[t]{c}0.2\end{tabular}}}}%
    \put(0,0){\includegraphics[width=\unitlength,page=3]{miscovplot1kD.pdf}}%
    \put(0.44204548,0.05081939){\color[rgb]{0.14901961,0.14901961,0.14901961}\makebox(0,0)[t]{\lineheight{0}\smash{\begin{tabular}[t]{c}0.4\end{tabular}}}}%
    \put(0,0){\includegraphics[width=\unitlength,page=4]{miscovplot1kD.pdf}}%
    \put(0.58295457,0.05081939){\color[rgb]{0.14901961,0.14901961,0.14901961}\makebox(0,0)[t]{\lineheight{0}\smash{\begin{tabular}[t]{c}0.6\end{tabular}}}}%
    \put(0,0){\includegraphics[width=\unitlength,page=5]{miscovplot1kD.pdf}}%
    \put(0.72386366,0.05081939){\color[rgb]{0.14901961,0.14901961,0.14901961}\makebox(0,0)[t]{\lineheight{0}\smash{\begin{tabular}[t]{c}0.8\end{tabular}}}}%
    \put(0,0){\includegraphics[width=\unitlength,page=6]{miscovplot1kD.pdf}}%
    \put(0.86477275,0.05081939){\color[rgb]{0.14901961,0.14901961,0.14901961}\makebox(0,0)[t]{\lineheight{0}\smash{\begin{tabular}[t]{c}1.0\end{tabular}}}}%
    \put(0.51250002,0.02113594){\color[rgb]{0.14901961,0.14901961,0.14901961}\makebox(0,0)[t]{\lineheight{0}\smash{\begin{tabular}[t]{c}Target coverage\end{tabular}}}}%
    \put(0,0){\includegraphics[width=\unitlength,page=7]{miscovplot1kD.pdf}}%
    \put(0.10980903,0.10050516){\color[rgb]{0.14901961,0.14901961,0.14901961}\makebox(0,0)[rt]{\lineheight{0}\smash{\begin{tabular}[t]{r}-0.100\end{tabular}}}}%
    \put(0,0){\includegraphics[width=\unitlength,page=8]{miscovplot1kD.pdf}}%
    \put(0.10980903,0.1661302){\color[rgb]{0.14901961,0.14901961,0.14901961}\makebox(0,0)[rt]{\lineheight{0}\smash{\begin{tabular}[t]{r}-0.075\end{tabular}}}}%
    \put(0,0){\includegraphics[width=\unitlength,page=9]{miscovplot1kD.pdf}}%
    \put(0.10980903,0.23175518){\color[rgb]{0.14901961,0.14901961,0.14901961}\makebox(0,0)[rt]{\lineheight{0}\smash{\begin{tabular}[t]{r}-0.050\end{tabular}}}}%
    \put(0,0){\includegraphics[width=\unitlength,page=10]{miscovplot1kD.pdf}}%
    \put(0.10980903,0.2973802){\color[rgb]{0.14901961,0.14901961,0.14901961}\makebox(0,0)[rt]{\lineheight{0}\smash{\begin{tabular}[t]{r}-0.025\end{tabular}}}}%
    \put(0,0){\includegraphics[width=\unitlength,page=11]{miscovplot1kD.pdf}}%
    \put(0.10980903,0.36300518){\color[rgb]{0.14901961,0.14901961,0.14901961}\makebox(0,0)[rt]{\lineheight{0}\smash{\begin{tabular}[t]{r}0.000\end{tabular}}}}%
    \put(0,0){\includegraphics[width=\unitlength,page=12]{miscovplot1kD.pdf}}%
    \put(0.10980903,0.42863019){\color[rgb]{0.14901961,0.14901961,0.14901961}\makebox(0,0)[rt]{\lineheight{0}\smash{\begin{tabular}[t]{r}0.025\end{tabular}}}}%
    \put(0,0){\includegraphics[width=\unitlength,page=13]{miscovplot1kD.pdf}}%
    \put(0.10980903,0.49425519){\color[rgb]{0.14901961,0.14901961,0.14901961}\makebox(0,0)[rt]{\lineheight{0}\smash{\begin{tabular}[t]{r}0.050\end{tabular}}}}%
    \put(0,0){\includegraphics[width=\unitlength,page=14]{miscovplot1kD.pdf}}%
    \put(0.10980903,0.5598802){\color[rgb]{0.14901961,0.14901961,0.14901961}\makebox(0,0)[rt]{\lineheight{0}\smash{\begin{tabular}[t]{r}0.075\end{tabular}}}}%
    \put(0,0){\includegraphics[width=\unitlength,page=15]{miscovplot1kD.pdf}}%
    \put(0.10980903,0.6255052){\color[rgb]{0.14901961,0.14901961,0.14901961}\makebox(0,0)[rt]{\lineheight{0}\smash{\begin{tabular}[t]{r}0.100\end{tabular}}}}%
    \put(0.0266588,0.37125002){\color[rgb]{0.14901961,0.14901961,0.14901961}\rotatebox{90}{\makebox(0,0)[t]{\lineheight{0}\smash{\begin{tabular}[t]{c}Miscoverage\end{tabular}}}}}%
    \put(0,0){\includegraphics[width=\unitlength,page=16]{miscovplot1kD.pdf}}%
    \put(0.20529514,0.26581232){\color[rgb]{0.14901961,0.14901961,0.14901961}\makebox(0,0)[lt]{\lineheight{0}\smash{\begin{tabular}[t]{l}True-post\end{tabular}}}}%
    \put(0,0){\includegraphics[width=\unitlength,page=17]{miscovplot1kD.pdf}}%
    \put(0.20529514,0.23395877){\color[rgb]{0.14901961,0.14901961,0.14901961}\makebox(0,0)[lt]{\lineheight{0}\smash{\begin{tabular}[t]{l}Approx-post\end{tabular}}}}%
    \put(0,0){\includegraphics[width=\unitlength,page=18]{miscovplot1kD.pdf}}%
    \put(0.20529514,0.20210518){\color[rgb]{0.14901961,0.14901961,0.14901961}\makebox(0,0)[lt]{\lineheight{0}\smash{\begin{tabular}[t]{l}BSC\end{tabular}}}}%
    \put(0,0){\includegraphics[width=\unitlength,page=19]{miscovplot1kD.pdf}}%
    \put(0.20529514,0.1702516){\color[rgb]{0.14901961,0.14901961,0.14901961}\makebox(0,0)[lt]{\lineheight{0}\smash{\begin{tabular}[t]{l}PBSC 1\end{tabular}}}}%
    \put(0,0){\includegraphics[width=\unitlength,page=20]{miscovplot1kD.pdf}}%
    \put(0.20529514,0.13839802){\color[rgb]{0.14901961,0.14901961,0.14901961}\makebox(0,0)[lt]{\lineheight{0}\smash{\begin{tabular}[t]{l}PBSC 2\end{tabular}}}}%
    \put(0,0){\includegraphics[width=\unitlength,page=21]{miscovplot1kD.pdf}}%
    \put(0.20529514,0.1065445){\color[rgb]{0.14901961,0.14901961,0.14901961}\makebox(0,0)[lt]{\lineheight{0}\smash{\begin{tabular}[t]{l}PBSC 3\end{tabular}}}}%
  \end{picture}%
\endgroup%

%% file: tmpplt/OUD1kbf.pdf_tex
\begingroup%
  \makeatletter%
  \providecommand\color[2][]{%
    \errmessage{(Inkscape) Color is used for the text in Inkscape, but the package 'color.sty' is not loaded}%
    \renewcommand\color[2][]{}%
  }%
  \providecommand\transparent[1]{%
    \errmessage{(Inkscape) Transparency is used (non-zero) for the text in Inkscape, but the package 'transparent.sty' is not loaded}%
    \renewcommand\transparent[1]{}%
  }%
  \providecommand\rotatebox[2]{#2}%
  \newcommand*\fsize{\dimexpr\f@size pt\relax}%
  \newcommand*\lineheight[1]{\fontsize{\fsize}{#1\fsize}\selectfont}%
  \ifx\svgwidth\undefined%
    \setlength{\unitlength}{476.64001465bp}%
    \ifx\svgscale\undefined%
      \relax%
    \else%
      \setlength{\unitlength}{\unitlength * \real{\svgscale}}%
    \fi%
  \else%
    \setlength{\unitlength}{\svgwidth}%
  \fi%
  \global\let\svgwidth\undefined%
  \global\let\svgscale\undefined%
  \makeatother%
  \begin{picture}(1,1.41238669)%
    \lineheight{1}%
    \setlength\tabcolsep{0pt}%
    \put(0,0){\includegraphics[width=\unitlength,page=1]{OUD1kbf.pdf}}%
    \put(0.14099809,0.89240845){\color[rgb]{0.14901961,0.14901961,0.14901961}\makebox(0,0)[t]{\lineheight{0}\smash{\begin{tabular}[t]{c}0\end{tabular}}}}%
    \put(0,0){\includegraphics[width=\unitlength,page=2]{OUD1kbf.pdf}}%
    \put(0.24238237,0.89240845){\color[rgb]{0.14901961,0.14901961,0.14901961}\makebox(0,0)[t]{\lineheight{0}\smash{\begin{tabular}[t]{c}20\end{tabular}}}}%
    \put(0,0){\includegraphics[width=\unitlength,page=3]{OUD1kbf.pdf}}%
    \put(0.34376667,0.89240845){\color[rgb]{0.14901961,0.14901961,0.14901961}\makebox(0,0)[t]{\lineheight{0}\smash{\begin{tabular}[t]{c}40\end{tabular}}}}%
    \put(0,0){\includegraphics[width=\unitlength,page=4]{OUD1kbf.pdf}}%
    \put(0.44515096,0.89240845){\color[rgb]{0.14901961,0.14901961,0.14901961}\makebox(0,0)[t]{\lineheight{0}\smash{\begin{tabular}[t]{c}60\end{tabular}}}}%
    \put(0,0){\includegraphics[width=\unitlength,page=5]{OUD1kbf.pdf}}%
    \put(0.11031386,0.92959166){\color[rgb]{0.14901961,0.14901961,0.14901961}\makebox(0,0)[rt]{\lineheight{0}\smash{\begin{tabular}[t]{r}0\end{tabular}}}}%
    \put(0,0){\includegraphics[width=\unitlength,page=6]{OUD1kbf.pdf}}%
    \put(0.11031386,0.97577895){\color[rgb]{0.14901961,0.14901961,0.14901961}\makebox(0,0)[rt]{\lineheight{0}\smash{\begin{tabular}[t]{r}10\end{tabular}}}}%
    \put(0,0){\includegraphics[width=\unitlength,page=7]{OUD1kbf.pdf}}%
    \put(0.11031386,1.02196623){\color[rgb]{0.14901961,0.14901961,0.14901961}\makebox(0,0)[rt]{\lineheight{0}\smash{\begin{tabular}[t]{r}20\end{tabular}}}}%
    \put(0,0){\includegraphics[width=\unitlength,page=8]{OUD1kbf.pdf}}%
    \put(0.11031386,1.06815355){\color[rgb]{0.14901961,0.14901961,0.14901961}\makebox(0,0)[rt]{\lineheight{0}\smash{\begin{tabular}[t]{r}30\end{tabular}}}}%
    \put(0,0){\includegraphics[width=\unitlength,page=9]{OUD1kbf.pdf}}%
    \put(0.11031386,1.11434084){\color[rgb]{0.14901961,0.14901961,0.14901961}\makebox(0,0)[rt]{\lineheight{0}\smash{\begin{tabular}[t]{r}40\end{tabular}}}}%
    \put(0,0){\includegraphics[width=\unitlength,page=10]{OUD1kbf.pdf}}%
    \put(0.11031386,1.16052813){\color[rgb]{0.14901961,0.14901961,0.14901961}\makebox(0,0)[rt]{\lineheight{0}\smash{\begin{tabular}[t]{r}50\end{tabular}}}}%
    \put(0,0){\includegraphics[width=\unitlength,page=11]{OUD1kbf.pdf}}%
    \put(0.11031386,1.20671542){\color[rgb]{0.14901961,0.14901961,0.14901961}\makebox(0,0)[rt]{\lineheight{0}\smash{\begin{tabular}[t]{r}60\end{tabular}}}}%
    \put(0,0){\includegraphics[width=\unitlength,page=12]{OUD1kbf.pdf}}%
    \put(0.36096047,0.94628166){\color[rgb]{0.14901961,0.14901961,0.14901961}\makebox(0,0)[lt]{\lineheight{0}\smash{\begin{tabular}[t]{l}True-post\end{tabular}}}}%
    \put(0,0){\includegraphics[width=\unitlength,page=13]{OUD1kbf.pdf}}%
    \put(0.56372632,0.89240845){\color[rgb]{0.14901961,0.14901961,0.14901961}\makebox(0,0)[t]{\lineheight{0}\smash{\begin{tabular}[t]{c}0\end{tabular}}}}%
    \put(0,0){\includegraphics[width=\unitlength,page=14]{OUD1kbf.pdf}}%
    \put(0.66375948,0.89240845){\color[rgb]{0.14901961,0.14901961,0.14901961}\makebox(0,0)[t]{\lineheight{0}\smash{\begin{tabular}[t]{c}20\end{tabular}}}}%
    \put(0,0){\includegraphics[width=\unitlength,page=15]{OUD1kbf.pdf}}%
    \put(0.76379265,0.89240845){\color[rgb]{0.14901961,0.14901961,0.14901961}\makebox(0,0)[t]{\lineheight{0}\smash{\begin{tabular}[t]{c}40\end{tabular}}}}%
    \put(0,0){\includegraphics[width=\unitlength,page=16]{OUD1kbf.pdf}}%
    \put(0.86382581,0.89240845){\color[rgb]{0.14901961,0.14901961,0.14901961}\makebox(0,0)[t]{\lineheight{0}\smash{\begin{tabular}[t]{c}60\end{tabular}}}}%
    \put(0,0){\includegraphics[width=\unitlength,page=17]{OUD1kbf.pdf}}%
    \put(0.53304112,0.92959166){\color[rgb]{0.14901961,0.14901961,0.14901961}\makebox(0,0)[rt]{\lineheight{0}\smash{\begin{tabular}[t]{r}0\end{tabular}}}}%
    \put(0,0){\includegraphics[width=\unitlength,page=18]{OUD1kbf.pdf}}%
    \put(0.53304112,0.97577895){\color[rgb]{0.14901961,0.14901961,0.14901961}\makebox(0,0)[rt]{\lineheight{0}\smash{\begin{tabular}[t]{r}10\end{tabular}}}}%
    \put(0,0){\includegraphics[width=\unitlength,page=19]{OUD1kbf.pdf}}%
    \put(0.53304112,1.02196623){\color[rgb]{0.14901961,0.14901961,0.14901961}\makebox(0,0)[rt]{\lineheight{0}\smash{\begin{tabular}[t]{r}20\end{tabular}}}}%
    \put(0,0){\includegraphics[width=\unitlength,page=20]{OUD1kbf.pdf}}%
    \put(0.53304112,1.06815355){\color[rgb]{0.14901961,0.14901961,0.14901961}\makebox(0,0)[rt]{\lineheight{0}\smash{\begin{tabular}[t]{r}30\end{tabular}}}}%
    \put(0,0){\includegraphics[width=\unitlength,page=21]{OUD1kbf.pdf}}%
    \put(0.53304112,1.11434084){\color[rgb]{0.14901961,0.14901961,0.14901961}\makebox(0,0)[rt]{\lineheight{0}\smash{\begin{tabular}[t]{r}40\end{tabular}}}}%
    \put(0,0){\includegraphics[width=\unitlength,page=22]{OUD1kbf.pdf}}%
    \put(0.53304112,1.16052813){\color[rgb]{0.14901961,0.14901961,0.14901961}\makebox(0,0)[rt]{\lineheight{0}\smash{\begin{tabular}[t]{r}50\end{tabular}}}}%
    \put(0,0){\includegraphics[width=\unitlength,page=23]{OUD1kbf.pdf}}%
    \put(0.53304112,1.20671542){\color[rgb]{0.14901961,0.14901961,0.14901961}\makebox(0,0)[rt]{\lineheight{0}\smash{\begin{tabular}[t]{r}60\end{tabular}}}}%
    \put(0,0){\includegraphics[width=\unitlength,page=24]{OUD1kbf.pdf}}%
    \put(0.75449575,0.94628166){\color[rgb]{0.14901961,0.14901961,0.14901961}\makebox(0,0)[lt]{\lineheight{0}\smash{\begin{tabular}[t]{l}Approx-post\end{tabular}}}}%
    \put(0,0){\includegraphics[width=\unitlength,page=25]{OUD1kbf.pdf}}%
    \put(0.14103144,0.50857161){\color[rgb]{0.14901961,0.14901961,0.14901961}\makebox(0,0)[t]{\lineheight{0}\smash{\begin{tabular}[t]{c}0\end{tabular}}}}%
    \put(0,0){\includegraphics[width=\unitlength,page=26]{OUD1kbf.pdf}}%
    \put(0.24275484,0.50857161){\color[rgb]{0.14901961,0.14901961,0.14901961}\makebox(0,0)[t]{\lineheight{0}\smash{\begin{tabular}[t]{c}20\end{tabular}}}}%
    \put(0,0){\includegraphics[width=\unitlength,page=27]{OUD1kbf.pdf}}%
    \put(0.34447823,0.50857161){\color[rgb]{0.14901961,0.14901961,0.14901961}\makebox(0,0)[t]{\lineheight{0}\smash{\begin{tabular}[t]{c}40\end{tabular}}}}%
    \put(0,0){\includegraphics[width=\unitlength,page=28]{OUD1kbf.pdf}}%
    \put(0.4462016,0.50857161){\color[rgb]{0.14901961,0.14901961,0.14901961}\makebox(0,0)[t]{\lineheight{0}\smash{\begin{tabular}[t]{c}60\end{tabular}}}}%
    \put(0,0){\includegraphics[width=\unitlength,page=29]{OUD1kbf.pdf}}%
    \put(0.11031386,0.54575479){\color[rgb]{0.14901961,0.14901961,0.14901961}\makebox(0,0)[rt]{\lineheight{0}\smash{\begin{tabular}[t]{r}0\end{tabular}}}}%
    \put(0,0){\includegraphics[width=\unitlength,page=30]{OUD1kbf.pdf}}%
    \put(0.11031386,0.59194208){\color[rgb]{0.14901961,0.14901961,0.14901961}\makebox(0,0)[rt]{\lineheight{0}\smash{\begin{tabular}[t]{r}10\end{tabular}}}}%
    \put(0,0){\includegraphics[width=\unitlength,page=31]{OUD1kbf.pdf}}%
    \put(0.11031386,0.63812943){\color[rgb]{0.14901961,0.14901961,0.14901961}\makebox(0,0)[rt]{\lineheight{0}\smash{\begin{tabular}[t]{r}20\end{tabular}}}}%
    \put(0,0){\includegraphics[width=\unitlength,page=32]{OUD1kbf.pdf}}%
    \put(0.11031386,0.68431672){\color[rgb]{0.14901961,0.14901961,0.14901961}\makebox(0,0)[rt]{\lineheight{0}\smash{\begin{tabular}[t]{r}30\end{tabular}}}}%
    \put(0,0){\includegraphics[width=\unitlength,page=33]{OUD1kbf.pdf}}%
    \put(0.11031386,0.730504){\color[rgb]{0.14901961,0.14901961,0.14901961}\makebox(0,0)[rt]{\lineheight{0}\smash{\begin{tabular}[t]{r}40\end{tabular}}}}%
    \put(0,0){\includegraphics[width=\unitlength,page=34]{OUD1kbf.pdf}}%
    \put(0.11031386,0.77669129){\color[rgb]{0.14901961,0.14901961,0.14901961}\makebox(0,0)[rt]{\lineheight{0}\smash{\begin{tabular}[t]{r}50\end{tabular}}}}%
    \put(0,0){\includegraphics[width=\unitlength,page=35]{OUD1kbf.pdf}}%
    \put(0.11031386,0.82287858){\color[rgb]{0.14901961,0.14901961,0.14901961}\makebox(0,0)[rt]{\lineheight{0}\smash{\begin{tabular}[t]{r}60\end{tabular}}}}%
    \put(0,0){\includegraphics[width=\unitlength,page=36]{OUD1kbf.pdf}}%
    \put(0.5643112,0.50857161){\color[rgb]{0.14901961,0.14901961,0.14901961}\makebox(0,0)[t]{\lineheight{0}\smash{\begin{tabular}[t]{c}0\end{tabular}}}}%
    \put(0,0){\includegraphics[width=\unitlength,page=37]{OUD1kbf.pdf}}%
    \put(0.66258325,0.50857161){\color[rgb]{0.14901961,0.14901961,0.14901961}\makebox(0,0)[t]{\lineheight{0}\smash{\begin{tabular}[t]{c}20\end{tabular}}}}%
    \put(0,0){\includegraphics[width=\unitlength,page=38]{OUD1kbf.pdf}}%
    \put(0.7608553,0.50857161){\color[rgb]{0.14901961,0.14901961,0.14901961}\makebox(0,0)[t]{\lineheight{0}\smash{\begin{tabular}[t]{c}40\end{tabular}}}}%
    \put(0,0){\includegraphics[width=\unitlength,page=39]{OUD1kbf.pdf}}%
    \put(0.85912729,0.50857161){\color[rgb]{0.14901961,0.14901961,0.14901961}\makebox(0,0)[t]{\lineheight{0}\smash{\begin{tabular}[t]{c}60\end{tabular}}}}%
    \put(0,0){\includegraphics[width=\unitlength,page=40]{OUD1kbf.pdf}}%
    \put(0.53304112,0.54575479){\color[rgb]{0.14901961,0.14901961,0.14901961}\makebox(0,0)[rt]{\lineheight{0}\smash{\begin{tabular}[t]{r}0\end{tabular}}}}%
    \put(0,0){\includegraphics[width=\unitlength,page=41]{OUD1kbf.pdf}}%
    \put(0.53304112,0.59194208){\color[rgb]{0.14901961,0.14901961,0.14901961}\makebox(0,0)[rt]{\lineheight{0}\smash{\begin{tabular}[t]{r}10\end{tabular}}}}%
    \put(0,0){\includegraphics[width=\unitlength,page=42]{OUD1kbf.pdf}}%
    \put(0.53304112,0.63812943){\color[rgb]{0.14901961,0.14901961,0.14901961}\makebox(0,0)[rt]{\lineheight{0}\smash{\begin{tabular}[t]{r}20\end{tabular}}}}%
    \put(0,0){\includegraphics[width=\unitlength,page=43]{OUD1kbf.pdf}}%
    \put(0.53304112,0.68431672){\color[rgb]{0.14901961,0.14901961,0.14901961}\makebox(0,0)[rt]{\lineheight{0}\smash{\begin{tabular}[t]{r}30\end{tabular}}}}%
    \put(0,0){\includegraphics[width=\unitlength,page=44]{OUD1kbf.pdf}}%
    \put(0.53304112,0.730504){\color[rgb]{0.14901961,0.14901961,0.14901961}\makebox(0,0)[rt]{\lineheight{0}\smash{\begin{tabular}[t]{r}40\end{tabular}}}}%
    \put(0,0){\includegraphics[width=\unitlength,page=45]{OUD1kbf.pdf}}%
    \put(0.53304112,0.77669129){\color[rgb]{0.14901961,0.14901961,0.14901961}\makebox(0,0)[rt]{\lineheight{0}\smash{\begin{tabular}[t]{r}50\end{tabular}}}}%
    \put(0,0){\includegraphics[width=\unitlength,page=46]{OUD1kbf.pdf}}%
    \put(0.53304112,0.82287858){\color[rgb]{0.14901961,0.14901961,0.14901961}\makebox(0,0)[rt]{\lineheight{0}\smash{\begin{tabular}[t]{r}60\end{tabular}}}}%
    \put(0,0){\includegraphics[width=\unitlength,page=47]{OUD1kbf.pdf}}%
    \put(0.1411953,0.12473471){\color[rgb]{0.14901961,0.14901961,0.14901961}\makebox(0,0)[t]{\lineheight{0}\smash{\begin{tabular}[t]{c}0\end{tabular}}}}%
    \put(0,0){\includegraphics[width=\unitlength,page=48]{OUD1kbf.pdf}}%
    \put(0.23229929,0.12473471){\color[rgb]{0.14901961,0.14901961,0.14901961}\makebox(0,0)[t]{\lineheight{0}\smash{\begin{tabular}[t]{c}20\end{tabular}}}}%
    \put(0,0){\includegraphics[width=\unitlength,page=49]{OUD1kbf.pdf}}%
    \put(0.32340328,0.12473471){\color[rgb]{0.14901961,0.14901961,0.14901961}\makebox(0,0)[t]{\lineheight{0}\smash{\begin{tabular}[t]{c}40\end{tabular}}}}%
    \put(0,0){\includegraphics[width=\unitlength,page=50]{OUD1kbf.pdf}}%
    \put(0.41450725,0.12473471){\color[rgb]{0.14901961,0.14901961,0.14901961}\makebox(0,0)[t]{\lineheight{0}\smash{\begin{tabular}[t]{c}60\end{tabular}}}}%
    \put(0,0){\includegraphics[width=\unitlength,page=51]{OUD1kbf.pdf}}%
    \put(0.11031386,0.16191795){\color[rgb]{0.14901961,0.14901961,0.14901961}\makebox(0,0)[rt]{\lineheight{0}\smash{\begin{tabular}[t]{r}0\end{tabular}}}}%
    \put(0,0){\includegraphics[width=\unitlength,page=52]{OUD1kbf.pdf}}%
    \put(0.11031386,0.20810524){\color[rgb]{0.14901961,0.14901961,0.14901961}\makebox(0,0)[rt]{\lineheight{0}\smash{\begin{tabular}[t]{r}10\end{tabular}}}}%
    \put(0,0){\includegraphics[width=\unitlength,page=53]{OUD1kbf.pdf}}%
    \put(0.11031386,0.25429253){\color[rgb]{0.14901961,0.14901961,0.14901961}\makebox(0,0)[rt]{\lineheight{0}\smash{\begin{tabular}[t]{r}20\end{tabular}}}}%
    \put(0,0){\includegraphics[width=\unitlength,page=54]{OUD1kbf.pdf}}%
    \put(0.11031386,0.30047981){\color[rgb]{0.14901961,0.14901961,0.14901961}\makebox(0,0)[rt]{\lineheight{0}\smash{\begin{tabular}[t]{r}30\end{tabular}}}}%
    \put(0,0){\includegraphics[width=\unitlength,page=55]{OUD1kbf.pdf}}%
    \put(0.11031386,0.34666717){\color[rgb]{0.14901961,0.14901961,0.14901961}\makebox(0,0)[rt]{\lineheight{0}\smash{\begin{tabular}[t]{r}40\end{tabular}}}}%
    \put(0,0){\includegraphics[width=\unitlength,page=56]{OUD1kbf.pdf}}%
    \put(0.11031386,0.39285445){\color[rgb]{0.14901961,0.14901961,0.14901961}\makebox(0,0)[rt]{\lineheight{0}\smash{\begin{tabular}[t]{r}50\end{tabular}}}}%
    \put(0,0){\includegraphics[width=\unitlength,page=57]{OUD1kbf.pdf}}%
    \put(0.11031386,0.43904174){\color[rgb]{0.14901961,0.14901961,0.14901961}\makebox(0,0)[rt]{\lineheight{0}\smash{\begin{tabular}[t]{r}60\end{tabular}}}}%
    \put(0,0){\includegraphics[width=\unitlength,page=58]{OUD1kbf.pdf}}%
    \put(0.5643169,0.12473471){\color[rgb]{0.14901961,0.14901961,0.14901961}\makebox(0,0)[t]{\lineheight{0}\smash{\begin{tabular}[t]{c}0\end{tabular}}}}%
    \put(0,0){\includegraphics[width=\unitlength,page=59]{OUD1kbf.pdf}}%
    \put(0.65661201,0.12473471){\color[rgb]{0.14901961,0.14901961,0.14901961}\makebox(0,0)[t]{\lineheight{0}\smash{\begin{tabular}[t]{c}20\end{tabular}}}}%
    \put(0,0){\includegraphics[width=\unitlength,page=60]{OUD1kbf.pdf}}%
    \put(0.74890713,0.12473471){\color[rgb]{0.14901961,0.14901961,0.14901961}\makebox(0,0)[t]{\lineheight{0}\smash{\begin{tabular}[t]{c}40\end{tabular}}}}%
    \put(0,0){\includegraphics[width=\unitlength,page=61]{OUD1kbf.pdf}}%
    \put(0.84120231,0.12473471){\color[rgb]{0.14901961,0.14901961,0.14901961}\makebox(0,0)[t]{\lineheight{0}\smash{\begin{tabular}[t]{c}60\end{tabular}}}}%
    \put(0,0){\includegraphics[width=\unitlength,page=62]{OUD1kbf.pdf}}%
    \put(0.53304112,0.16191795){\color[rgb]{0.14901961,0.14901961,0.14901961}\makebox(0,0)[rt]{\lineheight{0}\smash{\begin{tabular}[t]{r}0\end{tabular}}}}%
    \put(0,0){\includegraphics[width=\unitlength,page=63]{OUD1kbf.pdf}}%
    \put(0.53304112,0.20810524){\color[rgb]{0.14901961,0.14901961,0.14901961}\makebox(0,0)[rt]{\lineheight{0}\smash{\begin{tabular}[t]{r}10\end{tabular}}}}%
    \put(0,0){\includegraphics[width=\unitlength,page=64]{OUD1kbf.pdf}}%
    \put(0.53304112,0.25429253){\color[rgb]{0.14901961,0.14901961,0.14901961}\makebox(0,0)[rt]{\lineheight{0}\smash{\begin{tabular}[t]{r}20\end{tabular}}}}%
    \put(0,0){\includegraphics[width=\unitlength,page=65]{OUD1kbf.pdf}}%
    \put(0.53304112,0.30047981){\color[rgb]{0.14901961,0.14901961,0.14901961}\makebox(0,0)[rt]{\lineheight{0}\smash{\begin{tabular}[t]{r}30\end{tabular}}}}%
    \put(0,0){\includegraphics[width=\unitlength,page=66]{OUD1kbf.pdf}}%
    \put(0.53304112,0.34666717){\color[rgb]{0.14901961,0.14901961,0.14901961}\makebox(0,0)[rt]{\lineheight{0}\smash{\begin{tabular}[t]{r}40\end{tabular}}}}%
    \put(0,0){\includegraphics[width=\unitlength,page=67]{OUD1kbf.pdf}}%
    \put(0.53304112,0.39285445){\color[rgb]{0.14901961,0.14901961,0.14901961}\makebox(0,0)[rt]{\lineheight{0}\smash{\begin{tabular}[t]{r}50\end{tabular}}}}%
    \put(0,0){\includegraphics[width=\unitlength,page=68]{OUD1kbf.pdf}}%
    \put(0.53304112,0.43904174){\color[rgb]{0.14901961,0.14901961,0.14901961}\makebox(0,0)[rt]{\lineheight{0}\smash{\begin{tabular}[t]{r}60\end{tabular}}}}%
    \put(0,0){\includegraphics[width=\unitlength,page=69]{OUD1kbf.pdf}}%
    \put(0.5,0.01935981){\color[rgb]{0.14901961,0.14901961,0.14901961}\makebox(0,0)[t]{\lineheight{0}\smash{\begin{tabular}[t]{c}89\% credible interval\end{tabular}}}}%
    \put(0.03913001,0.5879339){\color[rgb]{0.14901961,0.14901961,0.14901961}\rotatebox{90}{\makebox(0,0)[lt]{\lineheight{0}\smash{\begin{tabular}[t]{l}Calibration sample\end{tabular}}}}}%
    \put(0,0){\includegraphics[width=\unitlength,page=70]{OUD1kbf.pdf}}%
    \put(0.41386712,0.56139582){\color[rgb]{0.14901961,0.14901961,0.14901961}\makebox(0,0)[lt]{\lineheight{0}\smash{\begin{tabular}[t]{l}BSC\end{tabular}}}}%
    \put(0,0){\includegraphics[width=\unitlength,page=71]{OUD1kbf.pdf}}%
    \put(0.80846451,0.56139582){\color[rgb]{0.14901961,0.14901961,0.14901961}\makebox(0,0)[lt]{\lineheight{0}\smash{\begin{tabular}[t]{l}PBSC 1\end{tabular}}}}%
    \put(0,0){\includegraphics[width=\unitlength,page=72]{OUD1kbf.pdf}}%
    \put(0.38579955,0.17755898){\color[rgb]{0.14901961,0.14901961,0.14901961}\makebox(0,0)[lt]{\lineheight{0}\smash{\begin{tabular}[t]{l}PBSC 2\end{tabular}}}}%
    \put(0,0){\includegraphics[width=\unitlength,page=73]{OUD1kbf.pdf}}%
    \put(0.80874648,0.17755898){\color[rgb]{0.14901961,0.14901961,0.14901961}\makebox(0,0)[lt]{\lineheight{0}\smash{\begin{tabular}[t]{l}PBSC 3\end{tabular}}}}%
  \end{picture}%
\endgroup%

%% file: tmpplt/PALT0bf1k.pdf_tex
\begingroup%
  \makeatletter%
  \providecommand\color[2][]{%
    \errmessage{(Inkscape) Color is used for the text in Inkscape, but the package 'color.sty' is not loaded}%
    \renewcommand\color[2][]{}%
  }%
  \providecommand\transparent[1]{%
    \errmessage{(Inkscape) Transparency is used (non-zero) for the text in Inkscape, but the package 'transparent.sty' is not loaded}%
    \renewcommand\transparent[1]{}%
  }%
  \providecommand\rotatebox[2]{#2}%
  \newcommand*\fsize{\dimexpr\f@size pt\relax}%
  \newcommand*\lineheight[1]{\fontsize{\fsize}{#1\fsize}\selectfont}%
  \ifx\svgwidth\undefined%
    \setlength{\unitlength}{476.64001465bp}%
    \ifx\svgscale\undefined%
      \relax%
    \else%
      \setlength{\unitlength}{\unitlength * \real{\svgscale}}%
    \fi%
  \else%
    \setlength{\unitlength}{\svgwidth}%
  \fi%
  \global\let\svgwidth\undefined%
  \global\let\svgscale\undefined%
  \makeatother%
  \begin{picture}(1,1.41238669)%
    \lineheight{1}%
    \setlength\tabcolsep{0pt}%
    \put(0,0){\includegraphics[width=\unitlength,page=1]{PALT0bf1k.pdf}}%
    \put(0.125,0.89240845){\makebox(0,0)[t]{\lineheight{0}\smash{\begin{tabular}[t]{c}1.0\end{tabular}}}}%
    \put(0,0){\includegraphics[width=\unitlength,page=2]{PALT0bf1k.pdf}}%
    \put(0.24242423,0.89240845){\makebox(0,0)[t]{\lineheight{0}\smash{\begin{tabular}[t]{c}1.5\end{tabular}}}}%
    \put(0,0){\includegraphics[width=\unitlength,page=3]{PALT0bf1k.pdf}}%
    \put(0.35984849,0.89240845){\makebox(0,0)[t]{\lineheight{0}\smash{\begin{tabular}[t]{c}2.0\end{tabular}}}}%
    \put(0,0){\includegraphics[width=\unitlength,page=4]{PALT0bf1k.pdf}}%
    \put(0.47727272,0.89240845){\makebox(0,0)[t]{\lineheight{0}\smash{\begin{tabular}[t]{c}2.5\end{tabular}}}}%
    \put(0,0){\includegraphics[width=\unitlength,page=5]{PALT0bf1k.pdf}}%
    \put(0.11031386,0.92960469){\makebox(0,0)[rt]{\lineheight{0}\smash{\begin{tabular}[t]{r}0\end{tabular}}}}%
    \put(0,0){\includegraphics[width=\unitlength,page=6]{PALT0bf1k.pdf}}%
    \put(0.11031386,0.98781999){\makebox(0,0)[rt]{\lineheight{0}\smash{\begin{tabular}[t]{r}200\end{tabular}}}}%
    \put(0,0){\includegraphics[width=\unitlength,page=7]{PALT0bf1k.pdf}}%
    \put(0.11031386,1.04603532){\makebox(0,0)[rt]{\lineheight{0}\smash{\begin{tabular}[t]{r}400\end{tabular}}}}%
    \put(0,0){\includegraphics[width=\unitlength,page=8]{PALT0bf1k.pdf}}%
    \put(0.11031386,1.10425062){\makebox(0,0)[rt]{\lineheight{0}\smash{\begin{tabular}[t]{r}600\end{tabular}}}}%
    \put(0,0){\includegraphics[width=\unitlength,page=9]{PALT0bf1k.pdf}}%
    \put(0.11031386,1.16246594){\makebox(0,0)[rt]{\lineheight{0}\smash{\begin{tabular}[t]{r}800\end{tabular}}}}%
    \put(0,0){\includegraphics[width=\unitlength,page=10]{PALT0bf1k.pdf}}%
    \put(0.11031386,1.22068125){\makebox(0,0)[rt]{\lineheight{0}\smash{\begin{tabular}[t]{r}1000\end{tabular}}}}%
    \put(0,0){\includegraphics[width=\unitlength,page=11]{PALT0bf1k.pdf}}%
    \put(0.36096047,0.94628166){\makebox(0,0)[lt]{\lineheight{0}\smash{\begin{tabular}[t]{l}True-post\end{tabular}}}}%
    \put(0,0){\includegraphics[width=\unitlength,page=12]{PALT0bf1k.pdf}}%
    \put(0.54772725,0.89240845){\makebox(0,0)[t]{\lineheight{0}\smash{\begin{tabular}[t]{c}1.0\end{tabular}}}}%
    \put(0,0){\includegraphics[width=\unitlength,page=13]{PALT0bf1k.pdf}}%
    \put(0.66515148,0.89240845){\makebox(0,0)[t]{\lineheight{0}\smash{\begin{tabular}[t]{c}1.5\end{tabular}}}}%
    \put(0,0){\includegraphics[width=\unitlength,page=14]{PALT0bf1k.pdf}}%
    \put(0.78257571,0.89240845){\makebox(0,0)[t]{\lineheight{0}\smash{\begin{tabular}[t]{c}2.0\end{tabular}}}}%
    \put(0,0){\includegraphics[width=\unitlength,page=15]{PALT0bf1k.pdf}}%
    \put(0.9,0.89240845){\makebox(0,0)[t]{\lineheight{0}\smash{\begin{tabular}[t]{c}2.5\end{tabular}}}}%
    \put(0,0){\includegraphics[width=\unitlength,page=16]{PALT0bf1k.pdf}}%
    \put(0.53304112,0.92960469){\makebox(0,0)[rt]{\lineheight{0}\smash{\begin{tabular}[t]{r}0\end{tabular}}}}%
    \put(0,0){\includegraphics[width=\unitlength,page=17]{PALT0bf1k.pdf}}%
    \put(0.53304112,0.98781999){\makebox(0,0)[rt]{\lineheight{0}\smash{\begin{tabular}[t]{r}200\end{tabular}}}}%
    \put(0,0){\includegraphics[width=\unitlength,page=18]{PALT0bf1k.pdf}}%
    \put(0.53304112,1.04603532){\makebox(0,0)[rt]{\lineheight{0}\smash{\begin{tabular}[t]{r}400\end{tabular}}}}%
    \put(0,0){\includegraphics[width=\unitlength,page=19]{PALT0bf1k.pdf}}%
    \put(0.53304112,1.10425062){\makebox(0,0)[rt]{\lineheight{0}\smash{\begin{tabular}[t]{r}600\end{tabular}}}}%
    \put(0,0){\includegraphics[width=\unitlength,page=20]{PALT0bf1k.pdf}}%
    \put(0.53304112,1.16246594){\makebox(0,0)[rt]{\lineheight{0}\smash{\begin{tabular}[t]{r}800\end{tabular}}}}%
    \put(0,0){\includegraphics[width=\unitlength,page=21]{PALT0bf1k.pdf}}%
    \put(0.53304112,1.22068125){\makebox(0,0)[rt]{\lineheight{0}\smash{\begin{tabular}[t]{r}1000\end{tabular}}}}%
    \put(0,0){\includegraphics[width=\unitlength,page=22]{PALT0bf1k.pdf}}%
    \put(0.75449575,0.94628166){\makebox(0,0)[lt]{\lineheight{0}\smash{\begin{tabular}[t]{l}Approx-post\end{tabular}}}}%
    \put(0,0){\includegraphics[width=\unitlength,page=23]{PALT0bf1k.pdf}}%
    \put(0.125,0.50857161){\makebox(0,0)[t]{\lineheight{0}\smash{\begin{tabular}[t]{c}1.0\end{tabular}}}}%
    \put(0,0){\includegraphics[width=\unitlength,page=24]{PALT0bf1k.pdf}}%
    \put(0.24242423,0.50857161){\makebox(0,0)[t]{\lineheight{0}\smash{\begin{tabular}[t]{c}1.5\end{tabular}}}}%
    \put(0,0){\includegraphics[width=\unitlength,page=25]{PALT0bf1k.pdf}}%
    \put(0.35984849,0.50857161){\makebox(0,0)[t]{\lineheight{0}\smash{\begin{tabular}[t]{c}2.0\end{tabular}}}}%
    \put(0,0){\includegraphics[width=\unitlength,page=26]{PALT0bf1k.pdf}}%
    \put(0.47727272,0.50857161){\makebox(0,0)[t]{\lineheight{0}\smash{\begin{tabular}[t]{c}2.5\end{tabular}}}}%
    \put(0,0){\includegraphics[width=\unitlength,page=27]{PALT0bf1k.pdf}}%
    \put(0.11031386,0.54576785){\makebox(0,0)[rt]{\lineheight{0}\smash{\begin{tabular}[t]{r}0\end{tabular}}}}%
    \put(0,0){\includegraphics[width=\unitlength,page=28]{PALT0bf1k.pdf}}%
    \put(0.11031386,0.60398315){\makebox(0,0)[rt]{\lineheight{0}\smash{\begin{tabular}[t]{r}200\end{tabular}}}}%
    \put(0,0){\includegraphics[width=\unitlength,page=29]{PALT0bf1k.pdf}}%
    \put(0.11031386,0.66219845){\makebox(0,0)[rt]{\lineheight{0}\smash{\begin{tabular}[t]{r}400\end{tabular}}}}%
    \put(0,0){\includegraphics[width=\unitlength,page=30]{PALT0bf1k.pdf}}%
    \put(0.11031386,0.72041375){\makebox(0,0)[rt]{\lineheight{0}\smash{\begin{tabular}[t]{r}600\end{tabular}}}}%
    \put(0,0){\includegraphics[width=\unitlength,page=31]{PALT0bf1k.pdf}}%
    \put(0.11031386,0.77862912){\makebox(0,0)[rt]{\lineheight{0}\smash{\begin{tabular}[t]{r}800\end{tabular}}}}%
    \put(0,0){\includegraphics[width=\unitlength,page=32]{PALT0bf1k.pdf}}%
    \put(0.11031386,0.83684442){\makebox(0,0)[rt]{\lineheight{0}\smash{\begin{tabular}[t]{r}1000\end{tabular}}}}%
    \put(0,0){\includegraphics[width=\unitlength,page=33]{PALT0bf1k.pdf}}%
    \put(0.41639407,0.56244483){\makebox(0,0)[lt]{\lineheight{0}\smash{\begin{tabular}[t]{l}BSC\end{tabular}}}}%
    \put(0,0){\includegraphics[width=\unitlength,page=34]{PALT0bf1k.pdf}}%
    \put(0.54772725,0.50857161){\makebox(0,0)[t]{\lineheight{0}\smash{\begin{tabular}[t]{c}1.0\end{tabular}}}}%
    \put(0,0){\includegraphics[width=\unitlength,page=35]{PALT0bf1k.pdf}}%
    \put(0.66515148,0.50857161){\makebox(0,0)[t]{\lineheight{0}\smash{\begin{tabular}[t]{c}1.5\end{tabular}}}}%
    \put(0,0){\includegraphics[width=\unitlength,page=36]{PALT0bf1k.pdf}}%
    \put(0.78257571,0.50857161){\makebox(0,0)[t]{\lineheight{0}\smash{\begin{tabular}[t]{c}2.0\end{tabular}}}}%
    \put(0,0){\includegraphics[width=\unitlength,page=37]{PALT0bf1k.pdf}}%
    \put(0.9,0.50857161){\makebox(0,0)[t]{\lineheight{0}\smash{\begin{tabular}[t]{c}2.5\end{tabular}}}}%
    \put(0,0){\includegraphics[width=\unitlength,page=38]{PALT0bf1k.pdf}}%
    \put(0.53304112,0.54576785){\makebox(0,0)[rt]{\lineheight{0}\smash{\begin{tabular}[t]{r}0\end{tabular}}}}%
    \put(0,0){\includegraphics[width=\unitlength,page=39]{PALT0bf1k.pdf}}%
    \put(0.53304112,0.60398315){\makebox(0,0)[rt]{\lineheight{0}\smash{\begin{tabular}[t]{r}200\end{tabular}}}}%
    \put(0,0){\includegraphics[width=\unitlength,page=40]{PALT0bf1k.pdf}}%
    \put(0.53304112,0.66219845){\makebox(0,0)[rt]{\lineheight{0}\smash{\begin{tabular}[t]{r}400\end{tabular}}}}%
    \put(0,0){\includegraphics[width=\unitlength,page=41]{PALT0bf1k.pdf}}%
    \put(0.53304112,0.72041375){\makebox(0,0)[rt]{\lineheight{0}\smash{\begin{tabular}[t]{r}600\end{tabular}}}}%
    \put(0,0){\includegraphics[width=\unitlength,page=42]{PALT0bf1k.pdf}}%
    \put(0.53304112,0.77862912){\makebox(0,0)[rt]{\lineheight{0}\smash{\begin{tabular}[t]{r}800\end{tabular}}}}%
    \put(0,0){\includegraphics[width=\unitlength,page=43]{PALT0bf1k.pdf}}%
    \put(0.53304112,0.83684442){\makebox(0,0)[rt]{\lineheight{0}\smash{\begin{tabular}[t]{r}1000\end{tabular}}}}%
    \put(0,0){\includegraphics[width=\unitlength,page=44]{PALT0bf1k.pdf}}%
    \put(0.80578582,0.56244483){\makebox(0,0)[lt]{\lineheight{0}\smash{\begin{tabular}[t]{l}SBSC 1\end{tabular}}}}%
    \put(0,0){\includegraphics[width=\unitlength,page=45]{PALT0bf1k.pdf}}%
    \put(0.125,0.12473471){\makebox(0,0)[t]{\lineheight{0}\smash{\begin{tabular}[t]{c}1.0\end{tabular}}}}%
    \put(0,0){\includegraphics[width=\unitlength,page=46]{PALT0bf1k.pdf}}%
    \put(0.24242423,0.12473471){\makebox(0,0)[t]{\lineheight{0}\smash{\begin{tabular}[t]{c}1.5\end{tabular}}}}%
    \put(0,0){\includegraphics[width=\unitlength,page=47]{PALT0bf1k.pdf}}%
    \put(0.35984849,0.12473471){\makebox(0,0)[t]{\lineheight{0}\smash{\begin{tabular}[t]{c}2.0\end{tabular}}}}%
    \put(0,0){\includegraphics[width=\unitlength,page=48]{PALT0bf1k.pdf}}%
    \put(0.47727272,0.12473471){\makebox(0,0)[t]{\lineheight{0}\smash{\begin{tabular}[t]{c}2.5\end{tabular}}}}%
    \put(0,0){\includegraphics[width=\unitlength,page=49]{PALT0bf1k.pdf}}%
    \put(0.11031386,0.16193101){\makebox(0,0)[rt]{\lineheight{0}\smash{\begin{tabular}[t]{r}0\end{tabular}}}}%
    \put(0,0){\includegraphics[width=\unitlength,page=50]{PALT0bf1k.pdf}}%
    \put(0.11031386,0.22014631){\makebox(0,0)[rt]{\lineheight{0}\smash{\begin{tabular}[t]{r}200\end{tabular}}}}%
    \put(0,0){\includegraphics[width=\unitlength,page=51]{PALT0bf1k.pdf}}%
    \put(0.11031386,0.27836161){\makebox(0,0)[rt]{\lineheight{0}\smash{\begin{tabular}[t]{r}400\end{tabular}}}}%
    \put(0,0){\includegraphics[width=\unitlength,page=52]{PALT0bf1k.pdf}}%
    \put(0.11031386,0.33657691){\makebox(0,0)[rt]{\lineheight{0}\smash{\begin{tabular}[t]{r}600\end{tabular}}}}%
    \put(0,0){\includegraphics[width=\unitlength,page=53]{PALT0bf1k.pdf}}%
    \put(0.11031386,0.39479228){\makebox(0,0)[rt]{\lineheight{0}\smash{\begin{tabular}[t]{r}800\end{tabular}}}}%
    \put(0,0){\includegraphics[width=\unitlength,page=54]{PALT0bf1k.pdf}}%
    \put(0.11031386,0.45300758){\makebox(0,0)[rt]{\lineheight{0}\smash{\begin{tabular}[t]{r}1000\end{tabular}}}}%
    \put(0,0){\includegraphics[width=\unitlength,page=55]{PALT0bf1k.pdf}}%
    \put(0.38305853,0.17860799){\makebox(0,0)[lt]{\lineheight{0}\smash{\begin{tabular}[t]{l}SBSC 2\end{tabular}}}}%
    \put(0,0){\includegraphics[width=\unitlength,page=56]{PALT0bf1k.pdf}}%
    \put(0.54772725,0.12473471){\makebox(0,0)[t]{\lineheight{0}\smash{\begin{tabular}[t]{c}1.0\end{tabular}}}}%
    \put(0,0){\includegraphics[width=\unitlength,page=57]{PALT0bf1k.pdf}}%
    \put(0.66515148,0.12473471){\makebox(0,0)[t]{\lineheight{0}\smash{\begin{tabular}[t]{c}1.5\end{tabular}}}}%
    \put(0,0){\includegraphics[width=\unitlength,page=58]{PALT0bf1k.pdf}}%
    \put(0.78257571,0.12473471){\makebox(0,0)[t]{\lineheight{0}\smash{\begin{tabular}[t]{c}2.0\end{tabular}}}}%
    \put(0,0){\includegraphics[width=\unitlength,page=59]{PALT0bf1k.pdf}}%
    \put(0.9,0.12473471){\makebox(0,0)[t]{\lineheight{0}\smash{\begin{tabular}[t]{c}2.5\end{tabular}}}}%
    \put(0,0){\includegraphics[width=\unitlength,page=60]{PALT0bf1k.pdf}}%
    \put(0.53304112,0.16193101){\makebox(0,0)[rt]{\lineheight{0}\smash{\begin{tabular}[t]{r}0\end{tabular}}}}%
    \put(0,0){\includegraphics[width=\unitlength,page=61]{PALT0bf1k.pdf}}%
    \put(0.53304112,0.22014631){\makebox(0,0)[rt]{\lineheight{0}\smash{\begin{tabular}[t]{r}200\end{tabular}}}}%
    \put(0,0){\includegraphics[width=\unitlength,page=62]{PALT0bf1k.pdf}}%
    \put(0.53304112,0.27836161){\makebox(0,0)[rt]{\lineheight{0}\smash{\begin{tabular}[t]{r}400\end{tabular}}}}%
    \put(0,0){\includegraphics[width=\unitlength,page=63]{PALT0bf1k.pdf}}%
    \put(0.53304112,0.33657691){\makebox(0,0)[rt]{\lineheight{0}\smash{\begin{tabular}[t]{r}600\end{tabular}}}}%
    \put(0,0){\includegraphics[width=\unitlength,page=64]{PALT0bf1k.pdf}}%
    \put(0.53304112,0.39479228){\makebox(0,0)[rt]{\lineheight{0}\smash{\begin{tabular}[t]{r}800\end{tabular}}}}%
    \put(0,0){\includegraphics[width=\unitlength,page=65]{PALT0bf1k.pdf}}%
    \put(0.53304112,0.45300758){\makebox(0,0)[rt]{\lineheight{0}\smash{\begin{tabular}[t]{r}1000\end{tabular}}}}%
    \put(0,0){\includegraphics[width=\unitlength,page=66]{PALT0bf1k.pdf}}%
    \put(0.80578582,0.17860799){\makebox(0,0)[lt]{\lineheight{0}\smash{\begin{tabular}[t]{l}SBSC 3\end{tabular}}}}%
    \put(0.5,0.06027119){\makebox(0,0)[t]{\lineheight{0}\smash{\begin{tabular}[t]{c}89\% credible interval\end{tabular}}}}%
    \put(0.03913001,0.5879339){\rotatebox{90}{\makebox(0,0)[lt]{\lineheight{0}\smash{\begin{tabular}[t]{l}Calibration sample\end{tabular}}}}}%
  \end{picture}%
\endgroup%

%% file: tmpplt/PALT1bf1k.pdf_tex
\begingroup%
  \makeatletter%
  \providecommand\color[2][]{%
    \errmessage{(Inkscape) Color is used for the text in Inkscape, but the package 'color.sty' is not loaded}%
    \renewcommand\color[2][]{}%
  }%
  \providecommand\transparent[1]{%
    \errmessage{(Inkscape) Transparency is used (non-zero) for the text in Inkscape, but the package 'transparent.sty' is not loaded}%
    \renewcommand\transparent[1]{}%
  }%
  \providecommand\rotatebox[2]{#2}%
  \newcommand*\fsize{\dimexpr\f@size pt\relax}%
  \newcommand*\lineheight[1]{\fontsize{\fsize}{#1\fsize}\selectfont}%
  \ifx\svgwidth\undefined%
    \setlength{\unitlength}{476.64001465bp}%
    \ifx\svgscale\undefined%
      \relax%
    \else%
      \setlength{\unitlength}{\unitlength * \real{\svgscale}}%
    \fi%
  \else%
    \setlength{\unitlength}{\svgwidth}%
  \fi%
  \global\let\svgwidth\undefined%
  \global\let\svgscale\undefined%
  \makeatother%
  \begin{picture}(1,1.41238669)%
    \lineheight{1}%
    \setlength\tabcolsep{0pt}%
    \put(0,0){\includegraphics[width=\unitlength,page=1]{PALT1bf1k.pdf}}%
    \put(0.125,0.89240845){\makebox(0,0)[t]{\lineheight{0}\smash{\begin{tabular}[t]{c}0.0\end{tabular}}}}%
    \put(0,0){\includegraphics[width=\unitlength,page=2]{PALT1bf1k.pdf}}%
    \put(0.19545454,0.89240845){\makebox(0,0)[t]{\lineheight{0}\smash{\begin{tabular}[t]{c}0.2\end{tabular}}}}%
    \put(0,0){\includegraphics[width=\unitlength,page=3]{PALT1bf1k.pdf}}%
    \put(0.26590909,0.89240845){\makebox(0,0)[t]{\lineheight{0}\smash{\begin{tabular}[t]{c}0.4\end{tabular}}}}%
    \put(0,0){\includegraphics[width=\unitlength,page=4]{PALT1bf1k.pdf}}%
    \put(0.33636364,0.89240845){\makebox(0,0)[t]{\lineheight{0}\smash{\begin{tabular}[t]{c}0.6\end{tabular}}}}%
    \put(0,0){\includegraphics[width=\unitlength,page=5]{PALT1bf1k.pdf}}%
    \put(0.40681818,0.89240845){\makebox(0,0)[t]{\lineheight{0}\smash{\begin{tabular}[t]{c}0.8\end{tabular}}}}%
    \put(0,0){\includegraphics[width=\unitlength,page=6]{PALT1bf1k.pdf}}%
    \put(0.47727272,0.89240845){\makebox(0,0)[t]{\lineheight{0}\smash{\begin{tabular}[t]{c}1.0\end{tabular}}}}%
    \put(0,0){\includegraphics[width=\unitlength,page=7]{PALT1bf1k.pdf}}%
    \put(0.11031386,0.92960469){\makebox(0,0)[rt]{\lineheight{0}\smash{\begin{tabular}[t]{r}0\end{tabular}}}}%
    \put(0,0){\includegraphics[width=\unitlength,page=8]{PALT1bf1k.pdf}}%
    \put(0.11031386,0.98781999){\makebox(0,0)[rt]{\lineheight{0}\smash{\begin{tabular}[t]{r}200\end{tabular}}}}%
    \put(0,0){\includegraphics[width=\unitlength,page=9]{PALT1bf1k.pdf}}%
    \put(0.11031386,1.04603532){\makebox(0,0)[rt]{\lineheight{0}\smash{\begin{tabular}[t]{r}400\end{tabular}}}}%
    \put(0,0){\includegraphics[width=\unitlength,page=10]{PALT1bf1k.pdf}}%
    \put(0.11031386,1.10425062){\makebox(0,0)[rt]{\lineheight{0}\smash{\begin{tabular}[t]{r}600\end{tabular}}}}%
    \put(0,0){\includegraphics[width=\unitlength,page=11]{PALT1bf1k.pdf}}%
    \put(0.11031386,1.16246594){\makebox(0,0)[rt]{\lineheight{0}\smash{\begin{tabular}[t]{r}800\end{tabular}}}}%
    \put(0,0){\includegraphics[width=\unitlength,page=12]{PALT1bf1k.pdf}}%
    \put(0.11031386,1.22068125){\makebox(0,0)[rt]{\lineheight{0}\smash{\begin{tabular}[t]{r}1000\end{tabular}}}}%
    \put(0,0){\includegraphics[width=\unitlength,page=13]{PALT1bf1k.pdf}}%
    \put(0.36096047,0.94628166){\makebox(0,0)[lt]{\lineheight{0}\smash{\begin{tabular}[t]{l}True-post\end{tabular}}}}%
    \put(0,0){\includegraphics[width=\unitlength,page=14]{PALT1bf1k.pdf}}%
    \put(0.54772725,0.89240845){\makebox(0,0)[t]{\lineheight{0}\smash{\begin{tabular}[t]{c}0.0\end{tabular}}}}%
    \put(0,0){\includegraphics[width=\unitlength,page=15]{PALT1bf1k.pdf}}%
    \put(0.61818179,0.89240845){\makebox(0,0)[t]{\lineheight{0}\smash{\begin{tabular}[t]{c}0.2\end{tabular}}}}%
    \put(0,0){\includegraphics[width=\unitlength,page=16]{PALT1bf1k.pdf}}%
    \put(0.68863633,0.89240845){\makebox(0,0)[t]{\lineheight{0}\smash{\begin{tabular}[t]{c}0.4\end{tabular}}}}%
    \put(0,0){\includegraphics[width=\unitlength,page=17]{PALT1bf1k.pdf}}%
    \put(0.75909086,0.89240845){\makebox(0,0)[t]{\lineheight{0}\smash{\begin{tabular}[t]{c}0.6\end{tabular}}}}%
    \put(0,0){\includegraphics[width=\unitlength,page=18]{PALT1bf1k.pdf}}%
    \put(0.8295454,0.89240845){\makebox(0,0)[t]{\lineheight{0}\smash{\begin{tabular}[t]{c}0.8\end{tabular}}}}%
    \put(0,0){\includegraphics[width=\unitlength,page=19]{PALT1bf1k.pdf}}%
    \put(0.9,0.89240845){\makebox(0,0)[t]{\lineheight{0}\smash{\begin{tabular}[t]{c}1.0\end{tabular}}}}%
    \put(0,0){\includegraphics[width=\unitlength,page=20]{PALT1bf1k.pdf}}%
    \put(0.53304112,0.92960469){\makebox(0,0)[rt]{\lineheight{0}\smash{\begin{tabular}[t]{r}0\end{tabular}}}}%
    \put(0,0){\includegraphics[width=\unitlength,page=21]{PALT1bf1k.pdf}}%
    \put(0.53304112,0.98781999){\makebox(0,0)[rt]{\lineheight{0}\smash{\begin{tabular}[t]{r}200\end{tabular}}}}%
    \put(0,0){\includegraphics[width=\unitlength,page=22]{PALT1bf1k.pdf}}%
    \put(0.53304112,1.04603532){\makebox(0,0)[rt]{\lineheight{0}\smash{\begin{tabular}[t]{r}400\end{tabular}}}}%
    \put(0,0){\includegraphics[width=\unitlength,page=23]{PALT1bf1k.pdf}}%
    \put(0.53304112,1.10425062){\makebox(0,0)[rt]{\lineheight{0}\smash{\begin{tabular}[t]{r}600\end{tabular}}}}%
    \put(0,0){\includegraphics[width=\unitlength,page=24]{PALT1bf1k.pdf}}%
    \put(0.53304112,1.16246594){\makebox(0,0)[rt]{\lineheight{0}\smash{\begin{tabular}[t]{r}800\end{tabular}}}}%
    \put(0,0){\includegraphics[width=\unitlength,page=25]{PALT1bf1k.pdf}}%
    \put(0.53304112,1.22068125){\makebox(0,0)[rt]{\lineheight{0}\smash{\begin{tabular}[t]{r}1000\end{tabular}}}}%
    \put(0,0){\includegraphics[width=\unitlength,page=26]{PALT1bf1k.pdf}}%
    \put(0.75449575,0.94628166){\makebox(0,0)[lt]{\lineheight{0}\smash{\begin{tabular}[t]{l}Approx-post\end{tabular}}}}%
    \put(0,0){\includegraphics[width=\unitlength,page=27]{PALT1bf1k.pdf}}%
    \put(0.125,0.50857161){\makebox(0,0)[t]{\lineheight{0}\smash{\begin{tabular}[t]{c}0.0\end{tabular}}}}%
    \put(0,0){\includegraphics[width=\unitlength,page=28]{PALT1bf1k.pdf}}%
    \put(0.19545454,0.50857161){\makebox(0,0)[t]{\lineheight{0}\smash{\begin{tabular}[t]{c}0.2\end{tabular}}}}%
    \put(0,0){\includegraphics[width=\unitlength,page=29]{PALT1bf1k.pdf}}%
    \put(0.26590909,0.50857161){\makebox(0,0)[t]{\lineheight{0}\smash{\begin{tabular}[t]{c}0.4\end{tabular}}}}%
    \put(0,0){\includegraphics[width=\unitlength,page=30]{PALT1bf1k.pdf}}%
    \put(0.33636364,0.50857161){\makebox(0,0)[t]{\lineheight{0}\smash{\begin{tabular}[t]{c}0.6\end{tabular}}}}%
    \put(0,0){\includegraphics[width=\unitlength,page=31]{PALT1bf1k.pdf}}%
    \put(0.40681818,0.50857161){\makebox(0,0)[t]{\lineheight{0}\smash{\begin{tabular}[t]{c}0.8\end{tabular}}}}%
    \put(0,0){\includegraphics[width=\unitlength,page=32]{PALT1bf1k.pdf}}%
    \put(0.47727272,0.50857161){\makebox(0,0)[t]{\lineheight{0}\smash{\begin{tabular}[t]{c}1.0\end{tabular}}}}%
    \put(0,0){\includegraphics[width=\unitlength,page=33]{PALT1bf1k.pdf}}%
    \put(0.11031386,0.54576785){\makebox(0,0)[rt]{\lineheight{0}\smash{\begin{tabular}[t]{r}0\end{tabular}}}}%
    \put(0,0){\includegraphics[width=\unitlength,page=34]{PALT1bf1k.pdf}}%
    \put(0.11031386,0.60398315){\makebox(0,0)[rt]{\lineheight{0}\smash{\begin{tabular}[t]{r}200\end{tabular}}}}%
    \put(0,0){\includegraphics[width=\unitlength,page=35]{PALT1bf1k.pdf}}%
    \put(0.11031386,0.66219845){\makebox(0,0)[rt]{\lineheight{0}\smash{\begin{tabular}[t]{r}400\end{tabular}}}}%
    \put(0,0){\includegraphics[width=\unitlength,page=36]{PALT1bf1k.pdf}}%
    \put(0.11031386,0.72041375){\makebox(0,0)[rt]{\lineheight{0}\smash{\begin{tabular}[t]{r}600\end{tabular}}}}%
    \put(0,0){\includegraphics[width=\unitlength,page=37]{PALT1bf1k.pdf}}%
    \put(0.11031386,0.77862912){\makebox(0,0)[rt]{\lineheight{0}\smash{\begin{tabular}[t]{r}800\end{tabular}}}}%
    \put(0,0){\includegraphics[width=\unitlength,page=38]{PALT1bf1k.pdf}}%
    \put(0.11031386,0.83684442){\makebox(0,0)[rt]{\lineheight{0}\smash{\begin{tabular}[t]{r}1000\end{tabular}}}}%
    \put(0,0){\includegraphics[width=\unitlength,page=39]{PALT1bf1k.pdf}}%
    \put(0.41639407,0.56244483){\makebox(0,0)[lt]{\lineheight{0}\smash{\begin{tabular}[t]{l}BSC\end{tabular}}}}%
    \put(0,0){\includegraphics[width=\unitlength,page=40]{PALT1bf1k.pdf}}%
    \put(0.54772725,0.50857161){\makebox(0,0)[t]{\lineheight{0}\smash{\begin{tabular}[t]{c}0.0\end{tabular}}}}%
    \put(0,0){\includegraphics[width=\unitlength,page=41]{PALT1bf1k.pdf}}%
    \put(0.61818179,0.50857161){\makebox(0,0)[t]{\lineheight{0}\smash{\begin{tabular}[t]{c}0.2\end{tabular}}}}%
    \put(0,0){\includegraphics[width=\unitlength,page=42]{PALT1bf1k.pdf}}%
    \put(0.68863633,0.50857161){\makebox(0,0)[t]{\lineheight{0}\smash{\begin{tabular}[t]{c}0.4\end{tabular}}}}%
    \put(0,0){\includegraphics[width=\unitlength,page=43]{PALT1bf1k.pdf}}%
    \put(0.75909086,0.50857161){\makebox(0,0)[t]{\lineheight{0}\smash{\begin{tabular}[t]{c}0.6\end{tabular}}}}%
    \put(0,0){\includegraphics[width=\unitlength,page=44]{PALT1bf1k.pdf}}%
    \put(0.8295454,0.50857161){\makebox(0,0)[t]{\lineheight{0}\smash{\begin{tabular}[t]{c}0.8\end{tabular}}}}%
    \put(0,0){\includegraphics[width=\unitlength,page=45]{PALT1bf1k.pdf}}%
    \put(0.9,0.50857161){\makebox(0,0)[t]{\lineheight{0}\smash{\begin{tabular}[t]{c}1.0\end{tabular}}}}%
    \put(0,0){\includegraphics[width=\unitlength,page=46]{PALT1bf1k.pdf}}%
    \put(0.53304112,0.54576785){\makebox(0,0)[rt]{\lineheight{0}\smash{\begin{tabular}[t]{r}0\end{tabular}}}}%
    \put(0,0){\includegraphics[width=\unitlength,page=47]{PALT1bf1k.pdf}}%
    \put(0.53304112,0.60398315){\makebox(0,0)[rt]{\lineheight{0}\smash{\begin{tabular}[t]{r}200\end{tabular}}}}%
    \put(0,0){\includegraphics[width=\unitlength,page=48]{PALT1bf1k.pdf}}%
    \put(0.53304112,0.66219845){\makebox(0,0)[rt]{\lineheight{0}\smash{\begin{tabular}[t]{r}400\end{tabular}}}}%
    \put(0,0){\includegraphics[width=\unitlength,page=49]{PALT1bf1k.pdf}}%
    \put(0.53304112,0.72041375){\makebox(0,0)[rt]{\lineheight{0}\smash{\begin{tabular}[t]{r}600\end{tabular}}}}%
    \put(0,0){\includegraphics[width=\unitlength,page=50]{PALT1bf1k.pdf}}%
    \put(0.53304112,0.77862912){\makebox(0,0)[rt]{\lineheight{0}\smash{\begin{tabular}[t]{r}800\end{tabular}}}}%
    \put(0,0){\includegraphics[width=\unitlength,page=51]{PALT1bf1k.pdf}}%
    \put(0.53304112,0.83684442){\makebox(0,0)[rt]{\lineheight{0}\smash{\begin{tabular}[t]{r}1000\end{tabular}}}}%
    \put(0,0){\includegraphics[width=\unitlength,page=52]{PALT1bf1k.pdf}}%
    \put(0.80578582,0.56244483){\makebox(0,0)[lt]{\lineheight{0}\smash{\begin{tabular}[t]{l}SBSC 1\end{tabular}}}}%
    \put(0,0){\includegraphics[width=\unitlength,page=53]{PALT1bf1k.pdf}}%
    \put(0.125,0.12473471){\makebox(0,0)[t]{\lineheight{0}\smash{\begin{tabular}[t]{c}0.0\end{tabular}}}}%
    \put(0,0){\includegraphics[width=\unitlength,page=54]{PALT1bf1k.pdf}}%
    \put(0.19545454,0.12473471){\makebox(0,0)[t]{\lineheight{0}\smash{\begin{tabular}[t]{c}0.2\end{tabular}}}}%
    \put(0,0){\includegraphics[width=\unitlength,page=55]{PALT1bf1k.pdf}}%
    \put(0.26590909,0.12473471){\makebox(0,0)[t]{\lineheight{0}\smash{\begin{tabular}[t]{c}0.4\end{tabular}}}}%
    \put(0,0){\includegraphics[width=\unitlength,page=56]{PALT1bf1k.pdf}}%
    \put(0.33636364,0.12473471){\makebox(0,0)[t]{\lineheight{0}\smash{\begin{tabular}[t]{c}0.6\end{tabular}}}}%
    \put(0,0){\includegraphics[width=\unitlength,page=57]{PALT1bf1k.pdf}}%
    \put(0.40681818,0.12473471){\makebox(0,0)[t]{\lineheight{0}\smash{\begin{tabular}[t]{c}0.8\end{tabular}}}}%
    \put(0,0){\includegraphics[width=\unitlength,page=58]{PALT1bf1k.pdf}}%
    \put(0.47727272,0.12473471){\makebox(0,0)[t]{\lineheight{0}\smash{\begin{tabular}[t]{c}1.0\end{tabular}}}}%
    \put(0,0){\includegraphics[width=\unitlength,page=59]{PALT1bf1k.pdf}}%
    \put(0.11031386,0.16193101){\makebox(0,0)[rt]{\lineheight{0}\smash{\begin{tabular}[t]{r}0\end{tabular}}}}%
    \put(0,0){\includegraphics[width=\unitlength,page=60]{PALT1bf1k.pdf}}%
    \put(0.11031386,0.22014631){\makebox(0,0)[rt]{\lineheight{0}\smash{\begin{tabular}[t]{r}200\end{tabular}}}}%
    \put(0,0){\includegraphics[width=\unitlength,page=61]{PALT1bf1k.pdf}}%
    \put(0.11031386,0.27836161){\makebox(0,0)[rt]{\lineheight{0}\smash{\begin{tabular}[t]{r}400\end{tabular}}}}%
    \put(0,0){\includegraphics[width=\unitlength,page=62]{PALT1bf1k.pdf}}%
    \put(0.11031386,0.33657691){\makebox(0,0)[rt]{\lineheight{0}\smash{\begin{tabular}[t]{r}600\end{tabular}}}}%
    \put(0,0){\includegraphics[width=\unitlength,page=63]{PALT1bf1k.pdf}}%
    \put(0.11031386,0.39479228){\makebox(0,0)[rt]{\lineheight{0}\smash{\begin{tabular}[t]{r}800\end{tabular}}}}%
    \put(0,0){\includegraphics[width=\unitlength,page=64]{PALT1bf1k.pdf}}%
    \put(0.11031386,0.45300758){\makebox(0,0)[rt]{\lineheight{0}\smash{\begin{tabular}[t]{r}1000\end{tabular}}}}%
    \put(0,0){\includegraphics[width=\unitlength,page=65]{PALT1bf1k.pdf}}%
    \put(0.38305853,0.17860799){\makebox(0,0)[lt]{\lineheight{0}\smash{\begin{tabular}[t]{l}SBSC 2\end{tabular}}}}%
    \put(0,0){\includegraphics[width=\unitlength,page=66]{PALT1bf1k.pdf}}%
    \put(0.54772725,0.12473471){\makebox(0,0)[t]{\lineheight{0}\smash{\begin{tabular}[t]{c}0.0\end{tabular}}}}%
    \put(0,0){\includegraphics[width=\unitlength,page=67]{PALT1bf1k.pdf}}%
    \put(0.61818179,0.12473471){\makebox(0,0)[t]{\lineheight{0}\smash{\begin{tabular}[t]{c}0.2\end{tabular}}}}%
    \put(0,0){\includegraphics[width=\unitlength,page=68]{PALT1bf1k.pdf}}%
    \put(0.68863633,0.12473471){\makebox(0,0)[t]{\lineheight{0}\smash{\begin{tabular}[t]{c}0.4\end{tabular}}}}%
    \put(0,0){\includegraphics[width=\unitlength,page=69]{PALT1bf1k.pdf}}%
    \put(0.75909086,0.12473471){\makebox(0,0)[t]{\lineheight{0}\smash{\begin{tabular}[t]{c}0.6\end{tabular}}}}%
    \put(0,0){\includegraphics[width=\unitlength,page=70]{PALT1bf1k.pdf}}%
    \put(0.8295454,0.12473471){\makebox(0,0)[t]{\lineheight{0}\smash{\begin{tabular}[t]{c}0.8\end{tabular}}}}%
    \put(0,0){\includegraphics[width=\unitlength,page=71]{PALT1bf1k.pdf}}%
    \put(0.9,0.12473471){\makebox(0,0)[t]{\lineheight{0}\smash{\begin{tabular}[t]{c}1.0\end{tabular}}}}%
    \put(0,0){\includegraphics[width=\unitlength,page=72]{PALT1bf1k.pdf}}%
    \put(0.53304112,0.16193101){\makebox(0,0)[rt]{\lineheight{0}\smash{\begin{tabular}[t]{r}0\end{tabular}}}}%
    \put(0,0){\includegraphics[width=\unitlength,page=73]{PALT1bf1k.pdf}}%
    \put(0.53304112,0.22014631){\makebox(0,0)[rt]{\lineheight{0}\smash{\begin{tabular}[t]{r}200\end{tabular}}}}%
    \put(0,0){\includegraphics[width=\unitlength,page=74]{PALT1bf1k.pdf}}%
    \put(0.53304112,0.27836161){\makebox(0,0)[rt]{\lineheight{0}\smash{\begin{tabular}[t]{r}400\end{tabular}}}}%
    \put(0,0){\includegraphics[width=\unitlength,page=75]{PALT1bf1k.pdf}}%
    \put(0.53304112,0.33657691){\makebox(0,0)[rt]{\lineheight{0}\smash{\begin{tabular}[t]{r}600\end{tabular}}}}%
    \put(0,0){\includegraphics[width=\unitlength,page=76]{PALT1bf1k.pdf}}%
    \put(0.53304112,0.39479228){\makebox(0,0)[rt]{\lineheight{0}\smash{\begin{tabular}[t]{r}800\end{tabular}}}}%
    \put(0,0){\includegraphics[width=\unitlength,page=77]{PALT1bf1k.pdf}}%
    \put(0.53304112,0.45300758){\makebox(0,0)[rt]{\lineheight{0}\smash{\begin{tabular}[t]{r}1000\end{tabular}}}}%
    \put(0,0){\includegraphics[width=\unitlength,page=78]{PALT1bf1k.pdf}}%
    \put(0.80578582,0.17860799){\makebox(0,0)[lt]{\lineheight{0}\smash{\begin{tabular}[t]{l}SBSC 3\end{tabular}}}}%
    \put(0.5,0.05712416){\makebox(0,0)[t]{\lineheight{0}\smash{\begin{tabular}[t]{c}89\% credible interval\end{tabular}}}}%
    \put(0.03913001,0.5879339){\rotatebox{90}{\makebox(0,0)[lt]{\lineheight{0}\smash{\begin{tabular}[t]{l}Calibration sample\end{tabular}}}}}%
  \end{picture}%
\endgroup%

%% file: tmpplt/PALT2bf1k.pdf_tex
\begingroup%
  \makeatletter%
  \providecommand\color[2][]{%
    \errmessage{(Inkscape) Color is used for the text in Inkscape, but the package 'color.sty' is not loaded}%
    \renewcommand\color[2][]{}%
  }%
  \providecommand\transparent[1]{%
    \errmessage{(Inkscape) Transparency is used (non-zero) for the text in Inkscape, but the package 'transparent.sty' is not loaded}%
    \renewcommand\transparent[1]{}%
  }%
  \providecommand\rotatebox[2]{#2}%
  \newcommand*\fsize{\dimexpr\f@size pt\relax}%
  \newcommand*\lineheight[1]{\fontsize{\fsize}{#1\fsize}\selectfont}%
  \ifx\svgwidth\undefined%
    \setlength{\unitlength}{476.64001465bp}%
    \ifx\svgscale\undefined%
      \relax%
    \else%
      \setlength{\unitlength}{\unitlength * \real{\svgscale}}%
    \fi%
  \else%
    \setlength{\unitlength}{\svgwidth}%
  \fi%
  \global\let\svgwidth\undefined%
  \global\let\svgscale\undefined%
  \makeatother%
  \begin{picture}(1,1.41238669)%
    \lineheight{1}%
    \setlength\tabcolsep{0pt}%
    \put(0,0){\includegraphics[width=\unitlength,page=1]{PALT2bf1k.pdf}}%
    \put(0.17532466,0.89240845){\makebox(0,0)[t]{\lineheight{0}\smash{\begin{tabular}[t]{c}0.4\end{tabular}}}}%
    \put(0,0){\includegraphics[width=\unitlength,page=2]{PALT2bf1k.pdf}}%
    \put(0.275974,0.89240845){\makebox(0,0)[t]{\lineheight{0}\smash{\begin{tabular}[t]{c}0.6\end{tabular}}}}%
    \put(0,0){\includegraphics[width=\unitlength,page=3]{PALT2bf1k.pdf}}%
    \put(0.37662336,0.89240845){\makebox(0,0)[t]{\lineheight{0}\smash{\begin{tabular}[t]{c}0.8\end{tabular}}}}%
    \put(0,0){\includegraphics[width=\unitlength,page=4]{PALT2bf1k.pdf}}%
    \put(0.47727272,0.89240845){\makebox(0,0)[t]{\lineheight{0}\smash{\begin{tabular}[t]{c}1.0\end{tabular}}}}%
    \put(0,0){\includegraphics[width=\unitlength,page=5]{PALT2bf1k.pdf}}%
    \put(0.11031386,0.92960469){\makebox(0,0)[rt]{\lineheight{0}\smash{\begin{tabular}[t]{r}0\end{tabular}}}}%
    \put(0,0){\includegraphics[width=\unitlength,page=6]{PALT2bf1k.pdf}}%
    \put(0.11031386,0.98781999){\makebox(0,0)[rt]{\lineheight{0}\smash{\begin{tabular}[t]{r}200\end{tabular}}}}%
    \put(0,0){\includegraphics[width=\unitlength,page=7]{PALT2bf1k.pdf}}%
    \put(0.11031386,1.04603532){\makebox(0,0)[rt]{\lineheight{0}\smash{\begin{tabular}[t]{r}400\end{tabular}}}}%
    \put(0,0){\includegraphics[width=\unitlength,page=8]{PALT2bf1k.pdf}}%
    \put(0.11031386,1.10425062){\makebox(0,0)[rt]{\lineheight{0}\smash{\begin{tabular}[t]{r}600\end{tabular}}}}%
    \put(0,0){\includegraphics[width=\unitlength,page=9]{PALT2bf1k.pdf}}%
    \put(0.11031386,1.16246594){\makebox(0,0)[rt]{\lineheight{0}\smash{\begin{tabular}[t]{r}800\end{tabular}}}}%
    \put(0,0){\includegraphics[width=\unitlength,page=10]{PALT2bf1k.pdf}}%
    \put(0.11031386,1.22068125){\makebox(0,0)[rt]{\lineheight{0}\smash{\begin{tabular}[t]{r}1000\end{tabular}}}}%
    \put(0,0){\includegraphics[width=\unitlength,page=11]{PALT2bf1k.pdf}}%
    \put(0.36096047,0.94628166){\makebox(0,0)[lt]{\lineheight{0}\smash{\begin{tabular}[t]{l}True-post\end{tabular}}}}%
    \put(0,0){\includegraphics[width=\unitlength,page=12]{PALT2bf1k.pdf}}%
    \put(0.59805193,0.89240845){\makebox(0,0)[t]{\lineheight{0}\smash{\begin{tabular}[t]{c}0.4\end{tabular}}}}%
    \put(0,0){\includegraphics[width=\unitlength,page=13]{PALT2bf1k.pdf}}%
    \put(0.69870129,0.89240845){\makebox(0,0)[t]{\lineheight{0}\smash{\begin{tabular}[t]{c}0.6\end{tabular}}}}%
    \put(0,0){\includegraphics[width=\unitlength,page=14]{PALT2bf1k.pdf}}%
    \put(0.79935064,0.89240845){\makebox(0,0)[t]{\lineheight{0}\smash{\begin{tabular}[t]{c}0.8\end{tabular}}}}%
    \put(0,0){\includegraphics[width=\unitlength,page=15]{PALT2bf1k.pdf}}%
    \put(0.9,0.89240845){\makebox(0,0)[t]{\lineheight{0}\smash{\begin{tabular}[t]{c}1.0\end{tabular}}}}%
    \put(0,0){\includegraphics[width=\unitlength,page=16]{PALT2bf1k.pdf}}%
    \put(0.53304112,0.92960469){\makebox(0,0)[rt]{\lineheight{0}\smash{\begin{tabular}[t]{r}0\end{tabular}}}}%
    \put(0,0){\includegraphics[width=\unitlength,page=17]{PALT2bf1k.pdf}}%
    \put(0.53304112,0.98781999){\makebox(0,0)[rt]{\lineheight{0}\smash{\begin{tabular}[t]{r}200\end{tabular}}}}%
    \put(0,0){\includegraphics[width=\unitlength,page=18]{PALT2bf1k.pdf}}%
    \put(0.53304112,1.04603532){\makebox(0,0)[rt]{\lineheight{0}\smash{\begin{tabular}[t]{r}400\end{tabular}}}}%
    \put(0,0){\includegraphics[width=\unitlength,page=19]{PALT2bf1k.pdf}}%
    \put(0.53304112,1.10425062){\makebox(0,0)[rt]{\lineheight{0}\smash{\begin{tabular}[t]{r}600\end{tabular}}}}%
    \put(0,0){\includegraphics[width=\unitlength,page=20]{PALT2bf1k.pdf}}%
    \put(0.53304112,1.16246594){\makebox(0,0)[rt]{\lineheight{0}\smash{\begin{tabular}[t]{r}800\end{tabular}}}}%
    \put(0,0){\includegraphics[width=\unitlength,page=21]{PALT2bf1k.pdf}}%
    \put(0.53304112,1.22068125){\makebox(0,0)[rt]{\lineheight{0}\smash{\begin{tabular}[t]{r}1000\end{tabular}}}}%
    \put(0,0){\includegraphics[width=\unitlength,page=22]{PALT2bf1k.pdf}}%
    \put(0.75449575,0.94628166){\makebox(0,0)[lt]{\lineheight{0}\smash{\begin{tabular}[t]{l}Approx-post\end{tabular}}}}%
    \put(0,0){\includegraphics[width=\unitlength,page=23]{PALT2bf1k.pdf}}%
    \put(0.17532466,0.50857161){\makebox(0,0)[t]{\lineheight{0}\smash{\begin{tabular}[t]{c}0.4\end{tabular}}}}%
    \put(0,0){\includegraphics[width=\unitlength,page=24]{PALT2bf1k.pdf}}%
    \put(0.275974,0.50857161){\makebox(0,0)[t]{\lineheight{0}\smash{\begin{tabular}[t]{c}0.6\end{tabular}}}}%
    \put(0,0){\includegraphics[width=\unitlength,page=25]{PALT2bf1k.pdf}}%
    \put(0.37662336,0.50857161){\makebox(0,0)[t]{\lineheight{0}\smash{\begin{tabular}[t]{c}0.8\end{tabular}}}}%
    \put(0,0){\includegraphics[width=\unitlength,page=26]{PALT2bf1k.pdf}}%
    \put(0.47727272,0.50857161){\makebox(0,0)[t]{\lineheight{0}\smash{\begin{tabular}[t]{c}1.0\end{tabular}}}}%
    \put(0,0){\includegraphics[width=\unitlength,page=27]{PALT2bf1k.pdf}}%
    \put(0.11031386,0.54576785){\makebox(0,0)[rt]{\lineheight{0}\smash{\begin{tabular}[t]{r}0\end{tabular}}}}%
    \put(0,0){\includegraphics[width=\unitlength,page=28]{PALT2bf1k.pdf}}%
    \put(0.11031386,0.60398315){\makebox(0,0)[rt]{\lineheight{0}\smash{\begin{tabular}[t]{r}200\end{tabular}}}}%
    \put(0,0){\includegraphics[width=\unitlength,page=29]{PALT2bf1k.pdf}}%
    \put(0.11031386,0.66219845){\makebox(0,0)[rt]{\lineheight{0}\smash{\begin{tabular}[t]{r}400\end{tabular}}}}%
    \put(0,0){\includegraphics[width=\unitlength,page=30]{PALT2bf1k.pdf}}%
    \put(0.11031386,0.72041375){\makebox(0,0)[rt]{\lineheight{0}\smash{\begin{tabular}[t]{r}600\end{tabular}}}}%
    \put(0,0){\includegraphics[width=\unitlength,page=31]{PALT2bf1k.pdf}}%
    \put(0.11031386,0.77862912){\makebox(0,0)[rt]{\lineheight{0}\smash{\begin{tabular}[t]{r}800\end{tabular}}}}%
    \put(0,0){\includegraphics[width=\unitlength,page=32]{PALT2bf1k.pdf}}%
    \put(0.11031386,0.83684442){\makebox(0,0)[rt]{\lineheight{0}\smash{\begin{tabular}[t]{r}1000\end{tabular}}}}%
    \put(0,0){\includegraphics[width=\unitlength,page=33]{PALT2bf1k.pdf}}%
    \put(0.41639407,0.56244483){\makebox(0,0)[lt]{\lineheight{0}\smash{\begin{tabular}[t]{l}BSC\end{tabular}}}}%
    \put(0,0){\includegraphics[width=\unitlength,page=34]{PALT2bf1k.pdf}}%
    \put(0.59805193,0.50857161){\makebox(0,0)[t]{\lineheight{0}\smash{\begin{tabular}[t]{c}0.4\end{tabular}}}}%
    \put(0,0){\includegraphics[width=\unitlength,page=35]{PALT2bf1k.pdf}}%
    \put(0.69870129,0.50857161){\makebox(0,0)[t]{\lineheight{0}\smash{\begin{tabular}[t]{c}0.6\end{tabular}}}}%
    \put(0,0){\includegraphics[width=\unitlength,page=36]{PALT2bf1k.pdf}}%
    \put(0.79935064,0.50857161){\makebox(0,0)[t]{\lineheight{0}\smash{\begin{tabular}[t]{c}0.8\end{tabular}}}}%
    \put(0,0){\includegraphics[width=\unitlength,page=37]{PALT2bf1k.pdf}}%
    \put(0.9,0.50857161){\makebox(0,0)[t]{\lineheight{0}\smash{\begin{tabular}[t]{c}1.0\end{tabular}}}}%
    \put(0,0){\includegraphics[width=\unitlength,page=38]{PALT2bf1k.pdf}}%
    \put(0.53304112,0.54576785){\makebox(0,0)[rt]{\lineheight{0}\smash{\begin{tabular}[t]{r}0\end{tabular}}}}%
    \put(0,0){\includegraphics[width=\unitlength,page=39]{PALT2bf1k.pdf}}%
    \put(0.53304112,0.60398315){\makebox(0,0)[rt]{\lineheight{0}\smash{\begin{tabular}[t]{r}200\end{tabular}}}}%
    \put(0,0){\includegraphics[width=\unitlength,page=40]{PALT2bf1k.pdf}}%
    \put(0.53304112,0.66219845){\makebox(0,0)[rt]{\lineheight{0}\smash{\begin{tabular}[t]{r}400\end{tabular}}}}%
    \put(0,0){\includegraphics[width=\unitlength,page=41]{PALT2bf1k.pdf}}%
    \put(0.53304112,0.72041375){\makebox(0,0)[rt]{\lineheight{0}\smash{\begin{tabular}[t]{r}600\end{tabular}}}}%
    \put(0,0){\includegraphics[width=\unitlength,page=42]{PALT2bf1k.pdf}}%
    \put(0.53304112,0.77862912){\makebox(0,0)[rt]{\lineheight{0}\smash{\begin{tabular}[t]{r}800\end{tabular}}}}%
    \put(0,0){\includegraphics[width=\unitlength,page=43]{PALT2bf1k.pdf}}%
    \put(0.53304112,0.83684442){\makebox(0,0)[rt]{\lineheight{0}\smash{\begin{tabular}[t]{r}1000\end{tabular}}}}%
    \put(0,0){\includegraphics[width=\unitlength,page=44]{PALT2bf1k.pdf}}%
    \put(0.80578582,0.56244483){\makebox(0,0)[lt]{\lineheight{0}\smash{\begin{tabular}[t]{l}SBSC 1\end{tabular}}}}%
    \put(0,0){\includegraphics[width=\unitlength,page=45]{PALT2bf1k.pdf}}%
    \put(0.17532466,0.12473471){\makebox(0,0)[t]{\lineheight{0}\smash{\begin{tabular}[t]{c}0.4\end{tabular}}}}%
    \put(0,0){\includegraphics[width=\unitlength,page=46]{PALT2bf1k.pdf}}%
    \put(0.275974,0.12473471){\makebox(0,0)[t]{\lineheight{0}\smash{\begin{tabular}[t]{c}0.6\end{tabular}}}}%
    \put(0,0){\includegraphics[width=\unitlength,page=47]{PALT2bf1k.pdf}}%
    \put(0.37662336,0.12473471){\makebox(0,0)[t]{\lineheight{0}\smash{\begin{tabular}[t]{c}0.8\end{tabular}}}}%
    \put(0,0){\includegraphics[width=\unitlength,page=48]{PALT2bf1k.pdf}}%
    \put(0.47727272,0.12473471){\makebox(0,0)[t]{\lineheight{0}\smash{\begin{tabular}[t]{c}1.0\end{tabular}}}}%
    \put(0,0){\includegraphics[width=\unitlength,page=49]{PALT2bf1k.pdf}}%
    \put(0.11031386,0.16193101){\makebox(0,0)[rt]{\lineheight{0}\smash{\begin{tabular}[t]{r}0\end{tabular}}}}%
    \put(0,0){\includegraphics[width=\unitlength,page=50]{PALT2bf1k.pdf}}%
    \put(0.11031386,0.22014631){\makebox(0,0)[rt]{\lineheight{0}\smash{\begin{tabular}[t]{r}200\end{tabular}}}}%
    \put(0,0){\includegraphics[width=\unitlength,page=51]{PALT2bf1k.pdf}}%
    \put(0.11031386,0.27836161){\makebox(0,0)[rt]{\lineheight{0}\smash{\begin{tabular}[t]{r}400\end{tabular}}}}%
    \put(0,0){\includegraphics[width=\unitlength,page=52]{PALT2bf1k.pdf}}%
    \put(0.11031386,0.33657691){\makebox(0,0)[rt]{\lineheight{0}\smash{\begin{tabular}[t]{r}600\end{tabular}}}}%
    \put(0,0){\includegraphics[width=\unitlength,page=53]{PALT2bf1k.pdf}}%
    \put(0.11031386,0.39479228){\makebox(0,0)[rt]{\lineheight{0}\smash{\begin{tabular}[t]{r}800\end{tabular}}}}%
    \put(0,0){\includegraphics[width=\unitlength,page=54]{PALT2bf1k.pdf}}%
    \put(0.11031386,0.45300758){\makebox(0,0)[rt]{\lineheight{0}\smash{\begin{tabular}[t]{r}1000\end{tabular}}}}%
    \put(0,0){\includegraphics[width=\unitlength,page=55]{PALT2bf1k.pdf}}%
    \put(0.38305853,0.17860799){\makebox(0,0)[lt]{\lineheight{0}\smash{\begin{tabular}[t]{l}SBSC 2\end{tabular}}}}%
    \put(0,0){\includegraphics[width=\unitlength,page=56]{PALT2bf1k.pdf}}%
    \put(0.59805193,0.12473471){\makebox(0,0)[t]{\lineheight{0}\smash{\begin{tabular}[t]{c}0.4\end{tabular}}}}%
    \put(0,0){\includegraphics[width=\unitlength,page=57]{PALT2bf1k.pdf}}%
    \put(0.69870129,0.12473471){\makebox(0,0)[t]{\lineheight{0}\smash{\begin{tabular}[t]{c}0.6\end{tabular}}}}%
    \put(0,0){\includegraphics[width=\unitlength,page=58]{PALT2bf1k.pdf}}%
    \put(0.79935064,0.12473471){\makebox(0,0)[t]{\lineheight{0}\smash{\begin{tabular}[t]{c}0.8\end{tabular}}}}%
    \put(0,0){\includegraphics[width=\unitlength,page=59]{PALT2bf1k.pdf}}%
    \put(0.9,0.12473471){\makebox(0,0)[t]{\lineheight{0}\smash{\begin{tabular}[t]{c}1.0\end{tabular}}}}%
    \put(0,0){\includegraphics[width=\unitlength,page=60]{PALT2bf1k.pdf}}%
    \put(0.53304112,0.16193101){\makebox(0,0)[rt]{\lineheight{0}\smash{\begin{tabular}[t]{r}0\end{tabular}}}}%
    \put(0,0){\includegraphics[width=\unitlength,page=61]{PALT2bf1k.pdf}}%
    \put(0.53304112,0.22014631){\makebox(0,0)[rt]{\lineheight{0}\smash{\begin{tabular}[t]{r}200\end{tabular}}}}%
    \put(0,0){\includegraphics[width=\unitlength,page=62]{PALT2bf1k.pdf}}%
    \put(0.53304112,0.27836161){\makebox(0,0)[rt]{\lineheight{0}\smash{\begin{tabular}[t]{r}400\end{tabular}}}}%
    \put(0,0){\includegraphics[width=\unitlength,page=63]{PALT2bf1k.pdf}}%
    \put(0.53304112,0.33657691){\makebox(0,0)[rt]{\lineheight{0}\smash{\begin{tabular}[t]{r}600\end{tabular}}}}%
    \put(0,0){\includegraphics[width=\unitlength,page=64]{PALT2bf1k.pdf}}%
    \put(0.53304112,0.39479228){\makebox(0,0)[rt]{\lineheight{0}\smash{\begin{tabular}[t]{r}800\end{tabular}}}}%
    \put(0,0){\includegraphics[width=\unitlength,page=65]{PALT2bf1k.pdf}}%
    \put(0.53304112,0.45300758){\makebox(0,0)[rt]{\lineheight{0}\smash{\begin{tabular}[t]{r}1000\end{tabular}}}}%
    \put(0,0){\includegraphics[width=\unitlength,page=66]{PALT2bf1k.pdf}}%
    \put(0.80578582,0.17860799){\makebox(0,0)[lt]{\lineheight{0}\smash{\begin{tabular}[t]{l}SBSC 3\end{tabular}}}}%
    \put(0.5,0.06656525){\makebox(0,0)[t]{\lineheight{0}\smash{\begin{tabular}[t]{c}89\% credible interval\end{tabular}}}}%
    \put(0.03913001,0.5879339){\rotatebox{90}{\makebox(0,0)[lt]{\lineheight{0}\smash{\begin{tabular}[t]{l}Calibration sample\end{tabular}}}}}%
  \end{picture}%
\endgroup%